\documentclass[aps, pra, groupedaddress, showkeys, twocolumn, floatfix, 10pt]{revtex4-2}
\usepackage{amssymb}
\usepackage{amsmath}
\usepackage{graphicx}
\usepackage[caption=false, subrefformat=parens,labelformat=parens,font=normalsize]{subfig}
\usepackage{booktabs}
\usepackage{longtable}
\usepackage{xcolor}
\usepackage{hyperref}
\newcommand*\mean[1]{\overline{#1}}
\newcommand*\meanNONDiag[1]{\widehat{#1}}

\newcommand{\dotsangle}{-15}
\newcommand{\dotsxshift}{.2ex}
\newcommand{\dotsyshift}{1ex}
\newcommand{\rdots}{\hspace{\dotsxshift}%
\raisebox{\dotsyshift}{\rotatebox{\dotsangle}{$\ddots$}}}

\begin{document}

\title{New Collectivity Measures for Financial Covariances and Correlations}

\author{Anton J. Heckens}
\email{anton.heckens@uni-due.de}
\author{Thomas Guhr}
\email{thomas.guhr@uni-due.de}
\affiliation{
	Fakult\"at f\"ur Physik, Universit\"at Duisburg-Essen, Duisburg, Germany		
}


\begin{abstract}
 Complex systems are usually non-stationary and their dynamics
  is often dominated by collective effects.  Collectivity, defined as
  coherent motion of the whole system or of some of its parts,
  manifests itself in the time-dependent structures of covariance and
  correlation matrices. The largest eigenvalue corresponds to the
  collective motion of the system as a whole, while the other large, isolated, eigenvalues indicate collectivity in parts of the system. In the case
  of finance, these are industrial sectors. By removing the collective
  motion of the system as a whole, the latter effects are much better revealed. 
  We measure a remaining collectivity to which we refer as average sector collectivity. 
  We identify
  collective signals around the Lehman Brothers crash and after the
  dot-com bubble burst.  For the Lehman Brothers crash, we find a
  potential precursor. We analyze 213 US stocks over a period
  of more than 30 years from 1990 to 2021.
  We plot the average sector collectivity versus the collectivity
  corresponding to the largest eigenvalue to study the whole market
  trajectory in a two dimensional space spanned by both collectivities. Therefore, we capture the average sector collectivity in a much more precise way. Additionally, we
  observe that larger values in the average sector collectivity are often
  accompanied by trend shifts in the mean covariances and mean
  correlations.  As of 2015/2016 the collectivity in the US stock
  markets changed fundamentally.
\end{abstract}

\maketitle

\section{\label{sec:Introduction}Introduction}

Non-stationarity is a characteristic feature of complex systems.  In
financial markets, this issue is of considerable importance for systemic
stability and it poses great challenges for risk
management~\cite{Mandelbrot1997,schwert1989does,LONGIN19953,mantegna1999introduction,bouchaud2003theory,Kwapien_2012,Kutner_2019}.

The measured time series without further processing yield covariance
and correlation matrices to which we refer as standard. The spectral
density typically features a bulk of smaller eigenvalues which are
compatible with a random matrix description~\cite{Marchenko_1967,Kwapien_2006} and
larger, isolated eigenvalues that contain system-specific
information~\cite{Laloux_1999,
	Noh_2000, Gopikrishnan_2001, Plerou_2002, Song_2011,
	MacMahon_2015,Stepanov_2015,Chetalova_2015,Benzaquen_2017,potters2020first}. The largest eigenvalue and the corresponding eigenvector are referred to as market mode in finance due to the collective motion of the system as a whole~\cite{MacMahon_2015,Allez_2012,Reigneron_2011}.  The other large outliers indicate collectivity in parts
of the system, in financial markets these are the industrial
sectors.
The dynamics is best identified with a moving time window to
obtain the time evolution of the correlation structures, including the
spectral properties. In Refs.~\cite{Heckens_2020,Heckens_2022}, we
put forward a new type of analysis by setting up a method to remove
the contributions specific for individual eigenvalues from the
standard covariance and correlation matrices. If the market mode
is removed, the remaining reduced-rank covariance
and correlation matrices allow for a much clearer analysis of the collectivity
in the industrial sectors.
Thus, our construction is likely to separate exogenous from endogenous effects.

We measured the average dynamical behavior of the reduced-rank
covariance matrices by the mean value of the matrix
elements~\cite{Heckens_2022}. The mean
reduced-rank covariances are potential measures for the average sector collectivity
as it represents a remaining collectivity to which each sector contributes as well as the collectivities induced by the relations between the sectors.  In
the analyzed time period from 1997 to 2012~\cite{Heckens_2022},
non-zero signals  emerge almost exclusively at time periods of endogenous
crises as in the vicinity of the Lehman Brothers crash and the dot-com
bubble burst. We identified these mean values as new measures for
endogenous
risk~\cite{sornette2003causes,bouchaud2010endogenous,Danielsson_2012,Bouchaud_2011}.

Moreover, we found~\cite{Heckens_2022} that the mean reduced-rank covariances are relatively large in crises periods
in which we detected corresponding market states.
In contrast to other market state definitions or so-called ``regimes'' in finance~\cite{Hamilton_1989,Marsili_2002,Procacci_2019} and measures such as the Sector Dominance Ratio (SDR)~\cite{Uechi_2015}
those market states are determined by clustering correlation
matrices~\cite{munnixIdentifyingStatesFinancial2012,Stepanov_2015,Rinn_2015,Chetalova_2015,Chetalova_2015_2,Stepanov_2015_MultiAsset,papenbrockHandlingRiskonRiskoff2015,jurczyk2017measuring,Pharasi_2018,Qiu_2018,pharasi2020market,pharasi2020dynamics,Heckens_2020,pharasi2021dynamics,marti2021review,Heckens_2022}.
The market operates in a state for a certain time, then jumps to
another one and also might return to a previous state.  This concept
of market states was applied in other fields like epileptic
seizures~\cite{rings2019traceability}, freeway
traffic~\cite{Wang_2020} and offshore wind
farms~\cite{bette2021nonstationarity}.

In the literature, the average sector collectivity is almost never taken into
account due to its seemingly negligible order of
magnitude~\cite{Borghesi_2007}.  One exception is Kenett et al.~\cite{kenett2011index} who separated the index
behavior from partial
correlations~\cite{shapira2009index,kenett2010dominating,Kenett_2015}. 
Partial correlations are essentially correlations computed from residuals after removing a common trend like an index from measured time series by linear regressions~\cite{anderson2003introduction,wiki:2021:PartialCorr}.
In Ref.~\cite{kenett2011index}, observables exclusively derived from correlation matrices were studied.
Our results are different using a spectral decomposition for standard covariance and correlation matrices.
Here, we analyze both collectivities, the one corresponding to the market mode and the average sector collectivity to study its potential economic significance for systemic risk~\cite{kritzman2011principal,Bisias_2012,zheng2012changes,Billio_2012,beguvsic2018information,Huang_2013,Musmeci_2016}.

This paper is organized as follows.
In Sec.~\ref{sec:DataSet}, we introduce the daily data set for our analysis. 
In Sec.~\ref{sec:NewCollectivityMeasures}, we derive the reduced-rank covariances matrices and discuss and define new measures for collectivity. We introduce new diagrams with regard to the collectivities in Sec.~\ref{sec:DataAnalysisAndRiskPhaseDiagrams}. 
We conclude in Sec.~\ref{sec:Conclusion}.

\section{\label{sec:DataSet}Data set}

We downloaded 213 US stocks closing prices $S_i(t)$ from Refinitiv Workspace for Students (formals EIKON)~\cite{Refinitiv} representing stocks from the S\&P~500 index. 
The ticker symbols of the stocks (including additional information) are listed in App.~\ref{sec:ListStocks}.
The period ranges from \mbox{1990-01-02} to \mbox{2021-08-06} (Year-Month-Day).
All 11 Global Industry Classification Standard (GICS) are represented by the chosen stocks~\cite{wiki:2020:GICS}.
We also downloaded daily closing data of the S\&P~500 index from Refinitiv Workspace~\cite{Refinitiv}.
From the closing prices $S_i(t)$ we calculate the \mbox{daily logarithmic returns ($\Delta t = 1$)}
\begin{equation} \label{eqn:LogReturn}
	G_i(t) =  \ln  \frac{S_i(t+\Delta t)}{S_i(t)} , \hspace{0.5cm} i = 1, 
	\ldots, K
	\,.
\end{equation}
We arrange the return time series in the data matrix $G$ as rows
\begin{equation} \label{eqn:DatamatrixG}
	G_{\text{tot}} = \begin{bmatrix}  G_1(1) & \dots & G_1(T_{\text{tot}})  \\
		\vdots & & \vdots \\
		G_i(1) & \dots & G_i(T_{\text{tot}}) \\
		\vdots & & \vdots \\
		G_{K}(1) & \dots & G_K(T_{\text{tot}}) 
	\end{bmatrix}
\end{equation}
with $K= 213$, being the number of stocks and $T_\text{tot} = 7961$,
being the total number of points in the return time series of a stock. 
Since we are interested in covariances and correlations, missing values are filled up by zeros in the returns (not in the prices).
Thus, non-zero signals in covariances and correlations are not caused by filled missing values.

In the following, we use a sliding window of $T_\text{sub} = 42$~trading days shifted by one trading day to create 7920~$K \times T_\text{sub}$  subblocks of the data matrix $G_\text{tot}$.
Consequently, we achieve a high time resolution of the temporal dynamics.

\section{\label{sec:NewCollectivityMeasures}New collectivity measures}

We introduce reduced-rank covariance matrices in Sec.~\ref{sec:ReducedRankCovarianceMatrices} and thoroughly discuss  mean values derived from these matrices as collectivity measures in Sec.~\ref{sec:DiscussingMeanValuesAsMeasuresForCollectivity}. 
We define new measures for collectivity in Sec.~\ref{sec:NewCollect}.

\subsection{\label{sec:ReducedRankCovarianceMatrices}Reduced-rank covariance matrices}

\begin{table}[!htb]
	\centering
	\caption{\label{tab:FinancialCrises}
		Historical events taken 
		from~Ref.~\cite{wiki:2021:LISTSTOCKSCRASHES} and one (estimated) precursor start (A) shown as dashed line in Fig.~\ref{subfig:Main:MeanValCovP2}.
	}
	\begin{tabular}{cl@{\hspace{1em}}c}
		\toprule
		Label &
		Crisis &
		\multicolumn{1}{c}{Date}
		\\
		&
		&
		\multicolumn{1}{c}{(Year-Month-Day)} \\
		\midrule
		(ER) & Early 1990s recession & 1990-07-15 \\
		(AC) & Asian financial crisis & 1997-10-27 \\
		(RC) & Russian financial crisis & 1998-08-17 \\
		(DC) & Dot-com bubble (before burst) & 2000-03-10 \\
		(MD) & Stock market downturn of 2002 & 2002-10-09 \\
		(A) & Precursor start & 2007-11-01 \\
		(LB) & Lehman Brothers crash & 2008-09-16 \\
		(ED) & European debt crisis & 2010-04-27 \\
		(AF) & August 2011 stock markets fall & 2011-08-01 \\
		(FC) & The Great Fall of China & 2015-08-18 \\
		(CO) & 2020 stock market crash & 2020-02-24 \\
		\bottomrule
	\end{tabular}
\end{table}

We define the standard covariance matrix as
\begin{equation} \label{eqn:CovarianceMatAADagger}
	\Sigma = \frac{1}{T} A A^{\dagger} \,,
\end{equation}
where the rows of the $K \times T_\text{sub}$  data matrix $A$ are normalized to mean value zero.
The symbol $\dagger$ indicates the transpose of a matrix.
The standard correlation matrix is defined as 
\begin{equation} \label{eqn:CorrelationMatMMDagger}
	C = \frac{1}{T} M M^{\dagger} \,,
\end{equation}
where the rows of  the $K \times T_\text{sub}$ matrix $M$ are normalized to mean value zero and standard deviation one.
For $\Sigma$, we calculate the mean (standard) covariance
\begin{equation} \label{eqn:MeanCovariance}
	\mean{\text{cov}} = \frac{1}{K^2} \sum_{i,l = 1}^K \Sigma_{il} \,,
\end{equation}
including all diagonal elements of $\Sigma$.
Analogously to Eq.~\eqref{eqn:MeanCovariance}, we compute the mean correlation $\mean{\text{corr}}$ including the unities on the main diagonal.
We split up the standard covariance matrix into two parts
\begin{align} \label{eqn:CovarianceMatSpectralDecomp}
	\Sigma = \sum_{i=1}^K \kappa_i u_i u^{\dagger}_i &= \kappa_K u_K u^{\dagger}_K + \sum_{i=1}^{K-1} \kappa_i u_i u^{\dagger}_i   \\
	&= \Sigma_{\text{BLE}} + \Sigma_{B} 
\end{align}
using the spectral decomposition with eigenvalues $\kappa_i$ and eigenvectors $u_i$. 
The covariance matrix $\Sigma_{\text{BLE}}$ belongs to the  largest eigenvalue $\kappa_K$ and the corresponding eigenvector $u_K$, \emph{i.e.} the market mode~\cite{MacMahon_2015,Allez_2012,Reigneron_2011}. The covariance matrix $\Sigma_{B}$ containing all other modes is dominated by the remaining large, isolated eigenvalues outside the bulk of the spectral density. For financial markets, these outliers correspond to the industrial sectors~\cite{Laloux_1999, Noh_2000, Gopikrishnan_2001, Plerou_2002, Song_2011, MacMahon_2015,  Stepanov_2015,Chetalova_2015,Benzaquen_2017,potters2020first}.   
The matrices $\Sigma_{\text{BLE}}$ and $\Sigma_B$ are both well-defined covariance matrices with a reduced rank.
For the sake of clarity, we only refer to $\Sigma_{{B}}$ as reduced-rank covariance matrix.

We apply the spectral decomposition to the standard correlation matrix
\begin{align} \label{eqn:CorrelationMatSpectralDecomp}
	C = \sum_{i=1}^K \lambda_i x_i x^{\dagger}_i &= \lambda_K x_K x^{\dagger}_K + \sum_{i=1}^{K-1} \lambda_i x_i x^{\dagger}_i   \\
	&= \Sigma_{\text{LLE}} + \Sigma_{L} \,,
\end{align}
with eigenvalues $\lambda_i$ and eigenvectors $x_i$.
We notice that the reduced-rank matrices $\Sigma_{\text{LLE}}$
and $\Sigma_L$ are not correlation matrices any more, because the separation destroys the proper normalization. However, they are well-defined covariance matrices.
We refer only to $\Sigma_{L}$ as reduced-rank covariance matrix.

\subsection{\label{sec:DiscussingMeanValuesAsMeasuresForCollectivity}Discussing mean values as measures for collectivity}

\begin{figure*}[!htb]
	\centering
	\begin{minipage}{1.0\textwidth}
		\subfloat%
		{\includegraphics[width=1.0\linewidth]{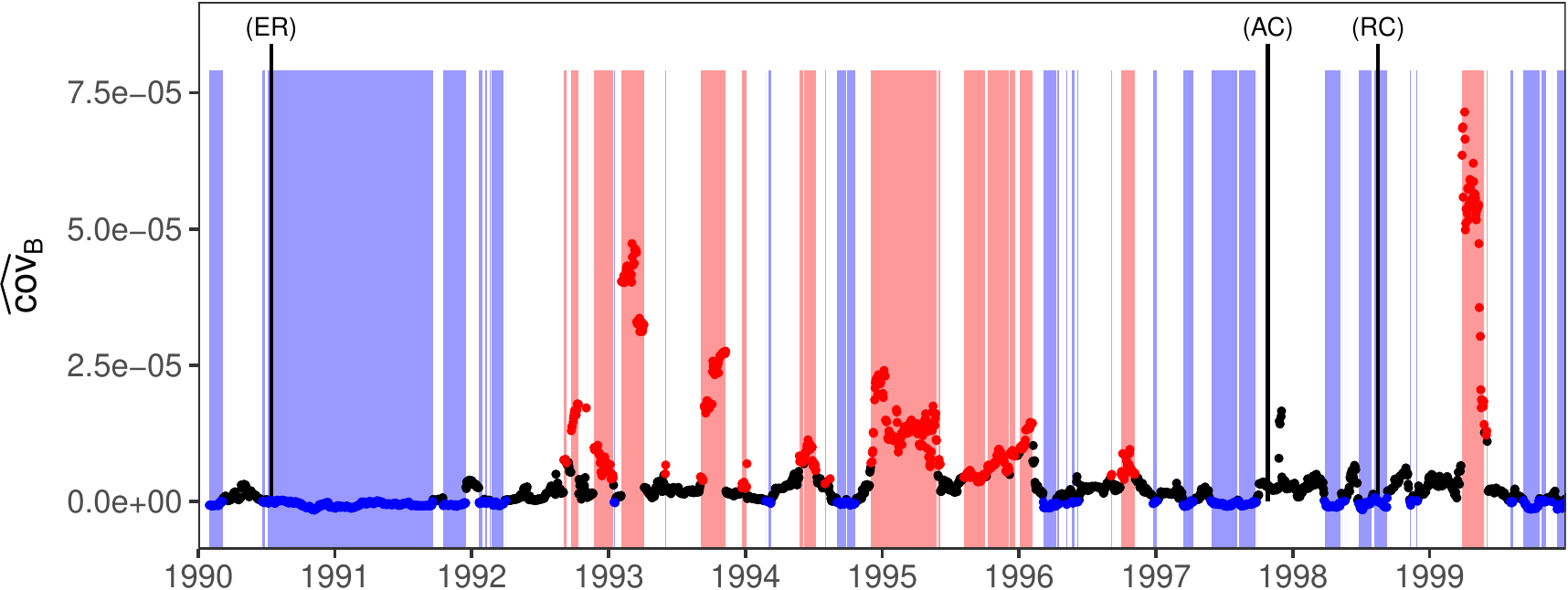}
		}\par
		\subfloat%
		{\includegraphics[width=1.0\textwidth]{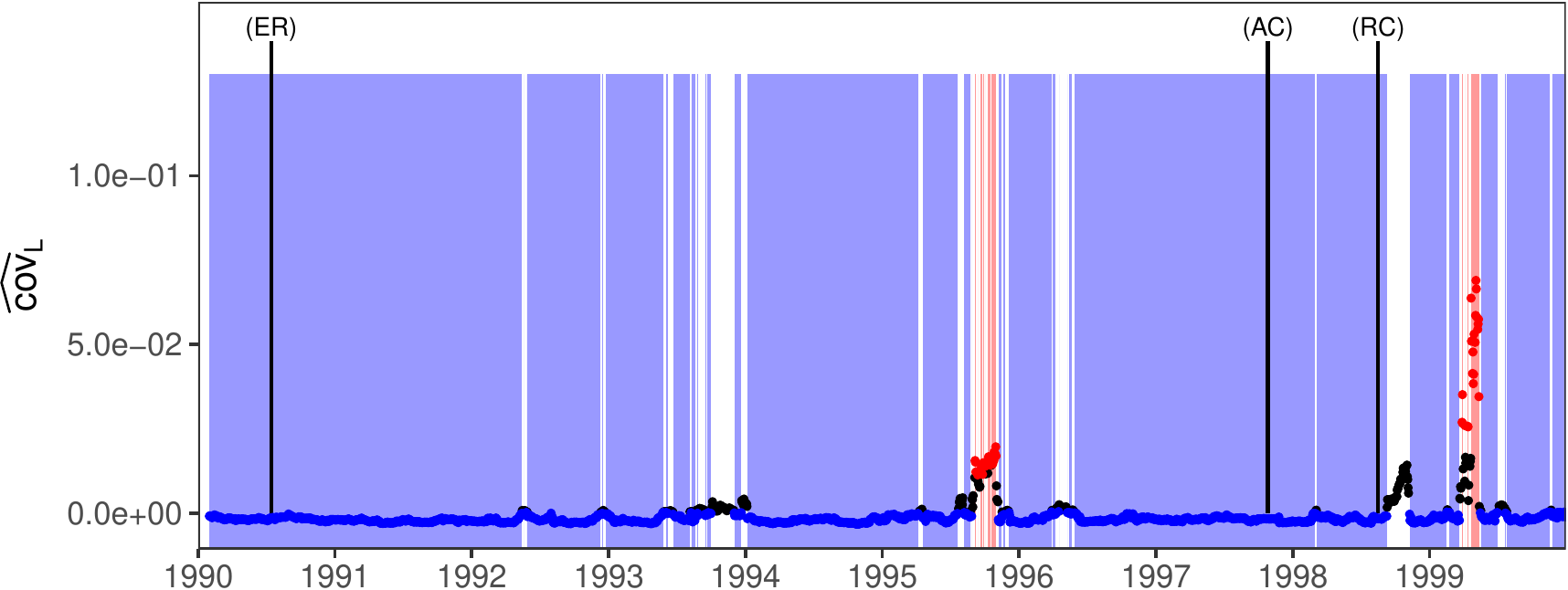}
		}%
	\end{minipage}%
	\caption{\label{subfig:Main:MeanValCovP1}Time evolutions (1990-1999) for the mean reduced-rank covariances $\meanNONDiag{\text{cov}}_{{B}}$ (top) and $\meanNONDiag{\text{cov}}_{{L}}$ (bottom).
		Three criteria (red/blue) for relative collectivity measures are described in Sec.~\ref{sec:AverageSectorCollectivity}. Historical events are listed in Tab.~\ref{tab:FinancialCrises}. Black dots belong to covariance matrices whose collectivities do not fulfill the three criteria.}
\end{figure*}
\begin{figure*}[!htb]
	\centering
	\begin{minipage}{1.0\textwidth}
		\subfloat%
		{\includegraphics[width=1.0\linewidth]{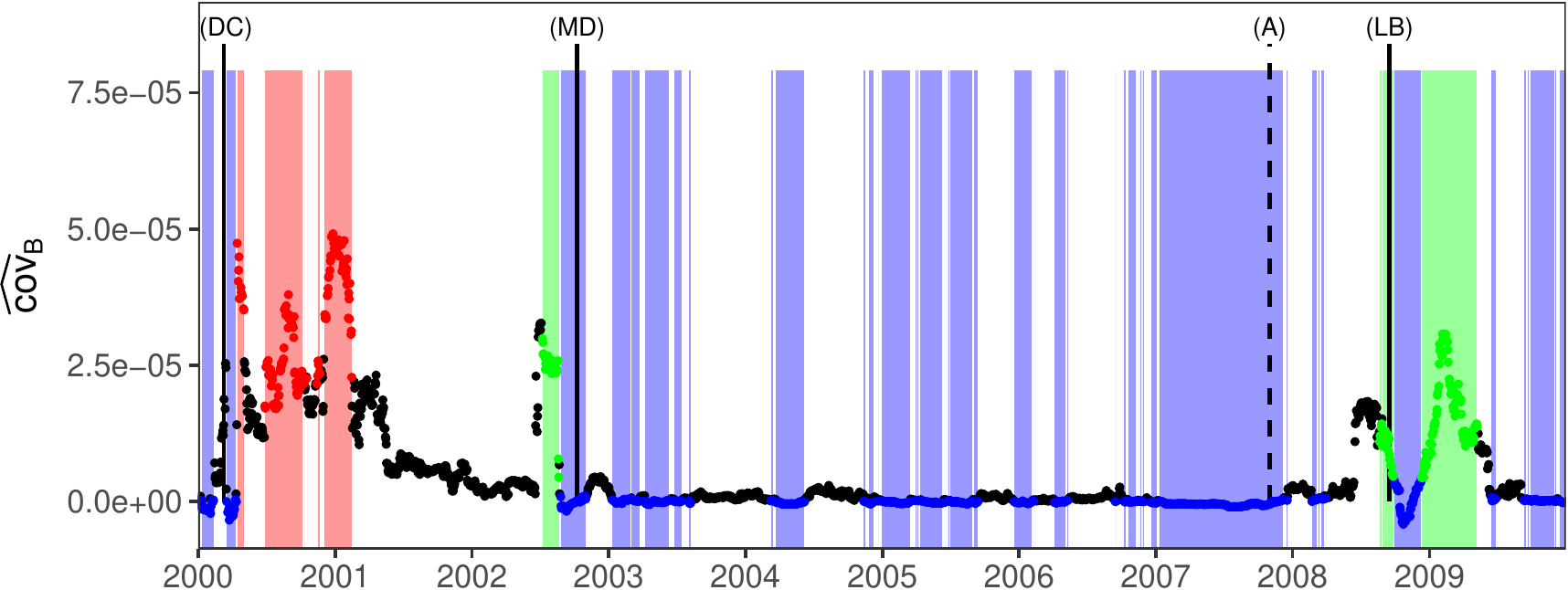}
		}\par
		\subfloat%
		{\includegraphics[width=1.0\textwidth]{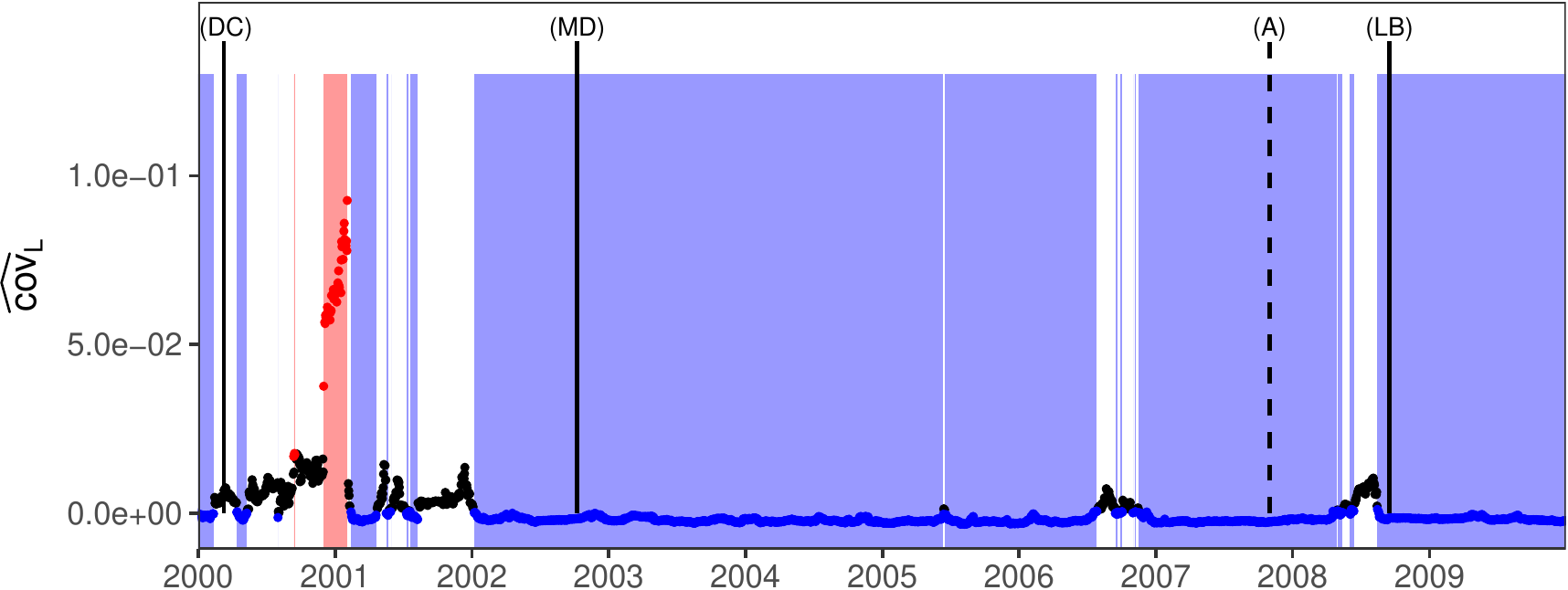}
		}%
	\end{minipage}%
	\caption{\label{subfig:Main:MeanValCovP2}Time evolutions (2000-2009) for the mean reduced-rank covariances $\meanNONDiag{\text{cov}}_{{B}}$ (top) and $\meanNONDiag{\text{cov}}_{{L}}$ (bottom).
		Three criteria (red/blue/green) for absolute and relative collectivity measures are described in Sec.~\ref{sec:AverageSectorCollectivity}. Historical events are listed in Tab.~\ref{tab:FinancialCrises}. Black dots belong to covariance matrices whose collectivities do not fulfill the three criteria.}
\end{figure*}
\begin{figure*}[!htb]
	\centering
	\begin{minipage}{1.0\textwidth}
		\subfloat%
		{\includegraphics[width=1.0\linewidth]{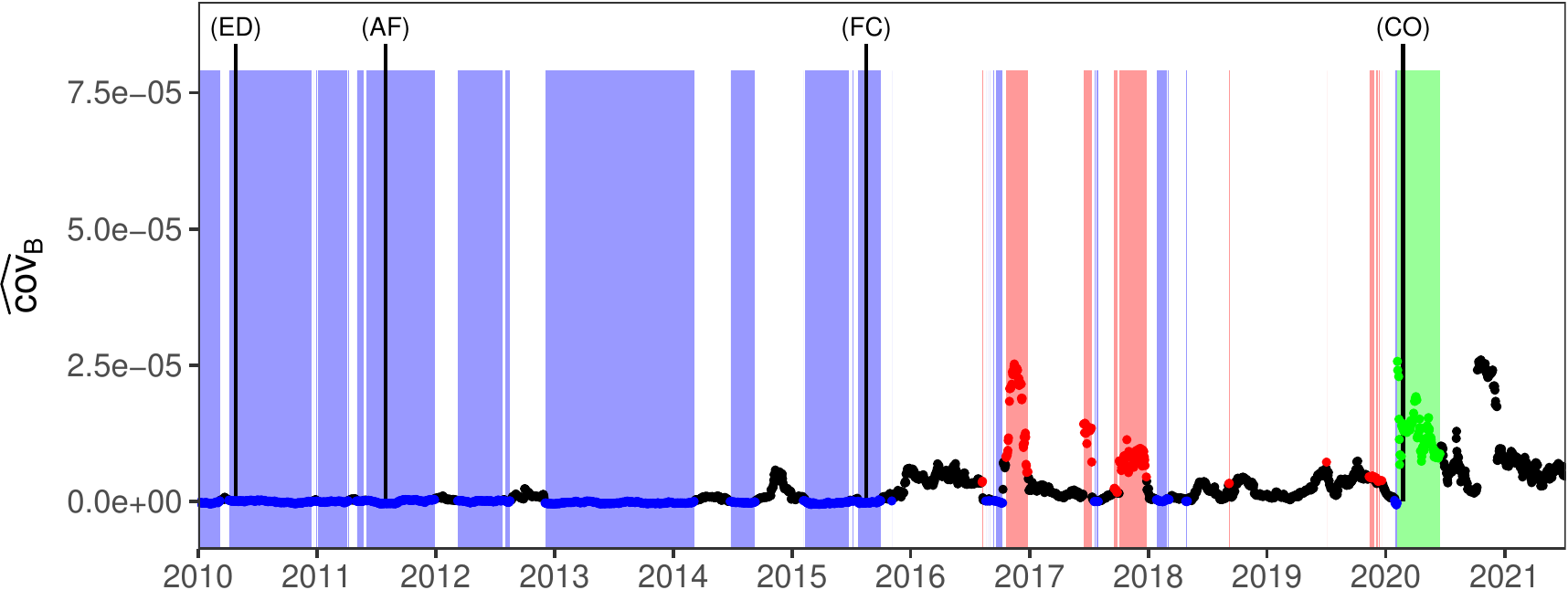}
		}\par
		\subfloat%
		{\includegraphics[width=1.0\textwidth]{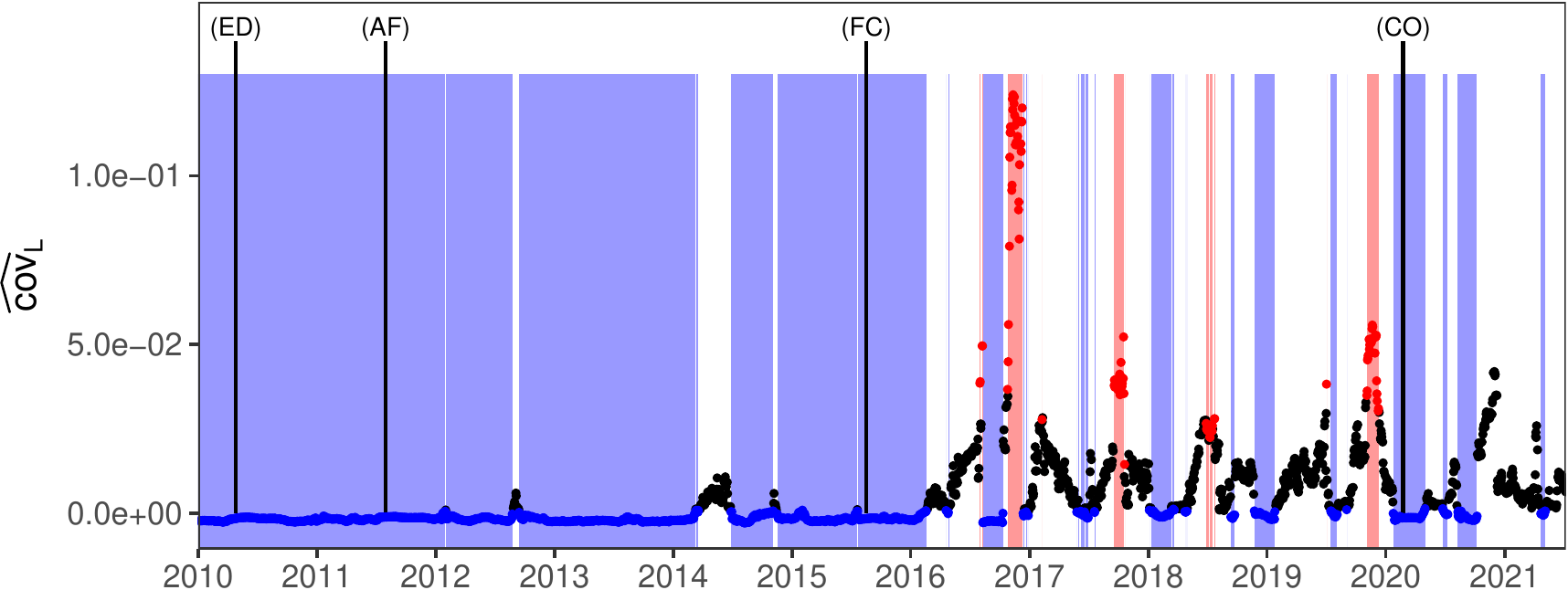}
		}%
	\end{minipage}%
	\caption{\label{subfig:Main:MeanValCovP3}Time evolutions (2010-2021) for the mean reduced-rank covariances $\meanNONDiag{\text{cov}}_{{B}}$ (top) and $\meanNONDiag{{cov}}_{\text{L}}$ (bottom).
		Three criteria (red/blue/green) for absolute and relative collectivity measures are described in Sec.~\ref{sec:AverageSectorCollectivity}. Historical events are listed in Tab.~\ref{tab:FinancialCrises}. Black dots belong to covariance matrices whose collectivities do not fulfill the three criteria.}
\end{figure*}
\begin{figure*}[!htb]
	\centering
	\begin{minipage}{0.5\textwidth}
		\subfloat[\label{subfig:RiskPhaseCovMean}]
		{\includegraphics[width=1.0\textwidth]{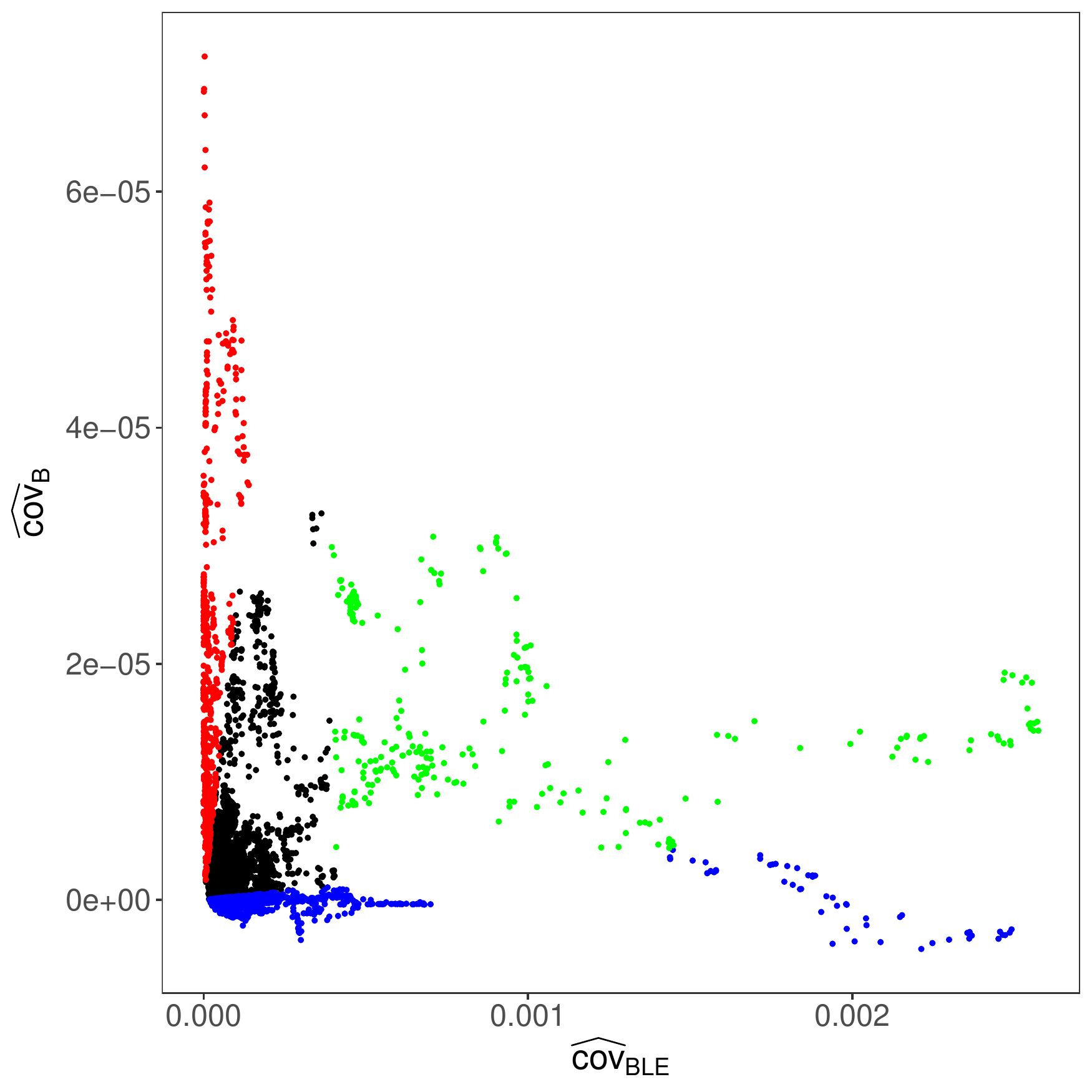}
		}
	\end{minipage}%
	\begin{minipage}{0.5\textwidth}
		\subfloat[\label{subfig:RiskPhaseCorrMean}]
		{\includegraphics[width=1.0\textwidth]{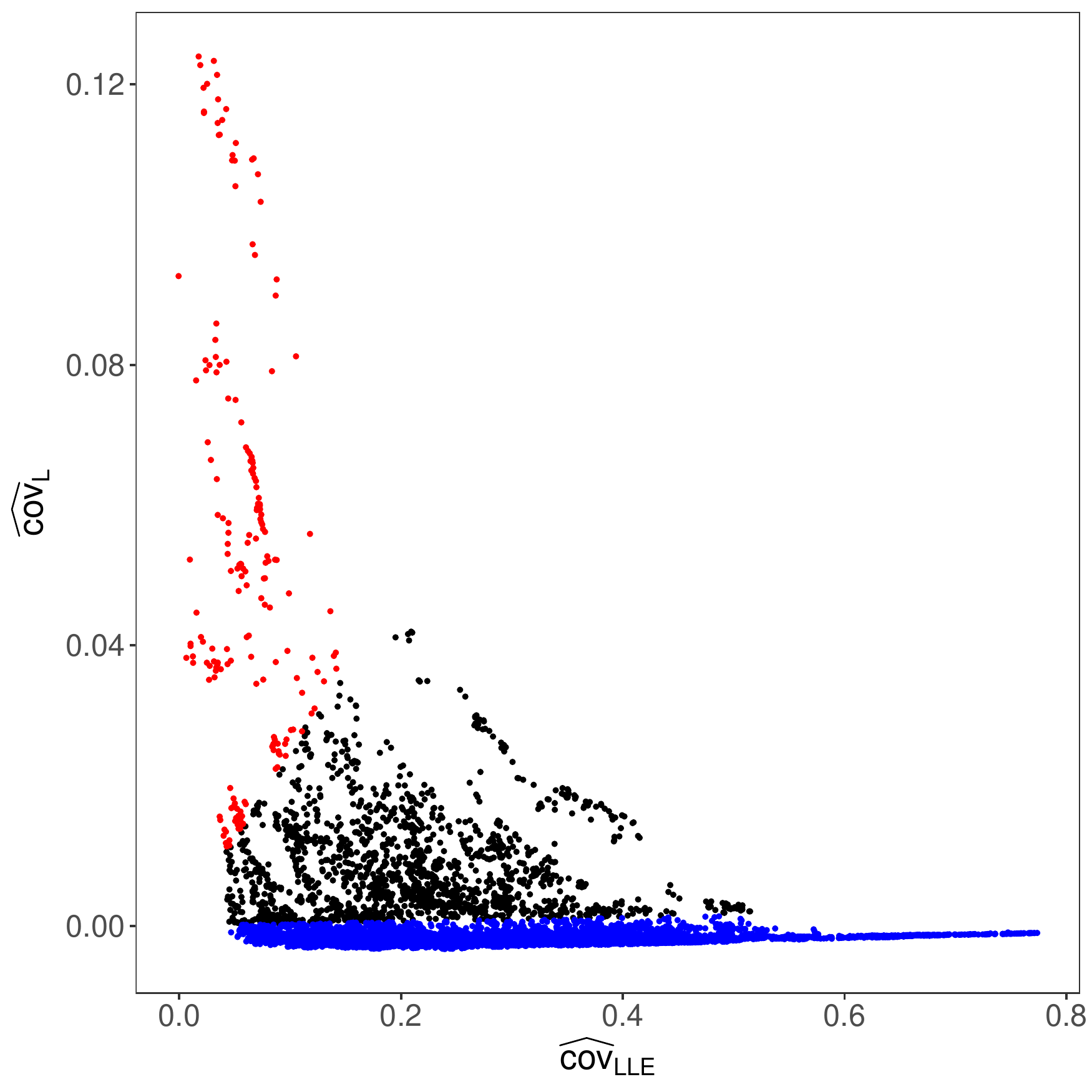}
		}
	\end{minipage}
	\caption{\label{subfig:Main:RiskPhaseMean}Risk-phase diagrams derived from \protect\subref{subfig:RiskPhaseCovMean} covariance matrix $\Sigma$ with $\meanNONDiag{\text{cov}}_{{B}}$ plotted versus $\meanNONDiag{\text{cov}}_{\text{BLE}}$ and \protect\subref{subfig:RiskPhaseCorrMean} correlation matrix $C$ with $\meanNONDiag{\text{cov}}_{{L}}$ plotted versus $\meanNONDiag{\text{cov}}_{\text{LLE}}$.  Each dot corresponds to one of the 7920 intervals of 42 trading days (see~Sec.~\ref{sec:DataSet}). 
		The colored dots (red/blue/green) correspond to the three criteria for absolute and relative collectivity measures described in Sec.~\ref{sec:AverageSectorCollectivity}.
		The green ones are only displayed in \protect\subref{subfig:RiskPhaseCovMean}.
		Black dots belong to covariance matrices whose collectivities do not fulfill the three criteria.}
\end{figure*}

For the market mode, the corresponding eigenvector has entries which are usually all positive and fluctuate only little about $1/\sqrt{K}$ for large covariance and correlation matrices, \textit{i.e.} all stocks move synchronously. The other eigenvectors corresponding to the other modes empirically show positive and negative elements indicating coherent movements of the stocks.
A high coherence means a strong collective behavior of all stocks.

A frequently used measure for collectivity is the inverse participation ratio (IPR)
\begin{equation} \label{eqn:InverseParticipationRatio}
	 I_i = \sum_{j=1}^K u_{ji}^4 \,, 
\end{equation}
summing over all $K$ components of eigenvector $u_i$~\cite{Plerou_2002,Plerou_1999,wang2021collective}.
There are two extreme scenarios.
For $u_{ji} = 1/\sqrt{K}$ we obtain $I_i = 1/K$ indicating extended vectors, $\textit{i.e.}$ all components contribute 
to $I_i$ equally. In the other extreme scenario, the localized vector $u_{ji} = \delta_{j1,i}$ corresponds to  $I_i =1$.
The value $I_i^{-1}$ is considered to be the number of stocks contributing to the $i$-th mode, \textit{i.e.}
$I_i$ is used as a threshold between important and noised vector components. 
Thus, the market mode has a IPR indicating that all components contribute approximately equally to $I_i$.

The IPR does not take the following points into account.
First, the dyadic matrices in Eqs.~\eqref{eqn:CovarianceMatSpectralDecomp} and~\eqref{eqn:CorrelationMatSpectralDecomp} are composed of two time-dependent quantities, the eigenvalues and the eigenvectors.
Second, effects of different eigenvectors and the corresponding collectivity are not taken into account.
Third, the covariances and correlations between different industrial sectors are not directly taken into account either, especially for the modes corresponding to industrial sectors.

To develop an intuition for the collectivity measures introduced above, we consider for a moment ensemble averages by replacing
\begin{equation} \label{eqn:RMWishart}
	\Sigma \longrightarrow \widetilde{\Sigma} = \frac{1}{T} \widetilde{A} {\widetilde{A}}^{\dagger} \,,
\end{equation}
where the $K \times T$ random matrices $\widetilde{A}$ are drawn from a random matrix probability density such as
\begin{equation} \label{eqn:RMDist}
	w(\widetilde{A}|\Sigma_0) = \frac{1}{\text{det}^{T/2} \left( 2 \pi \Sigma_0 \right) } \exp\left( - \frac{1}{2} \text{tr} \widetilde{A}^{\dagger} \Sigma_0^{-1} \widetilde{A} \right)\,.
\end{equation}
The following line of reasoning does not depend on the Gaussian form of \eqref{eqn:RMDist}, it also applies to heavy-tails random matrix probabilities $w(\widetilde{A}| \Sigma_0)$ such as the matrix variate $t$-distribution~\cite{gupta2018matrix}. The important ingredient is the structure covariance matrix $\Sigma_0$, in mathematical statistics often referred to as population covariance matrix.
By construction, the ensemble average of \eqref{eqn:RMWishart} with $w(\widetilde{A}|\Sigma_0)$ yields
\begin{equation} \label{eqn:RMWishartMean}
	 \langle \widetilde{\Sigma} \rangle_{\text{en}} = \frac{1}{T} \left\langle \widetilde{A} \widetilde{A}^{\dagger} \right\rangle_{\text{en}}  = \Sigma_0   \, 
\end{equation}
for all dimension $K,T$. Hence, the ensemble averaged mean covariance \eqref{eqn:MeanCovariance} is given by
\begin{equation} \label{eqn:RMMeanCovariance}
	  \langle  \mean{\text{cov}}  \rangle_{\text{en}}  = \frac{1}{K^2} \sum_{i,j = 1}^K \left\langle \Sigma_{ij} \right\rangle_{\text{en}} = \frac{1}{K^2} \sum_{i,j = 1}^K \Sigma_{0ij} \,.
\end{equation}
In a one-factor model~\cite{sharpe1963simplified,ROSS1976,Noh_2000,Guhr_2003,kenett2009rmt}, for example, the structure covariance matrix has block-diagonal form
\begin{equation} \label{eqn:BlockMatrix}
\Sigma_0	
=		
\setlength{\arraycolsep}{0pt}
\setlength{\delimitershortfall}{0pt}
\newcommand*{\myfbox}[1]{%
	\fcolorbox{gray}{gray!20!white}{$#1$}%
}
\begin{bmatrix}
	\,\myfbox{b_1} &  &  &\,  \\
	\, & \myfbox{b_2} &  & \, \\
	\, &  & \myfbox{b_3} & \, \\
	\, &  &   &  \rdots \, \\
\end{bmatrix} \,,
\end{equation}
where each of the blocks $b_1, b_2, b_3, \ldots$ indicates collectivity in parts of the system such as industrial sectors in finance.
The entries outside the blocks are zero.
The entries in these blocks are positive, implying that $\langle \mean{\text{cov}} \rangle$  according to \eqref{eqn:RMWishartMean} is the larger, the stronger the collectivity in the sectors. In the random matrix settings, this line of reasoning applies to arbitrary dimensions $K$, due to the ensemble average. However, if we now consider one individual covariance matrix from a one-factor model, the entries outside the blocks will not be zero as in \eqref{eqn:BlockMatrix},
they will be random numbers. They fluctuate around zero and thus average to zero in $\mean{\text{cov}}$ defined in \eqref{eqn:MeanCovariance}, provided the dimension $K$ is large.
This self-averaging for large $K$ allows one to apply the intuition gained from the random matrix model.
This line of reasoning remains valid, if we add the ``market mode'', \textit{i.e.} if we overlay \eqref{eqn:BlockMatrix} with a matrix with the same numerical value in all entries. The entries outside the blocks in \eqref{eqn:BlockMatrix} then fluctuate around a non-zero number whose value comes out the sharper, the larger $K$ when averaging a single matrix.

Nevertheless, the diagonal elements $\Sigma_{ii}$ of every covariance matrix are always positive and related to volatility
clustering \cite{Mandelbrot1997,DING_1993}.
To remove this effect and to be sensitive to block structure only, we properly redefine the collectivity measure.

\subsection{\label{sec:NewCollect}Definitions of new collectivity measures}

To be only sensitive to block structures, we define mean values for the covariance and correlation matrices introduced in Sec.~\ref{sec:ReducedRankCovarianceMatrices} without diagonal elements.
In the case of the standard covariance matrix $\Sigma$ the mean value now reads
\begin{equation} \label{eqn:MeanCovariance_NONDiag}
	\meanNONDiag{\text{cov}} = \frac{1}{K (K-1)} \sum_{i \neq j}^K \Sigma_{ij} \,.
\end{equation}
Using Eqs.~\eqref{eqn:CovarianceMatSpectralDecomp} and~\eqref{eqn:CorrelationMatSpectralDecomp}, we arrive at
\begin{equation} \label{eqn:MeanCovariancesEquation}
	\meanNONDiag{\text{cov}} = \meanNONDiag{\text{cov}}_{\text{BLE}} + \meanNONDiag{\text{cov}}_{B}
\end{equation}
and
\begin{equation} \label{eqn:MeanCorrelationEquation}
	\meanNONDiag{\text{corr}} = \meanNONDiag{\text{cov}}_{\text{LLE}} + \meanNONDiag{\text{cov}}_{L} \,.
\end{equation}
We refer to these quantities as \textit{absolute} measures for collectivity or absolute collectivity.
The mean values $\meanNONDiag{\text{cov}}_{\text{BLE}}$ and $\meanNONDiag{\text{cov}}_{\text{LLE}}$ are measures
for the collectivity corresponding to the market mode.
The mean values $\meanNONDiag{\text{cov}}_{\text{B}}$ and $\meanNONDiag{\text{cov}}_{\text{L}}$ measure the average sector collectivity within and between the sectors.

We also want to analyze how the mean covariance and correlation are distributed between the different dyadic matrices.
Therefore, it is useful to divide Eqs.~\eqref{eqn:MeanCovariancesEquation} and~\eqref{eqn:MeanCorrelationEquation} by the mean values $\meanNONDiag{\text{cov}}$ and $\meanNONDiag{\text{corr}}$,
\begin{equation} \label{eqn:MeanCovariancesEquationRel}
	1 = \frac{\meanNONDiag{\text{cov}}_{\text{BLE}}}{\meanNONDiag{\text{cov}}} + \frac{\meanNONDiag{\text{cov}}_{B}}{\meanNONDiag{\text{cov}}}
\end{equation}
and
\begin{equation} \label{eqn:MeanCorrelationEquationRel}
	1 = \frac{\meanNONDiag{\text{cov}}_{\text{LLE}}}{\meanNONDiag{\text{corr}} } + \frac{\meanNONDiag{\text{cov}}_{L}}{\meanNONDiag{\text{corr}} } \,.
\end{equation}
We refer to the  right-hand-side expressions as \textit{relative} measures
$\meanNONDiag{\text{cov}}_{\text{BLE}}   /  \meanNONDiag{\text{cov}}  $, $\meanNONDiag{\text{cov}}_{B}   /  \meanNONDiag{\text{cov}} $, $\meanNONDiag{\text{cov}}_{\text{LLE}}   /  \meanNONDiag{\text{corr}} $ and
$\meanNONDiag{\text{cov}}_{L} / \meanNONDiag{\text{corr}}$~  
for collectivity or relative collectivities.

\section{\label{sec:DataAnalysisAndRiskPhaseDiagrams}Data Analysis and Risk-Phase Diagrams}

We show the dynamics of the average sector collectivities in Sec.~\ref{sec:AverageSectorCollectivity}.
In Sec.~\ref{sec:RiskPhaseDiagramsAndTrajectories}, we introduce a graphical representation of the collectivity.
In Sec.~\ref{sec:RiskPhaseLinRegress}, we show risk-phase diagrams where we compute the average sector collectivity of correlation matrices obtained by two linear regression approaches. Moreover, we evaluate the collective behavior of dyadic matrices associated with eigenvalues smaller than the second largest eigenvalue by means of our new collectivity measures in Sec.~\ref{sec:HigherOrderEffects}.
We discuss the implications for the mean covariances and correlations in Sec.~\ref{sec:RelevanceRelCollec}.

\subsection{\label{sec:AverageSectorCollectivity}Dynamics of the average sector collectivity}

\begin{figure*}[!htb]
	\centering
	\begin{minipage}{0.5\textwidth}
		\subfloat[\label{subfig:RiskPhaseCovMeanPeriod}]
		{\includegraphics[width=1.0\textwidth]{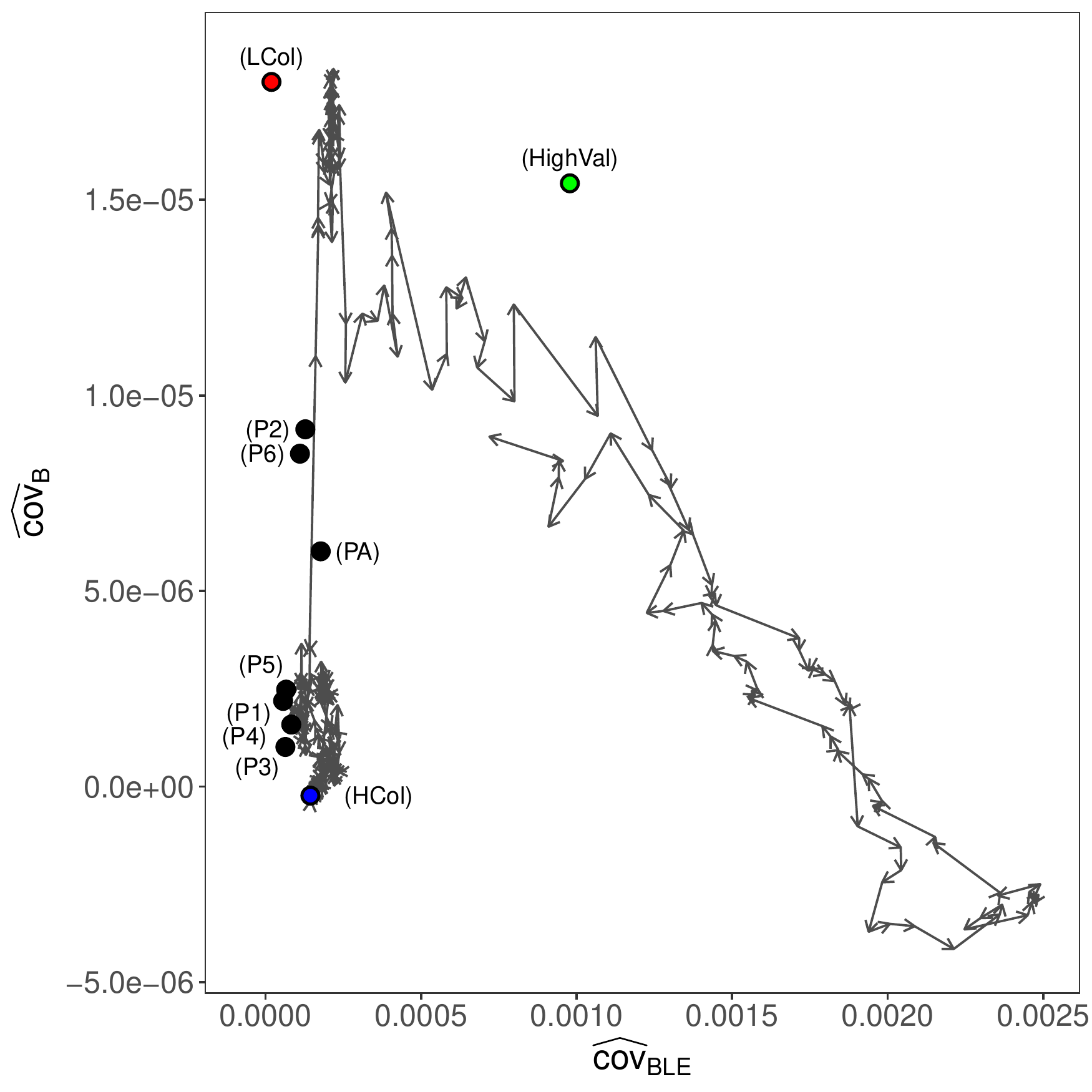}
		}
	\end{minipage}%
	\begin{minipage}{0.5\textwidth}
		\subfloat[\label{subfig:RiskPhaseCorrMeanPeriod}]
		{\includegraphics[width=1.0\textwidth]{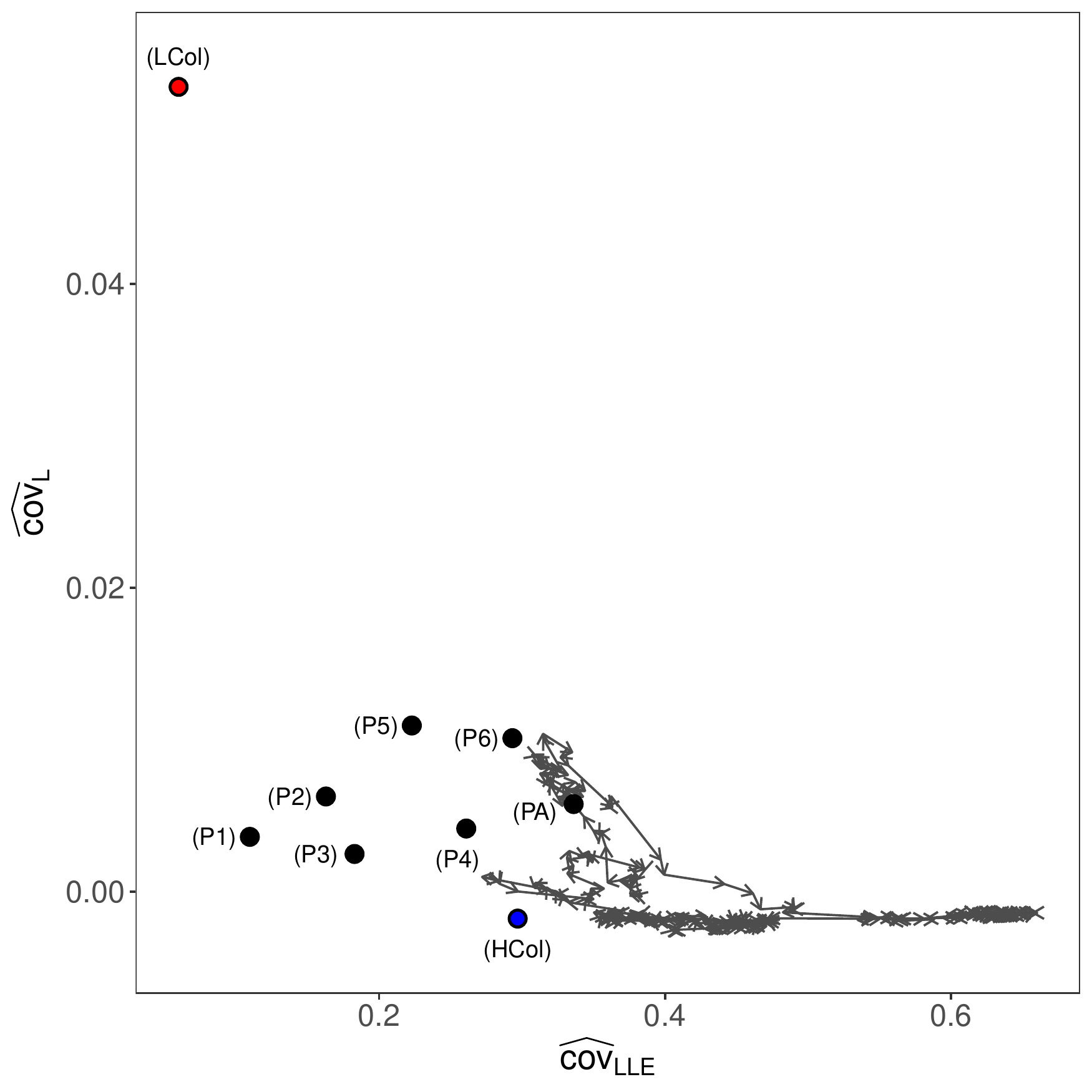}
		}
	\end{minipage}
	\caption{\label{subfig:Main:RiskPhaseMeanPeriod}Risk-phase diagrams derived from \protect\subref{subfig:RiskPhaseCovMeanPeriod} covariance matrix $\Sigma$ and \protect\subref{subfig:RiskPhaseCorrMeanPeriod} correlation matrix $C$.
		The colored dots (red, blue and green) correspond to three criteria for absolute and relative collectivity measures described in Sec.~\ref{sec:AverageSectorCollectivity}, additionally denoted by (LCol), (HighCol) and (HighVal). 
		The other seven dots belong to groups of covariance and correlation matrices described in Sec.~\ref{sec:AverageSectorCollectivity}.
		Market trajectories are shown from \mbox{2007-11-01} to \mbox{2008-12-31}. The arrows point into the directions into which the market moves next.}
\end{figure*}

In Figs.~\ref{subfig:Main:MeanValCovP1},~\ref{subfig:Main:MeanValCovP2} and~\ref{subfig:Main:MeanValCovP3}, the time evolutions of the mean values $\meanNONDiag{\text{cov}}_{{B}}$ and $\meanNONDiag{\text{cov}}_{{L}}$ are depicted for the time periods 1990-1999, 2000-2009 and 2010-2021.
We assign to each mean value a time stamp which corresponds to
the center of a 42 trading day interval and which is represented by a colored symbol.
In Tab.~\ref{tab:FinancialCrises}, we list historical events.
Event (A) denotes the estimated precursor start for the mean value 
$\meanNONDiag{\text{cov}}_{{B}}$ (cf.~Fig.~\ref{subfig:Main:MeanValCovP2}).
In Ref.~\cite{Heckens_2022}, we interpreted this signal and the corresponding signal in $\meanNONDiag{\text{cov}}_{{L}}$ as precursors for the Lehman Brothers crash inducing a spill-over effect from the financial sector in the US stock markets to the entire US stock market.

Our goal is to classify certain peaks or time periods with very small values 
in Figs.~\ref{subfig:Main:MeanValCovP1},~\ref{subfig:Main:MeanValCovP2} and~\ref{subfig:Main:MeanValCovP3}
by using three criteria corresponding to the three colors according to their different collective behavior.
The green stripes are only used in the time evolutions of $\meanNONDiag{\text{cov}}_{{B}}$.
The criteria for the blue stripes read
\begin{equation} \label{eqn:CovCollCondHigh}
	\frac{\meanNONDiag{\text{cov}}_{\text{BLE}}}{\meanNONDiag{\text{cov}}} > 0.997
\end{equation}
and
\begin{equation} \label{eqn:CorrCollCondHigh}
	\frac{\meanNONDiag{\text{cov}}_{\text{LLE}}}{\meanNONDiag{\text{corr}}} > 0.997 \,.
\end{equation}
These criteria are sensitive to very small mean values in Figs.~\ref{subfig:Main:MeanValCovP1},~\ref{subfig:Main:MeanValCovP2} and ~\ref{subfig:Main:MeanValCovP3}. Almost the entire collectivity is in the market mode.
The average sector collectivity derived from the reduced-rank covariance matrices  contributes only to a small amount to the entire collectivity.
A value larger than 0.997 corresponds to a high relative collectivity.
It never coincides with larger values in Figs.~\ref{subfig:Main:MeanValCovP1},~\ref{subfig:Main:MeanValCovP2} and ~\ref{subfig:Main:MeanValCovP3}. 
Remarkably, the blue stripes do coincide with almost all historical events listed in Tab.~\ref{tab:FinancialCrises}.

A fundamental change in the standard correlation matrices after the dot-com bubble burst period~\cite{kenett2011index,raddant2016phase,Raddant_2017,Pharasi_2018} is visible in Fig.~\ref{subfig:Main:MeanValCovP2}. 
However, we also observe that after the Great Fall of China (AF) the relative collectivity for the mean value $\meanNONDiag{\text{cov}}_{{L}}$ changes suddenly. In recent times, the mean values $\meanNONDiag{\text{cov}}_{{B}}$ and $\meanNONDiag{\text{cov}}_{{L}}$ are larger as well.
This indicates a fundamental change in the US stock markets as of 2015/2016, \textit{i.e.} before the crash of 2020 (CO).

The red stripes are found when 
\begin{equation} \label{eqn:CovCollCondLow}
	\frac{\meanNONDiag{\text{cov}}_{\text{BLE}}}{\meanNONDiag{\text{cov}}} < 0.8
\end{equation}
and
\begin{equation} \label{eqn:CorrCollCondLow}
	\frac{\meanNONDiag{\text{cov}}_{\text{LLE}}}{\meanNONDiag{\text{corr}}} < 0.8 \,.
\end{equation}
These criteria apply to time periods when peaks appear in Figs.~\ref{subfig:Main:MeanValCovP1},~\ref{subfig:Main:MeanValCovP2} and~\ref{subfig:Main:MeanValCovP3} which are connected to lower collectivities in the market mode.
We interpret values smaller than $0.8$ as a measure for a lower relative collectivity.
Those peaks are usually relatively sharp and large.

The green stripes emerge when the two conditions
\begin{align} \label{eqn:CovCollCondHigh_2}
  0.8 \leq	& \frac{\meanNONDiag{\text{cov}}_{\text{BLE}}}{\meanNONDiag{\text{cov}}} \leq 0.997 \,, \\
  \meanNONDiag{\text{cov}}_{\text{BLE}} &> 4.1 \cdot 10^{-4}
\end{align}
for $\meanNONDiag{\text{cov}}_{\text{BLE}} / \meanNONDiag{\text{cov}} $ and $\meanNONDiag{\text{cov}}_{\text{BLE}}$ are fulfilled.
They coincide with larger peaks
which cannot be captured by the criteria~Eqs.~\eqref{eqn:CovCollCondHigh} and \eqref{eqn:CovCollCondLow}.
For a representative subsample of 100 randomly selected stocks from all 213 stocks, the time evolutions for the average sector collectivity is similar to the results for 213 stocks (see~App.~\ref{sec:Subsamples}).

\begin{table}[!htb]
	\centering
	\caption{\label{tab:TimePeriods}
		Seven time periods taking into account historical events in Tab.~\ref{tab:FinancialCrises} and the time evolution of mean reduced-rank covariances $\meanNONDiag{\text{cov}}_{{B}}$ and $\meanNONDiag{\text{cov}}_{{L}}$ in Figs.~\ref{subfig:Main:MeanValCovP1},~\ref{subfig:Main:MeanValCovP2} and~\ref{subfig:Main:MeanValCovP3}.
	}
	\begin{tabular}{cl@{\hspace{1em}}c}
		\toprule
		Label &
		Decription &
		\multicolumn{1}{c}{Date}
		\\
		&
		&
		\multicolumn{1}{c}{(Year-Month-Day)} \\
		\midrule
		(P1) & Nineties & 1990-01-31 - 2000-02-08 \\
		(P2) & Post Dot-com bubble burst & 2000-02-09 - 2002-10-09 \\
		(P3) & Pre-Lehman crash & 2002-10-10 - 2007-10-31 \\
		(PA) & Precursor period & 2007-11-01 - 2008-08-14 \\
		(P4) & Post-Lehman crash & 2008-08-15 - 2015-08-18 \\
		(P5) & Post-China crisis & 2015-08-19 - 2020-01-22 \\
		(P6) & Post 2020 stock market crash  & 2020-01-23 - 2021-07-08 \\
		\bottomrule
	\end{tabular}
\end{table}

\subsection{\label{sec:RiskPhaseDiagramsAndTrajectories}Risk-phase diagrams and market trajectories}

\begin{figure}[!htb]
	\centering
	\includegraphics[width=1.0\columnwidth]{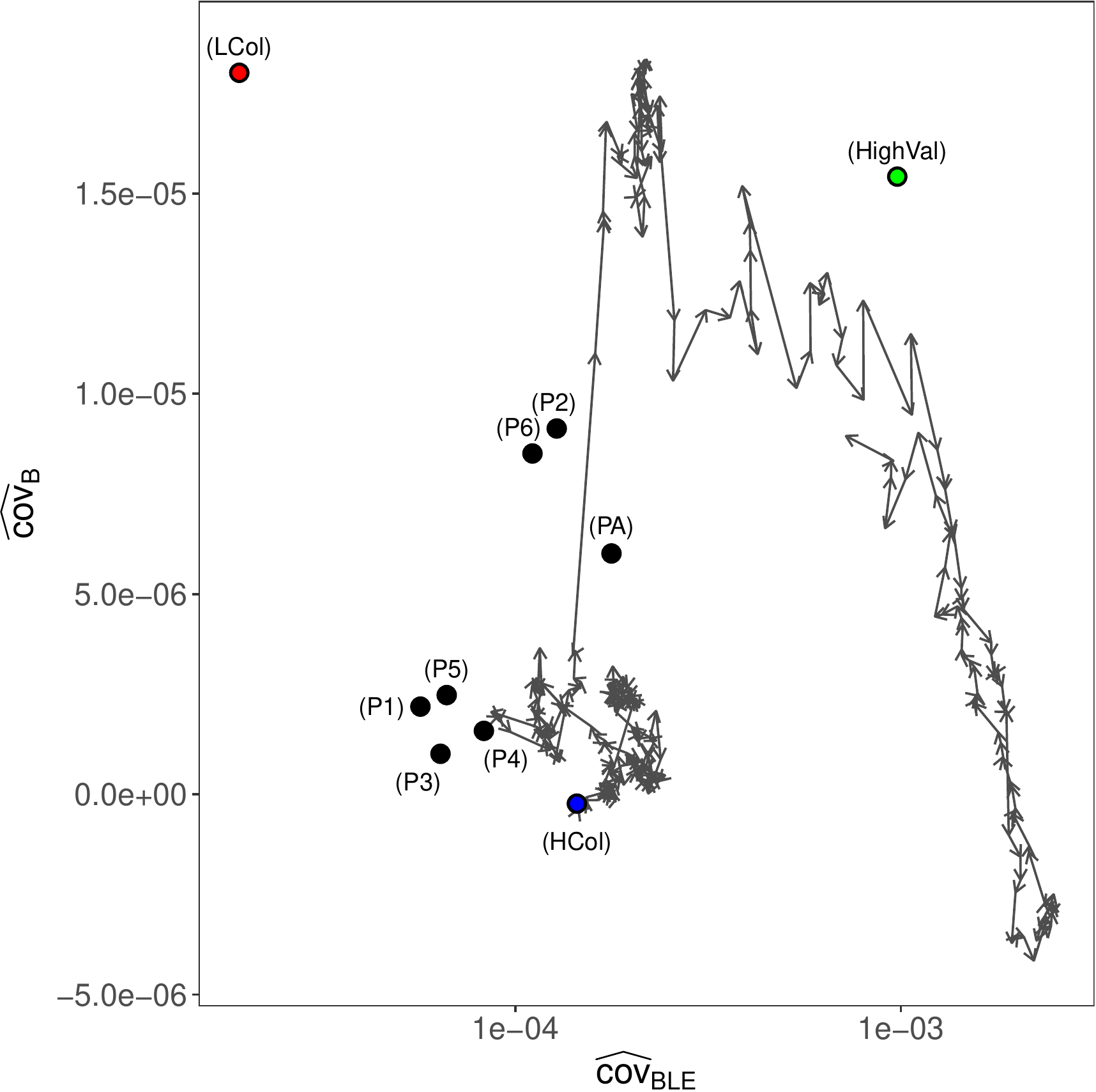}
	\caption{\label{fig:RiskPhaseCovLog} Risk-phase diagram plotted on a horizontal logarithmic scale.
		The colored dots (red, blue and green) correspond to three criteria for absolute and relative collectivity measures described in Sec.~\ref{sec:AverageSectorCollectivity}, additionally denoted by (LCol), (HighCol) and (HighVal). 
		The other seven dots belong to groups of covariance and correlation matrices described in Sec.~\ref{sec:RiskPhaseDiagramsAndTrajectories}. Market trajectories are shown from \mbox{2007-11-01} to \mbox{2008-12-31}. The arrows point into the directions into which the market moves next.}
	\label{fig:RiskPhaseLog}
\end{figure}
\begin{figure*}[!htb]
	\centering
	\begin{minipage}{0.5\textwidth}
		\subfloat[\label{subfig:RiskPhaseCovMean_LinR1}]
		{\includegraphics[width=1.0\textwidth]{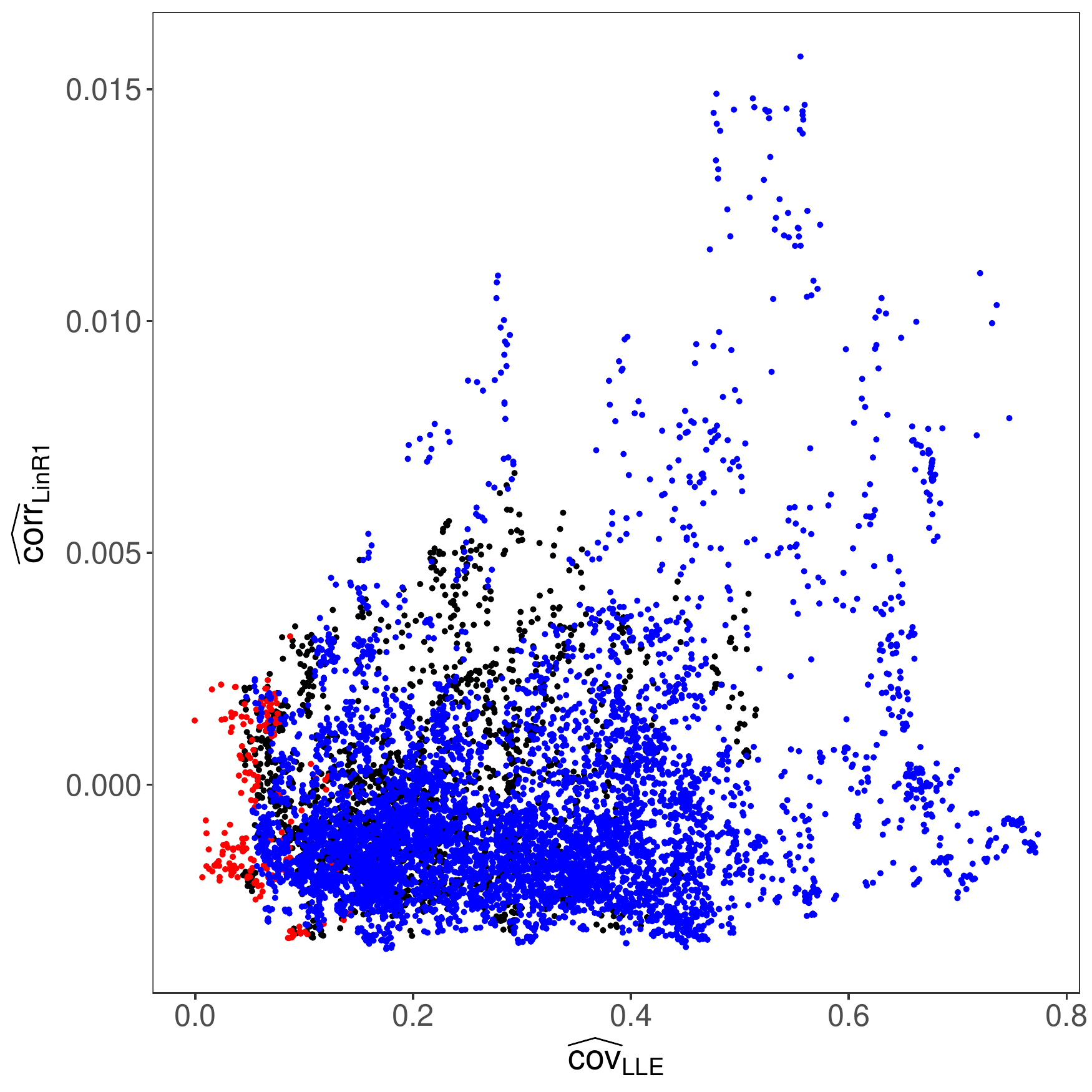}
		}
	\end{minipage}%
	\begin{minipage}{0.5\textwidth}
		\subfloat[\label{subfig:RiskPhaseCorrMean_LinR2}]
		{\includegraphics[width=1.0\textwidth]{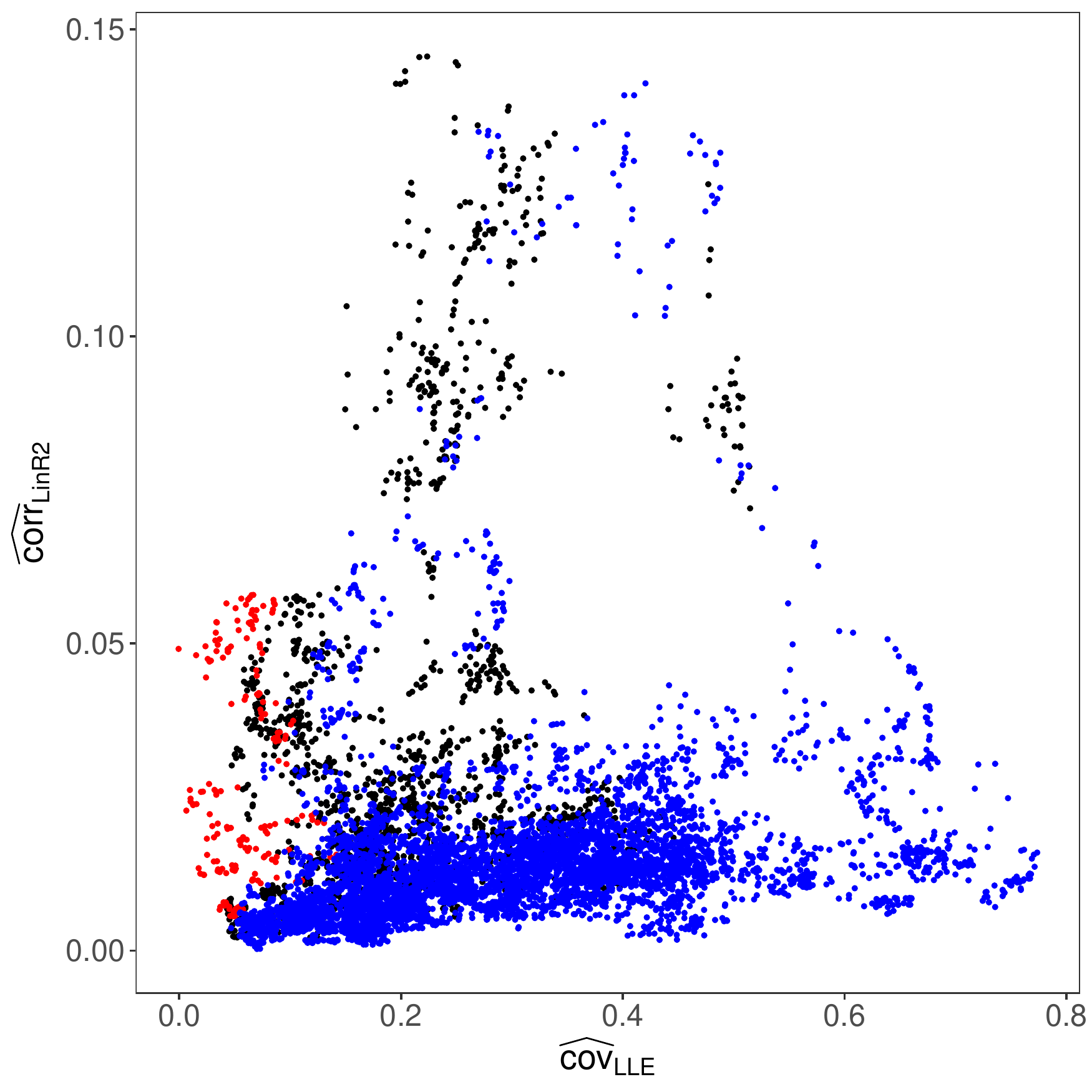}
		}
	\end{minipage}
	\caption{\label{subfig:Main:RiskPhaseMean_LinR}Risk-phase diagrams derived from \protect\subref{subfig:RiskPhaseCovMean} linear regression method in Eq.~\eqref{eqn:LinReg_MeditVariable_1} with $\meanNONDiag{\text{corr}}_{\text{LinR1}}$ plotted versus $\meanNONDiag{\text{cov}}_{\text{LLE}}$ and \protect\subref{subfig:RiskPhaseCorrMean} linear regression method in Eq.~\eqref{eqn:LinReg_MeditVariable_2} with $\meanNONDiag{\text{corr}}_{\text{LinR2}}$ plotted versus $\meanNONDiag{\text{cov}}_{\text{LLE}}$.  Each dot corresponds to one of the 7920 intervals of 42 trading days (see~Sec.~\ref{sec:DataSet}). 
		The colored dots (red/blue/green) correspond to the three criteria for absolute and relative collectivity measures described in Sec.~\ref{sec:AverageSectorCollectivity}.
		The green ones are only displayed in \protect\subref{subfig:RiskPhaseCovMean}.
		Black dots belong to matrices whose collectivities do not fulfill the three criteria.}
\end{figure*}
\begin{figure*}[!htb]
	\centering
	\begin{minipage}{0.5\textwidth}
		\subfloat[\label{subfig:RiskPhaseCovMean_B2}]
		{\includegraphics[width=1.0\textwidth]{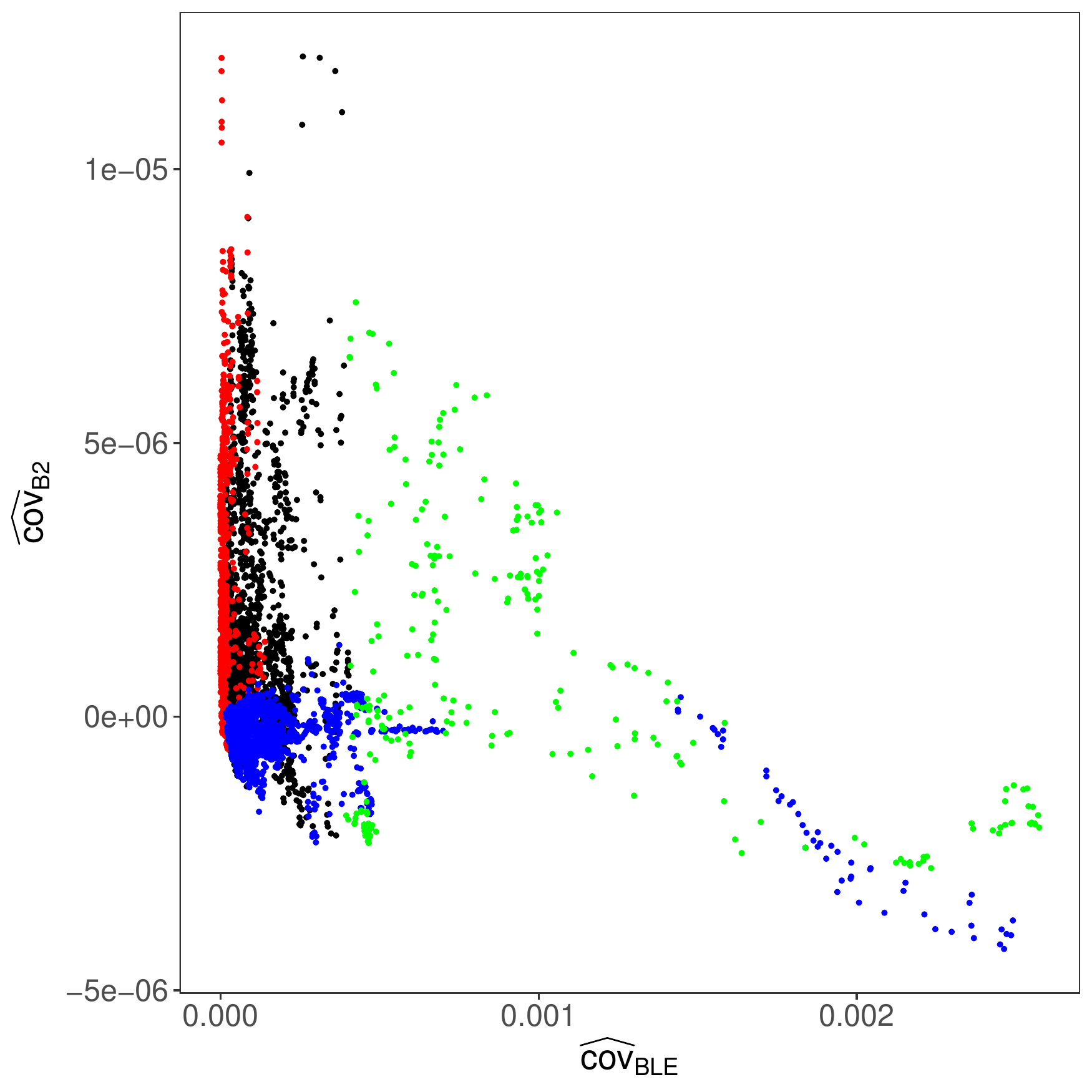}
		}
	\end{minipage}%
	\begin{minipage}{0.5\textwidth}
		\subfloat[\label{subfig:RiskPhaseCorrMean_L2}]
		{\includegraphics[width=1.0\textwidth]{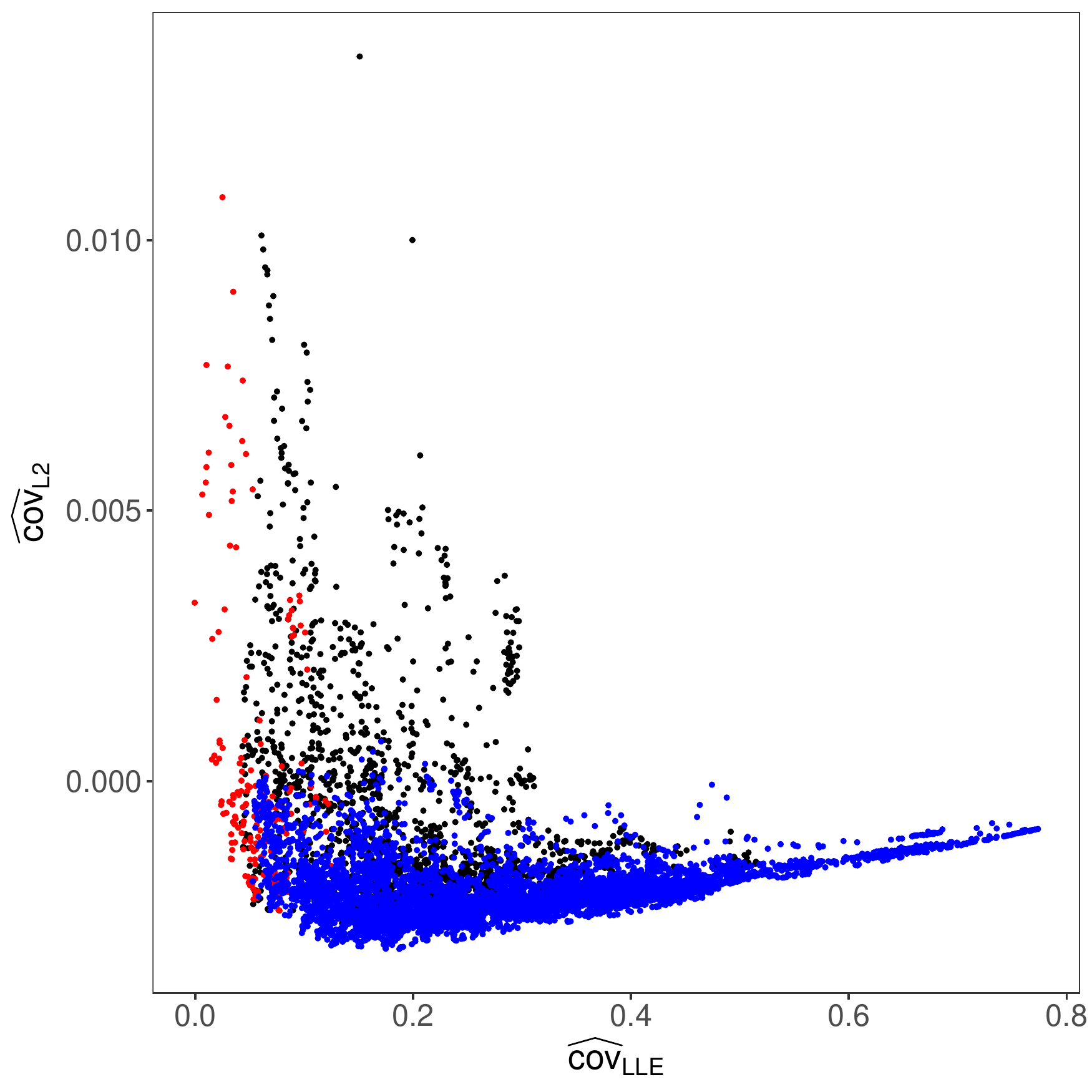}
		}
	\end{minipage}
	\caption{\label{subfig:Main:RiskPhaseMean_BL2}Risk-phase diagrams derived from \protect\subref{subfig:RiskPhaseCovMean} covariance matrix $\Sigma_{B2}$ in Eq.~\eqref{eqn:CovarianceMatSpectralDecomp_2} with $\meanNONDiag{\text{cov}}_{{B2}}$ plotted versus $\meanNONDiag{\text{cov}}_{\text{BLE}}$ and \protect\subref{subfig:RiskPhaseCorrMean} covariance matrix $\Sigma_{L2}$ in Eq.~\eqref{eqn:CorrelationMatSpectralDecomp_2} with $\meanNONDiag{\text{corr}}_{{L2}}$ plotted versus $\meanNONDiag{\text{cov}}_{\text{LLE}}$.  Each dot corresponds to one of the 7920 intervals of 42 trading days (see~Sec.~\ref{sec:DataSet}). 
		The colored dots (red/blue/green) correspond to the three criteria for absolute and relative collectivity measures described in Sec.~\ref{sec:AverageSectorCollectivity}.
		The green ones are only displayed in \protect\subref{subfig:RiskPhaseCovMean}.
		Black dots belong to matrices whose collectivities do not fulfill the three criteria.}
\end{figure*}
\begin{figure*}[!htb]
	\centering
	\begin{minipage}{1.0\textwidth}
		\subfloat%
		{\includegraphics[width=1.0\linewidth]{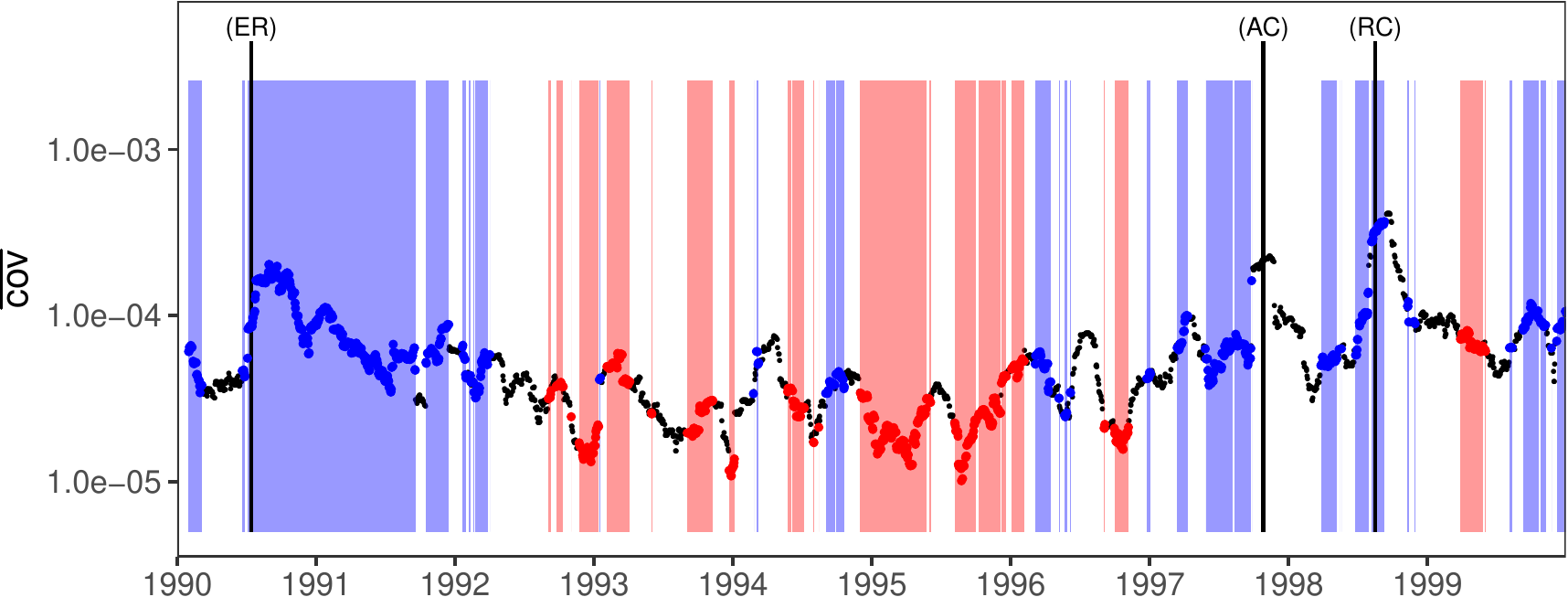}
		}\par
		\subfloat%
		{\includegraphics[width=1.0\textwidth]{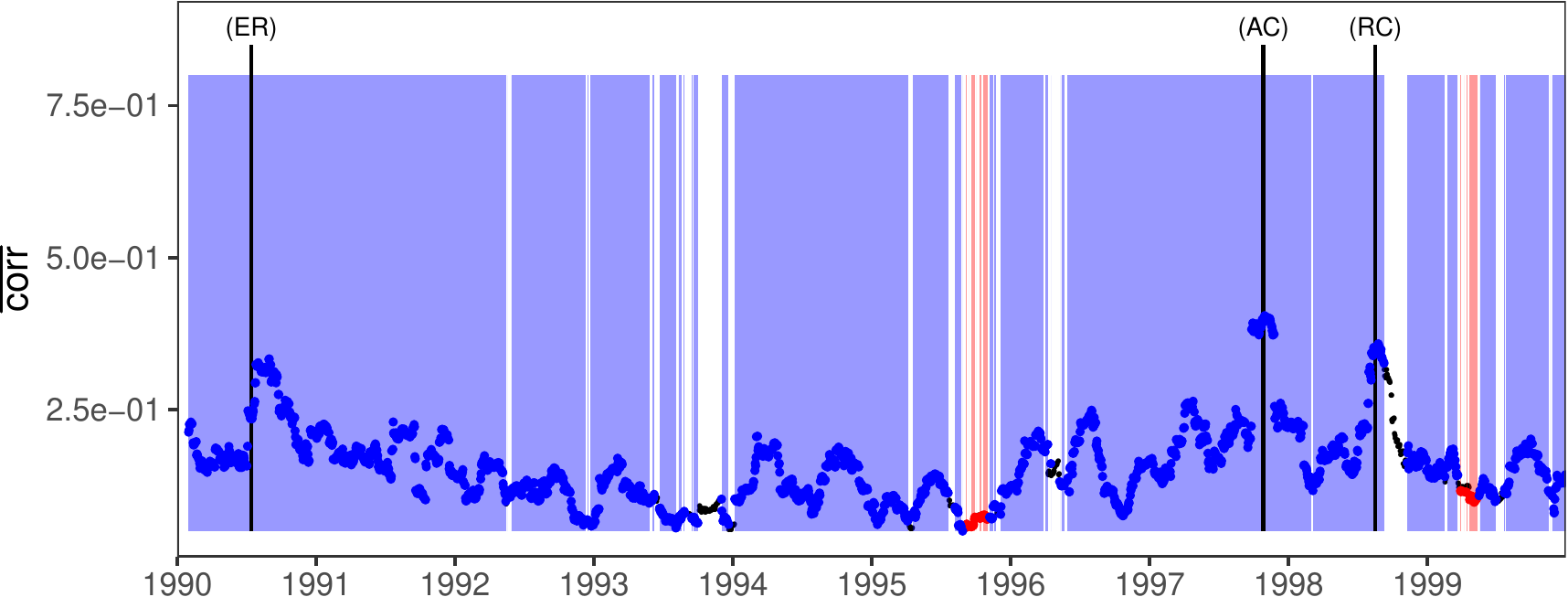}
		}
	\end{minipage}%
	\caption{\label{subfig:Main:MeanStandMeanCritP1}Time evolutions (1990-1999) for mean covariance $\mean{\text{cov}}$ (top, plotted on a logarithmic scale) and mean correlation $\mean{\text{corr}}$ (bottom). Three criteria (red/blue) for relative collectivity measures are described in Sec.~\ref{sec:AverageSectorCollectivity}. Historical events are listed in Tab.~\ref{tab:FinancialCrises}.}
\end{figure*}

\begin{figure*}[!htb]
	\centering
	\begin{minipage}{1.0\textwidth}
		\subfloat%
		{\includegraphics[width=1.0\linewidth]{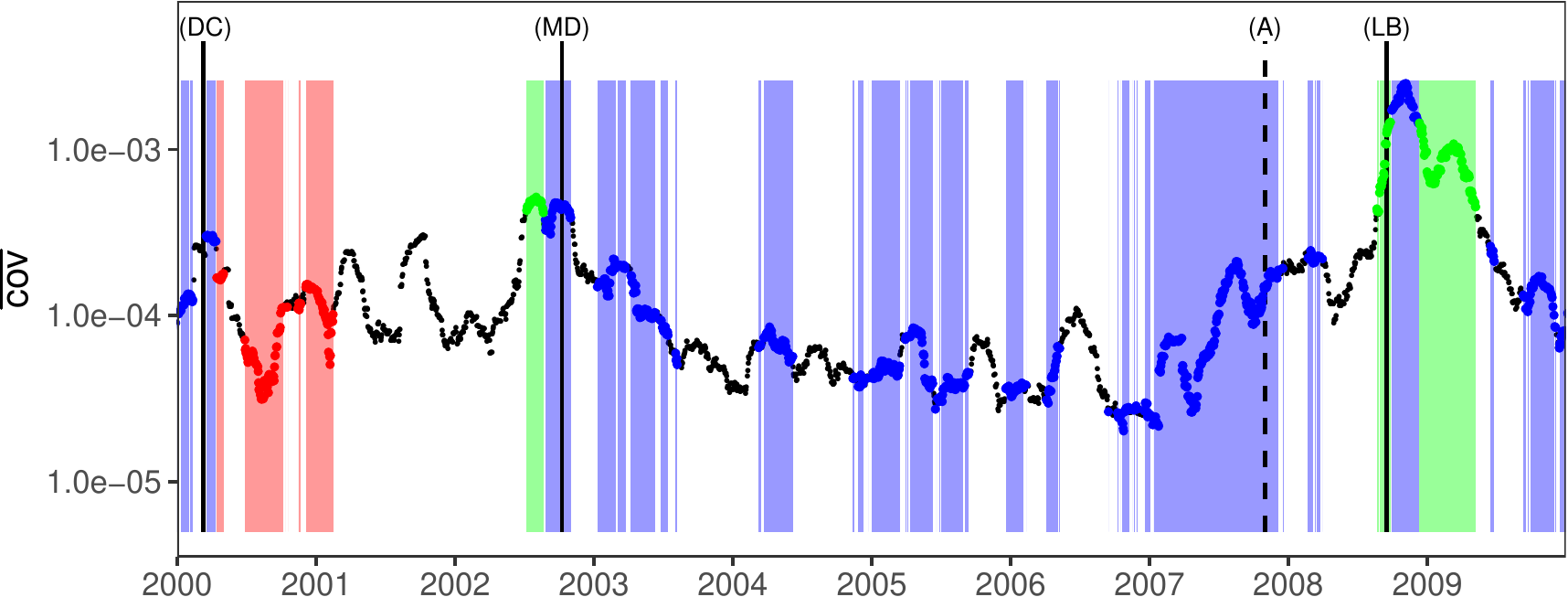}
		}\par
		\subfloat%
		{\includegraphics[width=1.0\textwidth]{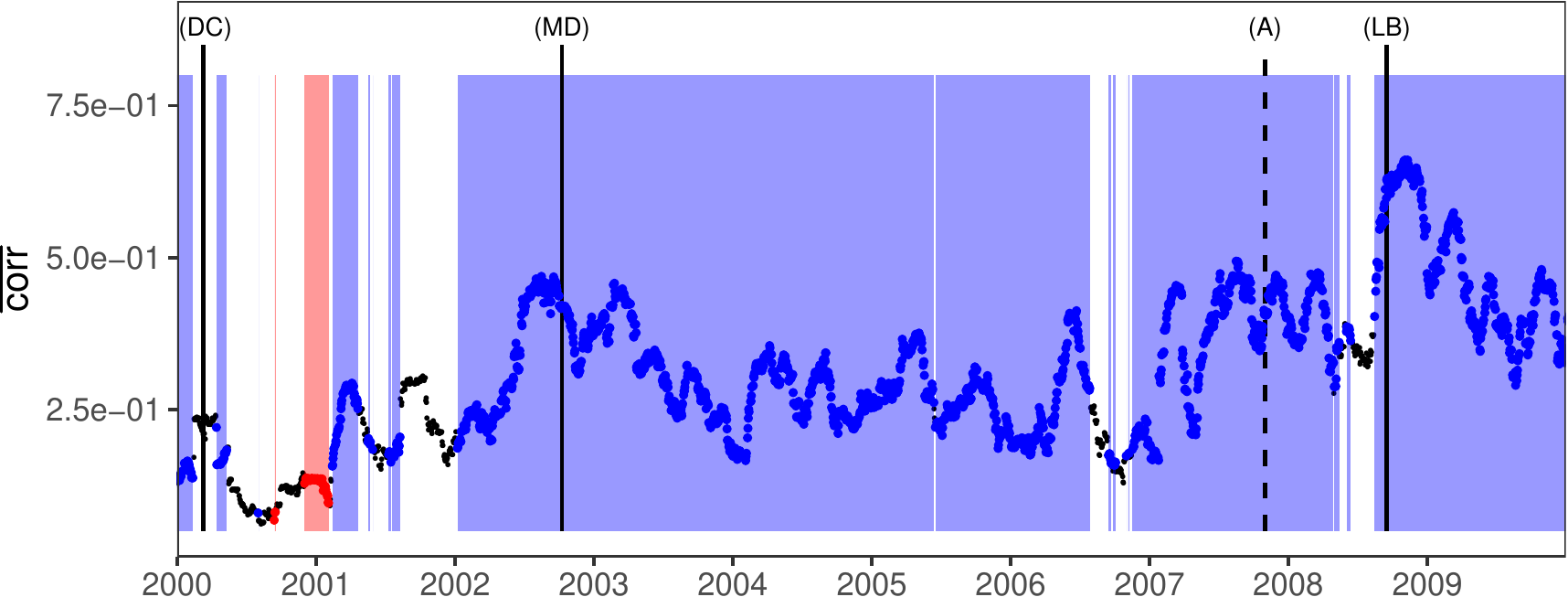}
		}
	\end{minipage}%
	\caption{\label{subfig:Main:MeanStandMeanCritP2}Time evolutions (2000-2009) for mean covariance $\mean{\text{cov}}$ (top, plotted on a logarithmic scale) and mean correlation $\mean{\text{corr}}$ (bottom). Three criteria (red/blue/green) for absolute and relative collectivity measures are described in Sec.~\ref{sec:AverageSectorCollectivity}. Historical events are listed in Tab.~\ref{tab:FinancialCrises}.}
\end{figure*}

\begin{figure*}[!htb]
	\centering
	\begin{minipage}{1.0\textwidth}
		\subfloat%
		{\includegraphics[width=1.0\linewidth]{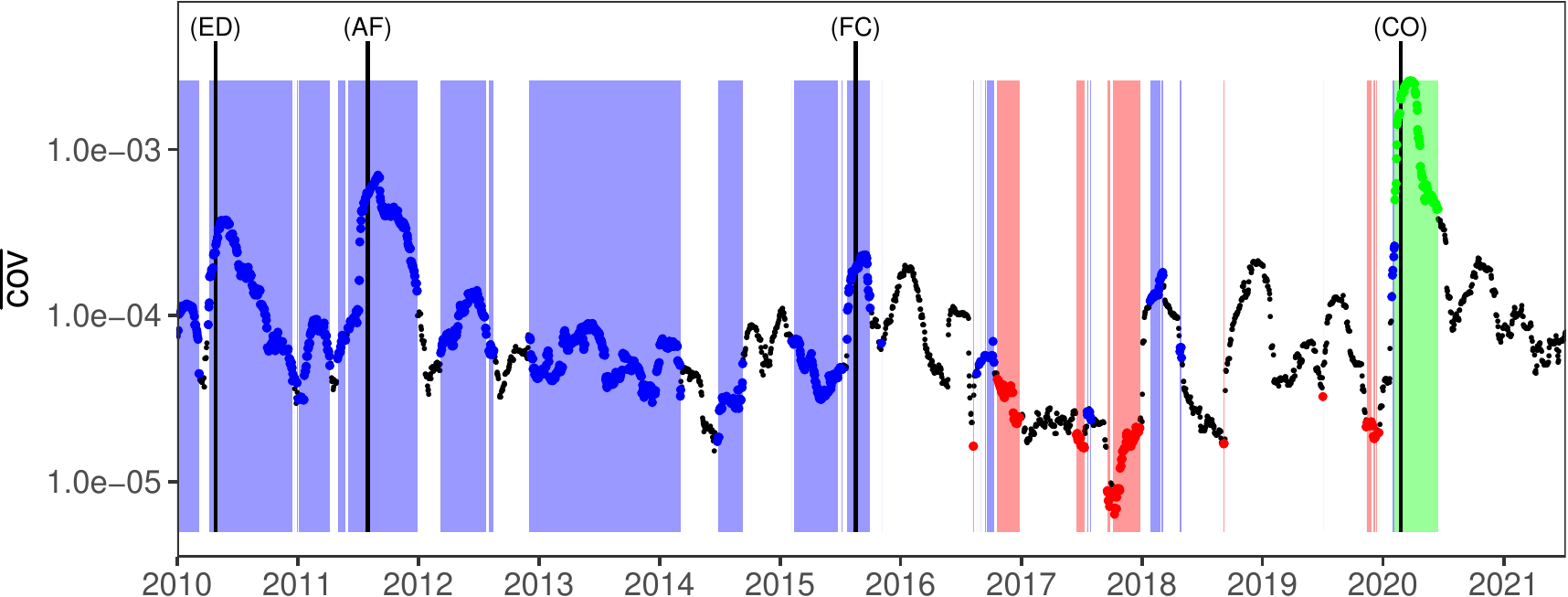}
		}\par
		\subfloat%
		{\includegraphics[width=1.0\textwidth]{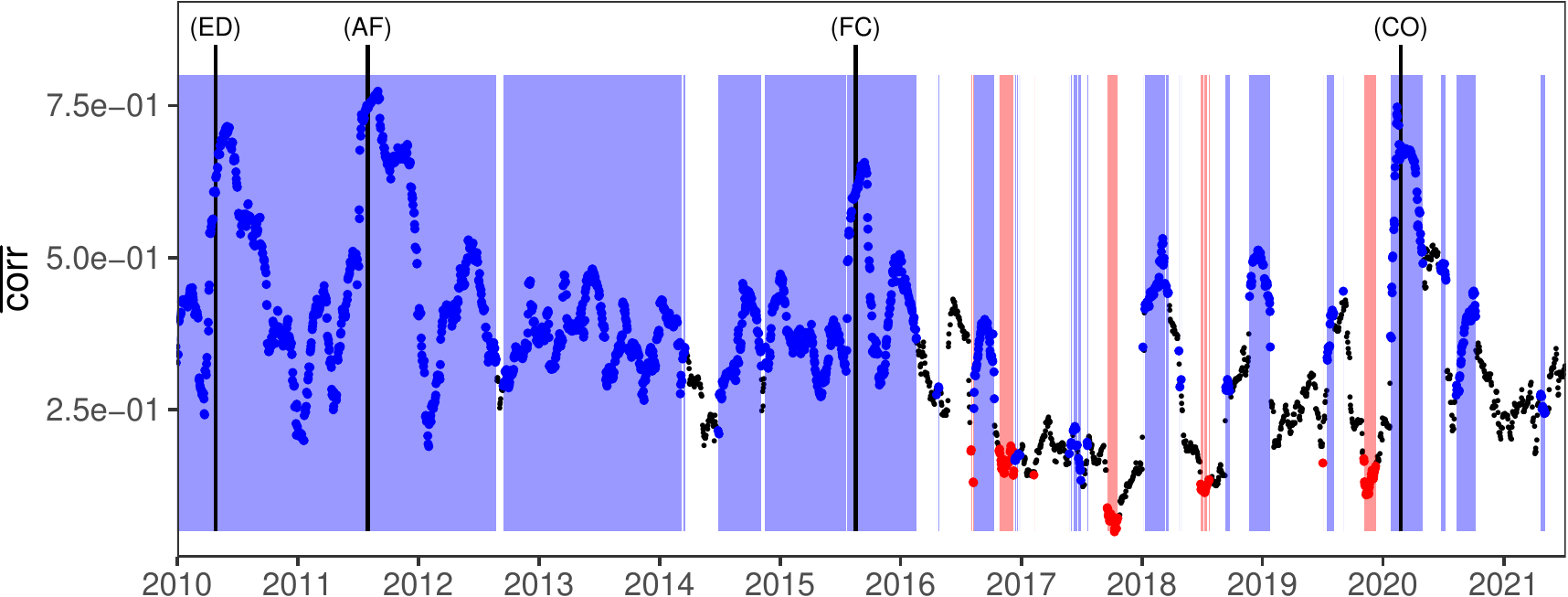}
		}
	\end{minipage}%
	\caption{\label{subfig:Main:MeanStandMeanCritP3}Time evolutions (2010-2021) for mean covariance $\mean{\text{cov}}$ (top, plotted on a logarithmic scale) and mean correlation $\mean{\text{corr}}$ (bottom). Three criteria (red/blue/green) for absolute and relative collectivity measures are described in Sec.~\ref{sec:AverageSectorCollectivity}. Historical events are listed in Tab.~\ref{tab:FinancialCrises}.}
\end{figure*}

Recently, a risk-phase diagram for stock markets was constructed by Chakraborti et al.~\cite{Chakraborti_2020} based on an eigen-entropy measure~\cite{DROZDZ2000440,IZRAILEV1990299},
originally introduced in the framework of quantum chaotic systems~\cite{IZRAILEV198813}.
Kenett et al.~\cite{kenett2011index} plotted the average sector collectivity versus the 
mean standard correlation.
The average sector collectivity was determined by taking the average of all partial correlation matrix elements~\cite{shapira2009index,kenett2010dominating,Kenett_2015}. Essentially, for each partial correlation coefficient a linear regressions was used to remove influences of an index from the standard correlation coefficients~\cite{anderson2003introduction,wiki:2021:PartialCorr}.
Signals in the vicinity of the dot-com bubble burst were detected.
These authors found that the market fundamentally changed its collective behavior at the beginning of 2002.

Here, we proceed differently. The absolute and relative collectivities are derived from the covariance and correlation matrix, \textit{i.e.} we use two different approaches.
Instead of linear regressions, we obtain the average sector collectivities from reduced-rank covariance matrices after using spectral decompositions.
Furthermore, we use mean values derived from covariance matrices corresponding to the market mode in both risk-phase diagrams.
Our results show an overall richer peak structure over a 30 year period than in Ref.~\cite{kenett2011index}. For instance, we detect precursor signals prior to (LB) in Fig.~\ref{subfig:Main:MeanValCovP2} (cf.~Ref.~\cite{Heckens_2022}).
Differences between our study and Ref.~\cite{kenett2011index} manifest themselves by an altered spectral density for partial correlation matrices~\cite{kenett2009rmt}.

In our risk-phase diagrams in Fig.~\ref{subfig:Main:RiskPhaseMean} we plot the mean covariances 
$\meanNONDiag{\text{cov}}_{\text{BLE}}$ versus $\meanNONDiag{\text{cov}}_{{B}}$ in Fig.~\subref*{subfig:RiskPhaseCovMean} and $\meanNONDiag{\text{cov}}_{\text{LLE}}$ versus $\meanNONDiag{\text{cov}}_{{L}}$ in Fig.~\subref*{subfig:RiskPhaseCorrMean}.
Each dot corresponds to an interval of 42 trading days  (see~Sec.~\ref{sec:DataSet}).
We reuse the colors for the three different criteria in Sec.~\ref{sec:AverageSectorCollectivity}.

Dots corresponding to the blue stripes from Figs.~\ref{subfig:Main:MeanValCovP1},~\ref{subfig:Main:MeanValCovP2} and~\ref{subfig:Main:MeanValCovP3}
are distributed along the horizontal axis, whereas dots corresponding to the red stripes can be found along the vertical axis.
The red dots emerge for low values $\meanNONDiag{\text{cov}}_{\text{BLE}}$ and $\meanNONDiag{\text{cov}}_{\text{LLE}}$.
The alignment of the blue dots along the horizontal axis indicates a high absolute and relative market collectivity in the risk-phase diagram, whereas the alignment of the red ones along the vertical axis shows a low absolute and relative market collectivity.

For a representative subsample of 100 randomly selected stocks from all 213 stocks, the risk-phase diagrams are similar compared to those for 213 stocks (see~App.~\ref{sec:Subsamples}).

We now calculate the mean values of all absolute collectivities for the covariance and correlation matrices 
whose collectivities fulfill one of the three criteria.
In Fig.~\ref{subfig:Main:RiskPhaseMeanPeriod}, we plot the resulting mean absolute collectivities in risk-phase diagrams as well.
In this way, we depict the centers of the dots
from Fig.~\ref{subfig:Main:RiskPhaseMean} corresponding to covariance and correlation matrices whose collectivities fulfill one of the three criteria.
Furthermore, we sort the covariance and correlation matrices whose collectivities do not fulfill the three criteria into
seven groups according to the seven time periods listed in Tab.~\ref{tab:TimePeriods}.
For each of these seven groups of covariance and correlation matrices we calculate the mean values of all absolute collectivities.
For instance, this allows us to make statements about the average dynamics 
of the precursor period (PA).
Additionally, we added the trajectories of the market for the potential precursor period prior to the Lehman Brothers crash from \mbox{2007-11-01} to \mbox{2008-12-31} to Figs.~\subref*{subfig:RiskPhaseCovMeanPeriod} and Fig.~\subref*{subfig:RiskPhaseCorrMeanPeriod}.
The arrows point into the directions into which the market moves next.
For a clearer visual representation and separation
 of the different dots in Fig.~\subref*{subfig:RiskPhaseCovMeanPeriod}, we plot $\meanNONDiag{\text{cov}}_{\text{BLE}}$ in Fig.~\ref{fig:RiskPhaseCovLog} on a logarithmic scale. 

Time period (PA) is accompanied by relatively large values in $\meanNONDiag{\text{cov}}_{\text{BLE}}$
and  $\meanNONDiag{\text{cov}}_{\text{LLE}}$, \textit{i.e.} the market was already at a high risk level.
The risk-phase diagrams in Fig.~\ref{subfig:Main:RiskPhaseMeanPeriod} show that $\meanNONDiag{\text{cov}}_{{B}}$ and $\meanNONDiag{\text{cov}}_{{L}}$ induced by predominantly endogenous effects are potential measures for systemic risk (cf.~Ref.~\cite{Heckens_2022}).
In the vicinity of the dots corresponding to the time period (PA), we find (P2) and (P6) in Fig.~\subref*{subfig:RiskPhaseCovMeanPeriod} and
time periods (P4) and (P6) in Fig.~\subref*{subfig:RiskPhaseCorrMeanPeriod}. Time periods (P2) and (P6) are accompanied by larger mean values $\meanNONDiag{\text{cov}}_{\text{BLE}}$ in Fig.~\ref{fig:RiskPhaseCovLog}. (P4) and (P6) feature larger mean values $\meanNONDiag{\text{cov}}_{\text{LLE}}$ in Fig.~\ref{subfig:RiskPhaseCorrMeanPeriod}.
Intriguingly, this indicates that recently, the market is at a high risk level on average in $\meanNONDiag{\text{cov}}_{\text{BLE}}$
and  $\meanNONDiag{\text{cov}}_{\text{LLE}}$ after the crash of 2020 (CO).

\subsection{\label{sec:RiskPhaseLinRegress}Risk-phase diagrams for correlation matrices obtained by linear regression methods}

The market-wide behavior of all stocks is often removed by linear regression methods~\cite{Plerou_2002,Borghesi_2007,shapira2009index,kenett2010dominating,Ross_2014,Kenett_2015}.
For each stock $i$ and each interval of length $T_{\text{sub}} = 42$~trading days, we use the linear regression model
\begin{equation} \label{eqn:LinReg}
	G_i(t) = \alpha_i + \beta_i I(t) + \epsilon_i(t)
\end{equation}
to determine the residuals $\epsilon_i(t)$ with stock-specific constants $\alpha_i$ and $\beta_i$.
In the language of partial correlations, $I(t)$ is referred to as mediating variable.
Our first choice is
\begin{equation} \label{eqn:LinReg_MeditVariable_1}
	I_1(t) = \frac{1}{K} \sum_{i=1}^K G_i(t) ,
\end{equation}
which can be interpreted as ``average market return'' or ``center of mass''.
As second mediating variable, we use
the logarithmic returns of the daily closing prices from the S\&P~500 index
\begin{equation} \label{eqn:LinReg_MeditVariable_2}
	I_2(t) = G^{\text{SP500}}(t) \,.
\end{equation}
We obtain 7920 $K\times T_{\text{sub}}$ data matrices whose elements are the residuals $\epsilon_i(t)$.
Each row of those data matrices contains the residuals for one stock $i$. 
Finally, we compute the correlation matrices for each of these data matrices.
Analogously to the definition of absolute collectivities in Eq.~\eqref{eqn:MeanCovariance_NONDiag}, we define
the absolute collectivities $\meanNONDiag{\text{corr}}_{\text{LinR1}}$ for Eq.~\eqref{eqn:LinReg_MeditVariable_1}
and $\meanNONDiag{\text{corr}}_{\text{LinR2}}$ for Eq.~\eqref{eqn:LinReg_MeditVariable_2}.

In Fig.~\ref{subfig:Main:RiskPhaseMean_LinR}, we show the risk-phase diagrams for both linear regression approaches.
We plot $\meanNONDiag{\text{cov}}_{\text{LLE}}$ on the horizontal axis.
The first linear regression method removes more collectivity than the second one.
For the second mediating variable, we find much larger values for the average sector collectivity as of 2020 after event (CO).
In contrast to the risk-phase diagrams in Fig.~\ref{subfig:Main:RiskPhaseMean}, where we used the spectral decompositions for standard covariance and correlation matrices, we do not see clear structures along both axes.
Especially the absolute collectivities along the vertical axes are strongly suppressed.
Closer inspection shows that the criterion in Eq.~\eqref{eqn:CorrCollCondLow} almost never works properly for the correlation matrices obtained
by linear regression methods. The peaks in the average sector collectivities are at inconsistent positions. Details are
given in App.~\ref{sec:TimeEvoLinRegress}.

\subsection{\label{sec:HigherOrderEffects}Higher collective order effects in reduced-rank correlation matrices}

The question arises whether the other dyadic matrices corresponding to eigenvalues smaller than the second one show a collective behavior as well. In Ref.~\cite{Heckens_2020} we proposed a method which allows us to detect such collective signals.
By subtracting the dyadic matrix corresponding to the second largest eigenvalues from the reduced-rank covariance matrices $\Sigma_B$ and $\Sigma_L$, we obtain different reduced-rank covariance matrices
\begin{align} \label{eqn:CovarianceMatSpectralDecomp_2}
	\Sigma_{B2}  = \Sigma_{B} - \kappa_{K-1} \, u_{K-1} \ u^{\dagger}_{K-1} 
\end{align}
for the covariance approach and
\begin{align} \label{eqn:CorrelationMatSpectralDecomp_2}
	\Sigma_{L2}  = \Sigma_{L} - \kappa_{L-1} \, u_{L-1} \ u^{\dagger}_{L-1} 
\end{align}
for the correlation approach.
In Fig.~\ref{subfig:Main:RiskPhaseMean_BL2}, we plot the associated absolute collectivities $\meanNONDiag{\text{cov}}_{B2}$ and $\meanNONDiag{\text{cov}}_{L2}$ on the vertical axis of the risk-phase diagrams.
In both cases, the average sector collectivities are smaller than in Fig.~\ref{subfig:Main:RiskPhaseMean}. The criterion in Eq.~\eqref{eqn:CorrCollCondLow} still coincides with peaks in the time evolutions in App.~\ref{sec:TimeEvoHighCollec}. 
The criterion in Eq.~\eqref{eqn:CorrCollCondLow} still works well, the peaks in the time evolutions are consistent with the ones previously found. For the correlation approach, those peaks vanish almost completely. For details, see App.~\ref{sec:TimeEvoHighCollec}.

\subsection{\label{sec:RelevanceRelCollec}Relevance of the relative collectivity for
standard mean covariances and correlations}

In Figs.~\ref{subfig:Main:MeanStandMeanCritP1},~\ref{subfig:Main:MeanStandMeanCritP2} and~\ref{subfig:Main:MeanStandMeanCritP3}, the time evolutions of the mean covariance $\mean{\text{cov}}$ and mean correlation $\mean{\text{corr}}$ (including diagonal elements, see Eq.~\eqref{eqn:MeanCovariance}) are displayed and compared with the criteria from Sec.~\ref{sec:AverageSectorCollectivity}.
We see that large values in the relative collectivity usually coincide with larger values in 
$\mean{\text{cov}}$ and $\mean{\text{corr}}$. Small values for the relative collectivity usually coincide with smaller values of $\mean{\text{cov}}$ and $\mean{\text{corr}}$.

Most importantly, we find that
smaller relative collectivities often indicate a trend shift to larger mean values of $\mean{\text{cov}}$ and $\mean{\text{corr}}$. 
Larger relative collectivities often emerge when trend shifts to smaller values of $\mean{\text{cov}}$ and $\mean{\text{corr}}$ occur (cf.~Fig.~\ref{subfig:Main:MeanStandMeanCritP3}).
Although high and low relative collectivities do not always coincide with trend shifts, it happens quite often, especially within the last decade.

The trend shifts for larger relative collectivities appear more frequently when a crisis takes place or builds up.
If we understand the potential precursor in Fig.~\ref{subfig:Main:MeanStandMeanCritP2} (cf.~Ref.~\cite{Heckens_2022}) as a trend shift to larger mean standard covariances and correlations corresponding to the Lehman Brothers crash (LB) as well (spill-over effect), there seems to be a generic mechanism at work for the mean values $\meanNONDiag{\text{cov}}_{{B}}$ and $\meanNONDiag{\text{cov}}_{{L}}$.

\section{\label{sec:Conclusion}Conclusions}

We introduced and discussed averages of covariance and correlation matrices as new measures for collectivity capable of quantifying collectivity in industrial sectors.
We critically discussed the role of the diagonal elements of covariance and correlation matrices. This led us
to a definition for absolute collectivities without the diagonal elements.
The average sector collectivity describes the average dynamics of the reduced-rank covariance matrices derived from standard covariance and correlation matrices.
We constructed a new kind of risk-phase diagram taking into account the absolute and relative measures for collectivity.
We followed the market trajectory plotting the average sector collectivity versus the market mode collectivity.

We discovered a new phase in the financial markets which emerges for small values of mean standard covariances
$\meanNONDiag{\text{cov}}_{\text{BLE}}$ and $\meanNONDiag{\text{cov}}_{\text{LLE}}$.
We found that larger peaks in the mean values $\meanNONDiag{\text{cov}}_{{B}}$ and $\meanNONDiag{\text{cov}}_{{L}}$ as absolute collectivities indicate trend shifts in the time evolutions of the mean covariance $\mean{\text{cov}}$ and mean correlation $\mean{\text{corr}}$ to larger mean values.
Comparing the relative collectivities with the time evolutions of the mean covariance $\mean{\text{cov}}$ and mean correlation $\mean{\text{corr}}$,
large values for the relative collectivity indicate trend shifts to smaller values in $\mean{\text{cov}}$
and $\mean{\text{corr}}$, 
whereas smaller values for the relative collectivity indicate trend shifts to larger values.

Both linear regression methods tend to suppress
the average sector collectivity along the vertical axis in the risk-phase diagrams. The risk-phase diagrams with values derived by spectral decomposition show a
much clearer structure along both axes. Additionally, we analyzed higher collective order effects by subtracting the dyadic matrices corresponding to the second largest
eigenvalues and evaluating the average sector collectivities for the resulting reduced-rank covariance matrices. 
We found a remaining average sector collectivity in
the covariance approach. For the correlation approach,
the collectivity in the resulting reduced-rank covariance matrices is usually relatively small compared to
the one in the dyadic matrix associated with the second largest eigenvalue.

There might be an interpretation for the trend shift to larger mean values in $\mean{\text{cov}}$
and $\mean{\text{corr}}$.
In Refs.~\cite{Heckens_2020,Heckens_2022}, we observed in reduced-rank correlation matrices anti-correlations between the financial sector and the other ten industrial sectors prior to the Lehman Brothers crash.
Other signals which are not part of the new phase, for instance the potential precursor prior to the Lehman Brothers crash, indicate trend shifts as well.
Some traders seemed to have reorganized their portfolios prior to the crash from the financial sector to others possibly anticipating a potential drastic decline of the US stock market prior to the Lehman crash.
This reshuffling induced a collectivity that may be interpreted as a ``sector crash''.

We can divide the entire time period into two periods, one before and one after the China crisis of 2015.
In the first period, especially the two endogenous crises, the dot-com bubble burst and the period around the Lehman crash show larger peaks in $\meanNONDiag{\text{cov}}_{\text{B}}$. 
Larger values in the absolute collectivities are rather likely due to endogenous effects
since the predominantly exogenous market collectivity is removed from the standard covariance and correlation matrices (cf.~Ref.~\cite{Heckens_2022}). Our hypothesis of traders reorganizing
their portfolios adds to our previous observations of endogenous, induced effects in the reduced-rank covariance and correlation matrices (cf.~Refs.~\cite{Heckens_2020,Heckens_2022}).
Prior to the Lehman Brothers crash, the potential precursor in the mean reduced-rank covariances is accompanied by relatively large mean covariances corresponding to the market mode. This corroborates
our previous result that signals in $\overline{\text{cov}}_{{B}}$ and $\mean{\text{cov}}_{{L}}$ indicate systemic risk~\cite{Heckens_2022}.

In the period after the China crisis, these signals emerged more often.
The period after the crash of 2020 features relatively large values for the average sector and market collectivity compared to other periods analyzed in our study.

Overall, larger values for the market collectivity together with a larger average sector collectivity indicate
structural instabilities such as potential trend shifts in the risk-phase diagram. 
The average sector collectivity contributes to the understanding of systemic risk and refines its estimation.

\clearpage
\bibliography{Lit-8.bib}

\appendix

\clearpage
\onecolumngrid
\section{\label{sec:ListStocks}List of selected stocks}

{\tiny
	\begin{longtable}{rrllp{4cm}}
		\caption[]{Overview of the 213 selected stocks of the S\&P 500 index (cf.~{\cite{Refinitiv}}).} 
		
		\label{tab:OverviewSP500} \\
		\toprule
		Number & Symbol & Security & Sector & Sub-Industry\\
		\midrule\endfirsthead
		\caption*{Continuation: Overview of the 213 selected stocks of the S\&P 500 
			index (cf.~{\cite{Refinitiv}}).}  \\
		\toprule
		Number & Symbol & Security & Sector & Sub-Industry\\
		\midrule\endhead
1&SLB&Schlumberger NV&Energy&Oil \& Gas Equipment \& Services \\
2&BKR&Baker Hughes Co&Energy&Oil \& Gas Equipment \& Services \\
3&HAL&Halliburton Co&Energy&Oil \& Gas Equipment \& Services \\
4&OXY&Occidental Petroleum Corp&Energy&Integrated Oil \& Gas \\
5&XOM&Exxon Mobil Corp&Energy&Integrated Oil \& Gas \\
6&CVX&Chevron Corp&Energy&Integrated Oil \& Gas \\
7&DVN&Devon Energy Corp&Energy&Oil \& Gas Exploration \& Production \\
8&COP&Conocophillips&Energy&Oil \& Gas Exploration \& Production \\
9&MRO&Marathon Oil Corp&Energy&Oil \& Gas Exploration \& Production \\
10&APA&APA Corp (US)&Energy&Oil \& Gas Exploration \& Production \\
11&HES&Hess Corp&Energy&Oil \& Gas Exploration \& Production \\
12&EOG&EOG Resources Inc&Energy&Oil \& Gas Exploration \& Production \\
13&WMB&Williams Companies Inc&Energy&Oil \& Gas Storage \& Transportation \\
14&OKE&ONEOK Inc&Energy&Oil \& Gas Storage \& Transportation \\
15&FMC&FMC Corp&Materials&Fertilizers \& Agricultural Chemicals \\
16&APD&Air Products and Chemicals Inc&Materials&Industrial Gases \\
17&SHW&Sherwin-Williams Co&Materials&Specialty Chemicals \\
18&ECL&Ecolab Inc&Materials&Specialty Chemicals \\
19&IFF&International Flavors \& Fragrances Inc&Materials&Specialty Chemicals \\
20&PPG&PPG Industries Inc&Materials&Specialty Chemicals \\
21&VMC&Vulcan Materials Co&Materials&Construction Materials \\
22&BLL&Ball Corp&Materials&Metal \& Glass Containers \\
23&SEE&Sealed Air Corp&Materials&Paper Packaging \\
24&IP&International Paper Co&Materials&Paper Packaging \\
25&AVY&Avery Dennison Corp&Materials&Paper Packaging \\
26&NEM&Newmont Corporation&Materials&Gold \\
27&NUE&Nucor Corp&Materials&Steel \\
28&BA&Boeing Co&Industrials&Aerospace \& Defense \\
29&GD&General Dynamics Corp&Industrials&Aerospace \& Defense \\
30&TXT&Textron Inc&Industrials&Aerospace \& Defense \\
31&HWM&Howmet Aerospace Inc&Industrials&Aerospace \& Defense \\
32&NOC&Northrop Grumman Corp&Industrials&Aerospace \& Defense \\
33&RTX&Raytheon Technologies Corp&Industrials&Aerospace \& Defense \\
34&LHX&L3harris Technologies Inc&Industrials&Aerospace \& Defense \\
35&MAS&Masco Corp&Industrials&Building Products \\
36&TT&Trane Technologies PLC&Industrials&Building Products \\
37&EMR&Emerson Electric Co&Industrials&Electrical Components \& Equipment \\
38&ROK&Rockwell Automation Inc&Industrials&Electrical Components \& Equipment \\
39&ETN&Eaton Corporation PLC&Industrials&Electrical Components \& Equipment \\
40&GE&General Electric Co&Industrials&Industrial Conglomerates \\
41&MMM&3M Co&Industrials&Industrial Conglomerates \\
42&HON&Honeywell International Inc&Industrials&Industrial Conglomerates \\
43&DE&Deere \& Co&Industrials&Agricultural \& Farm Machinery \\
44&PCAR&Paccar Inc&Industrials&Construction Machinery \& Heavy Trucks \\
45&CMI&Cummins Inc&Industrials&Construction Machinery \& Heavy Trucks \\
46&CAT&Caterpillar Inc&Industrials&Construction Machinery \& Heavy Trucks \\
47&SNA&Snap-On Inc&Industrials&Industrial Machinery \\
48&IEX&IDEX Corp&Industrials&Industrial Machinery \\
49&PH&Parker-Hannifin Corp&Industrials&Industrial Machinery \\
50&DOV&Dover Corp&Industrials&Industrial Machinery \\
51&ITW&Illinois Tool Works Inc&Industrials&Industrial Machinery \\
52&SWK&Stanley Black \& Decker Inc&Industrials&Industrial Machinery \\
53&GWW&W W Grainger Inc&Industrials&Trading Companies \& Distributors \\
54&FAST&Fastenal Co&Industrials&Trading Companies \& Distributors \\
55&CTAS&Cintas Corp&Industrials&Diversified Support Services \\
56&ROL&Rollins Inc&Industrials&Environmental \& Facilities Services \\
57&EFX&Equifax Inc&Industrials&Research \& Consulting Services \\
58&J&Jacobs Engineering Group Inc&Industrials&Research \& Consulting Services \\
59&EXPD&Expeditors International of Washington Inc&Industrials&Air Freight \& Logistics \\
60&FDX&FedEx Corp&Industrials&Air Freight \& Logistics \\
61&ALK&Alaska Air Group Inc&Industrials&Airlines \\
62&LUV&Southwest Airlines Co&Industrials&Airlines \\
63&NSC&Norfolk Southern Corp&Industrials&Railroads \\
64&CSX&CSX Corp&Industrials&Railroads \\
65&KSU&Kansas City Southern&Industrials&Railroads \\
66&UNP&Union Pacific Corp&Industrials&Railroads \\
67&JBHT&J B Hunt Transport Services Inc&Industrials&Trucking \\
68&F&Ford Motor Co&Consumer Discretionary&Automobile Manufacturers \\
69&LEG&Leggett \& Platt Inc&Consumer Discretionary&Home Furnishings \\
70&LEN&Lennar Corp&Consumer Discretionary&Homebuilding \\
71&PHM&Pultegroup Inc&Consumer Discretionary&Homebuilding \\
72&WHR&Whirlpool Corp&Consumer Discretionary&Household Appliances \\
73&NWL&Newell Brands Inc&Consumer Discretionary&Housewares \& Specialties \\
74&HAS&Hasbro Inc&Consumer Discretionary&Leisure Products \\
75&PVH&PVH Corp&Consumer Discretionary&Apparel, Accessories \& Luxury Goods \\
76&VFC&VF Corp&Consumer Discretionary&Apparel, Accessories \& Luxury Goods \\
77&MGM&MGM Resorts International&Consumer Discretionary&Casinos \& Gaming \\
78&CCL&Carnival Corp&Consumer Discretionary&Hotels, Resorts \& Cruise Lines \\
79&MCD&Mcdonald's Corp&Consumer Discretionary&Restaurants \\
80&GPC&Genuine Parts Co&Consumer Discretionary&Distributors \\
81&TGT&Target Corp&Consumer Discretionary&General Merchandise Stores \\
82&ROST&Ross Stores Inc&Consumer Discretionary&Apparel Retail \\
83&TJX&TJX Companies Inc&Consumer Discretionary&Apparel Retail \\
84&GPS&Gap Inc&Consumer Discretionary&Apparel Retail \\
85&BBY&Best Buy Co Inc&Consumer Discretionary&Computer \& Electronics Retail \\
86&HD&Home Depot Inc&Consumer Discretionary&Home Improvement Retail \\
87&LOW&Lowe's Companies Inc&Consumer Discretionary&Home Improvement Retail \\
88&BBWI&Bath \& Body Works Inc&Consumer Discretionary&Specialty Stores \\
89&WBA&Walgreens Boots Alliance Inc&Consumer Staples&Drug Retail \\
90&SYY&Sysco Corp&Consumer Staples&Food Distributors \\
91&KR&Kroger Co&Consumer Staples&Food Retail \\
92&WMT&Walmart Inc&Consumer Staples&Hypermarkets \& Super Centers \\
93&PEP&PepsiCo Inc&Consumer Staples&Soft Drinks \\
94&KO&Coca-Cola Co&Consumer Staples&Soft Drinks \\
95&ADM&Archer-Daniels-Midland Co&Consumer Staples&Agricultural Products \\
96&HSY&Hershey Co&Consumer Staples&Packaged Foods \& Meats \\
97&K&Kellogg Co&Consumer Staples&Packaged Foods \& Meats \\
98&GIS&General Mills Inc&Consumer Staples&Packaged Foods \& Meats \\
99&CAG&Conagra Brands Inc&Consumer Staples&Packaged Foods \& Meats \\
100&CPB&Campbell Soup Co&Consumer Staples&Packaged Foods \& Meats \\
101&HRL&Hormel Foods Corp&Consumer Staples&Packaged Foods \& Meats \\
102&MO&Altria Group Inc&Consumer Staples&Tobacco \\
103&CL&Colgate-Palmolive Co&Consumer Staples&Household Products \\
104&CLX&Clorox Co&Consumer Staples&Household Products \\
105&PG&Procter \& Gamble Co&Consumer Staples&Household Products \\
106&KMB&Kimberly-Clark Corp&Consumer Staples&Household Products \\
107&ABMD&ABIOMED Inc&Health Care&Health Care Equipment \\
108&BAX&Baxter International Inc&Health Care&Health Care Equipment \\
109&BDX&Becton Dickinson and Co&Health Care&Health Care Equipment \\
110&DHR&Danaher Corp&Health Care&Health Care Equipment \\
111&MDT&Medtronic PLC&Health Care&Health Care Equipment \\
112&TFX&Teleflex Inc&Health Care&Health Care Equipment \\
113&ABT&Abbott Laboratories&Health Care&Health Care Equipment \\
114&WST&West Pharmaceutical Services Inc&Health Care&Health Care Supplies \\
115&CVS&CVS Health Corp&Health Care&Health Care  Services \\
116&CI&Cigna Corp&Health Care&Health Care  Services \\
117&HUM&Humana Inc&Health Care&Managed Health Care \\
118&CERN&Cerner Corp&Health Care&Health Care Technology \\
119&AMGN&Amgen Inc&Health Care&Biotechnology \\
120&BIO&Bio Rad Laboratories Inc&Health Care&Life Sciences Tools \& Services \\
121&TMO&Thermo Fisher Scientific Inc&Health Care&Life Sciences Tools \& Services \\
122&PKI&PerkinElmer Inc&Health Care&Life Sciences Tools \& Services \\
123&LLY&Eli Lilly and Co&Health Care&Pharmaceuticals \\
124&VTRS&Viatris Inc&Health Care&Pharmaceuticals \\
125&JNJ&Johnson \& Johnson&Health Care&Pharmaceuticals \\
126&MRK&Merck \& Co Inc&Health Care&Pharmaceuticals \\
127&PFE&Pfizer Inc&Health Care&Pharmaceuticals \\
128&BMY&Bristol-Myers Squibb Co&Health Care&Pharmaceuticals \\
129&JPM&JPMorgan Chase \& Co&Financials&Diversified Banks \\
130&WFC&Wells Fargo \& Co&Financials&Diversified Banks \\
131&C&Citigroup Inc&Financials&Diversified Banks \\
132&BAC&Bank of America Corp&Financials&Diversified Banks \\
133&HBAN&Huntington Bancshares Inc&Financials&Regional Banks \\
134&MTB&M\&T Bank Corp&Financials&Regional Banks \\
135&PBCT&People's United Financial Inc&Financials&Regional Banks \\
136&PNC&PNC Financial Services Group Inc&Financials&Regional Banks \\
137&FITB&Fifth Third Bancorp&Financials&Regional Banks \\
138&ZION&Zions Bancorporation NA&Financials&Regional Banks \\
139&SIVB&SVB Financial Group&Financials&Regional Banks \\
140&BEN&Franklin Resources Inc&Financials&Asset Management \& Custody Banks \\
141&TROW&T Rowe Price Group Inc&Financials&Asset Management \& Custody Banks \\
142&BK&Bank of New York Mellon Corp&Financials&Asset Management \& Custody Banks \\
143&RJF&Raymond James Financial Inc&Financials&Investment Banking \& Brokerage \\
144&SCHW&Charles Schwab Corp&Financials&Investment Banking \& Brokerage \\
145&AXP&American Express Co&Financials&Consumer Finance \\
146&AJG&Arthur J Gallagher \& Co&Financials&Insurance Brokers \\
147&MMC&Marsh \& McLennan Companies Inc&Financials&Insurance Brokers \\
148&AON&Aon PLC&Financials&Insurance Brokers \\
149&GL&Globe Life Inc&Financials&Life \& Health Insurance \\
150&UNM&Unum Group&Financials&Life \& Health Insurance \\
151&AFL&Aflac Inc&Financials&Life \& Health Insurance \\
152&LNC&Lincoln National Corp&Financials&Life \& Health Insurance \\
153&AIG&American International Group Inc&Financials&Multi-line Insurance \\
154&CINF&Cincinnati Financial Corp&Financials&Property \& Casualty Insurance \\
155&L&Loews Corp&Financials&Property \& Casualty Insurance \\
156&PGR&Progressive Corp&Financials&Property \& Casualty Insurance \\
157&WELL&Welltower Inc&Real Estate&Health Care REITs \\
158&PEAK&Healthpeak Properties Inc&Real Estate&Health Care REITs \\
159&HST&Host Hotels \& Resorts Inc&Real Estate&Hotel \& Resort REITs \\
160&DRE&Duke Realty Corp&Real Estate&Industrial REITs \\
161&FRT&Federal Realty Investment Trust&Real Estate&Retail REITs \\
162&WY&Weyerhaeuser Co&Real Estate&Specialized REITs \\
163&PSA&Public Storage&Real Estate&Specialized REITs \\
164&AMAT&Applied Materials Inc&Information Technology&Semiconductor Equipment \\
165&TER&Teradyne Inc&Information Technology&Semiconductor Equipment \\
166&KLAC&KLA Corp&Information Technology&Semiconductor Equipment \\
167&LRCX&Lam Research Corp&Information Technology&Semiconductor Equipment \\
168&AMD&Advanced Micro Devices Inc&Information Technology&Semiconductors \\
169&TXN&Texas Instruments Inc&Information Technology&Semiconductors \\
170&ADI&Analog Devices Inc&Information Technology&Semiconductors \\
171&INTC&Intel Corp&Information Technology&Semiconductors \\
172&PAYX&Paychex Inc&Information Technology&Data Processing \& Outsourced Services \\
173&FISV&Fiserv Inc&Information Technology&Data Processing \& Outsourced Services \\
174&ADP&Automatic Data Processing Inc&Information Technology&Data Processing \& Outsourced Services \\
175&IBM&International Business Machines Corp&Information Technology&IT Consulting \& Other Services \\
176&DXC&DXC Technology Co&Information Technology&IT Consulting \& Other Services \\
177&PTC&PTC Inc&Information Technology&Application Software \\
178&ADBE&Adobe Inc&Information Technology&Application Software \\
179&ADSK&Autodesk Inc&Information Technology&Application Software \\
180&NLOK&NortonLifeLock Inc&Information Technology&Systems Software \\
181&ORCL&Oracle Corp&Information Technology&Systems Software \\
182&MSFT&Microsoft Corp&Information Technology&Systems Software \\
183&MSI&Motorola Solutions Inc&Information Technology&Communications Equipment \\
184&GLW&Corning Inc&Information Technology&Electronic Components \\
185&WDC&Western Digital Corp&Information Technology&Technology Hardware, Storage \& Peripherals \\
186&HPQ&HP Inc&Information Technology&Technology Hardware, Storage \& Peripherals \\
187&AAPL&Apple Inc&Information Technology&Technology Hardware, Storage \& Peripherals \\
188&EA&Electronic Arts Inc&Communication Services&Interactive Home Entertainment \\
189&DIS&Walt Disney Co&Communication Services&Movies \& Entertainment \\
190&IPG&Interpublic Group of Companies Inc&Communication Services&Advertising \\
191&CMCSA&Comcast Corp&Communication Services&Cable \& Satellite \\
192&LUMN&Lumen Technologies Inc&Communication Services&Alternative Carriers \\
193&T&AT\&T Inc&Communication Services&Integrated Telecommunication Services \\
194&VZ&Verizon Communications Inc&Communication Services&Integrated Telecommunication Services \\
195&EVRG&Evergy Inc&Utilities&Electric Utilities \\
196&LNT&Alliant Energy Corp&Utilities&Electric Utilities \\
197&EXC&Exelon Corp&Utilities&Electric Utilities \\
198&DUK&Duke Energy Corp&Utilities&Electric Utilities \\
199&PNW&Pinnacle West Capital Corp&Utilities&Electric Utilities \\
200&NEE&Nextera Energy Inc&Utilities&Electric Utilities \\
201&ES&Eversource Energy&Utilities&Electric Utilities \\
202&AEP&American Electric Power Company Inc&Utilities&Electric Utilities \\
203&EIX&Edison International&Utilities&Electric Utilities \\
204&XEL&Xcel Energy Inc&Utilities&Electric Utilities \\
205&ETR&Entergy Corp&Utilities&Electric Utilities \\
206&PPL&PPL Corp&Utilities&Electric Utilities \\
207&SO&Southern Co&Utilities&Electric Utilities \\
208&ATO&Atmos Energy Corp&Utilities&Gas Utilities \\
209&ED&Consolidated Edison Inc&Utilities&Multi-Utilities \\
210&D&Dominion Energy Inc&Utilities&Multi-Utilities \\
211&CMS&CMS Energy Corp&Utilities&Multi-Utilities \\
212&PEG&Public Service Enterprise Group Inc&Utilities&Multi-Utilities \\
213&DTE&DTE Energy Co&Utilities&Multi-Utilities \\
		\bottomrule
	\end{longtable}
}

\newpage

\section{\label{sec:Subsamples}Analysis of subsamples}

\begin{figure*}[!htb]
	\centering
	\begin{minipage}{1.0\textwidth}
		\subfloat%
		{\includegraphics[width=1.0\linewidth]{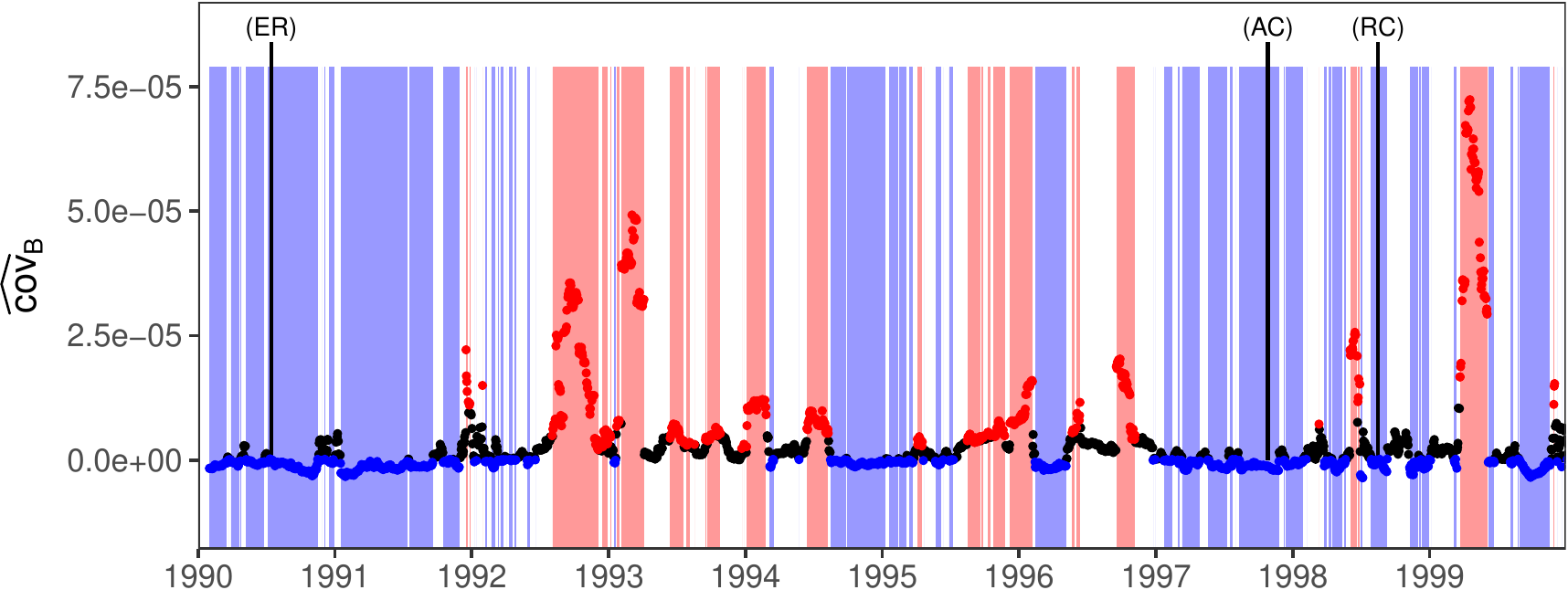}
		}\par
		\subfloat%
		{\includegraphics[width=1.0\textwidth]{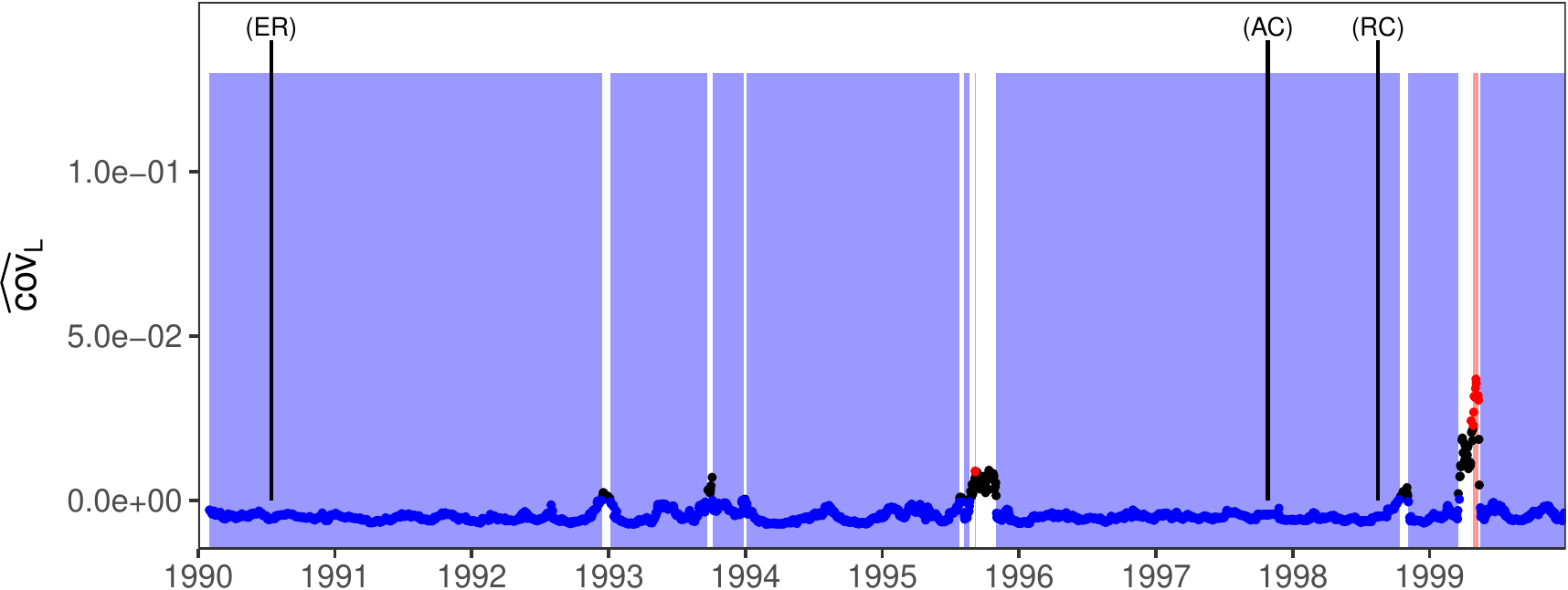}
		}%
	\end{minipage}%
	\caption{\label{subfig:Main:MeanValCovP1_Sub100}For $100$ randomly selected stocks from all 213 stocks: Time evolutions (1990-1999) for the mean reduced-rank covariances $\meanNONDiag{\text{cov}}_{{B}}$ (top) and $\meanNONDiag{\text{cov}}_{{L}}$ (bottom).
		Three criteria (red/blue) for relative collectivity measures are described in Sec.~\ref{sec:AverageSectorCollectivity}. Historical events are listed in Tab.~\ref{tab:FinancialCrises}. Black dots belong to covariance matrices whose collectivities do not fulfill the three criteria.}
\end{figure*}
\begin{figure*}[!htb]
	\centering
	\begin{minipage}{1.0\textwidth}
		\subfloat%
		{\includegraphics[width=1.0\linewidth]{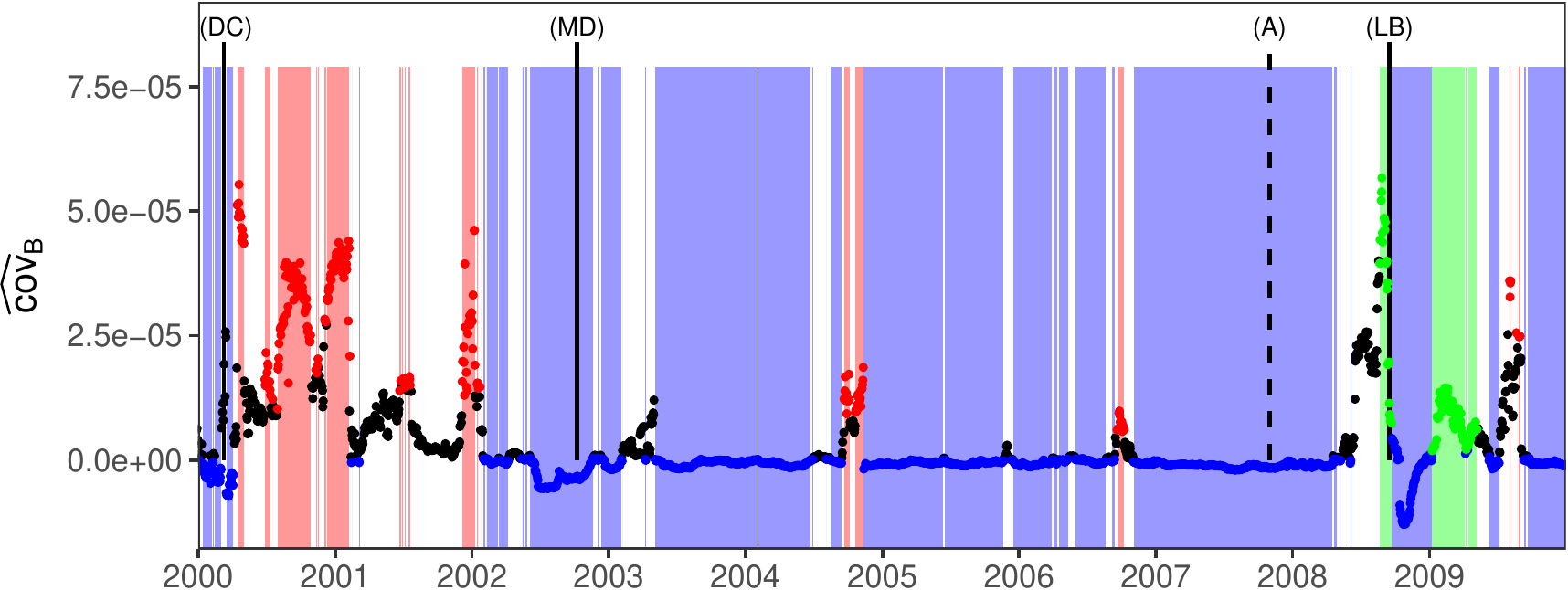}
		}\par
		\subfloat%
		{\includegraphics[width=1.0\textwidth]{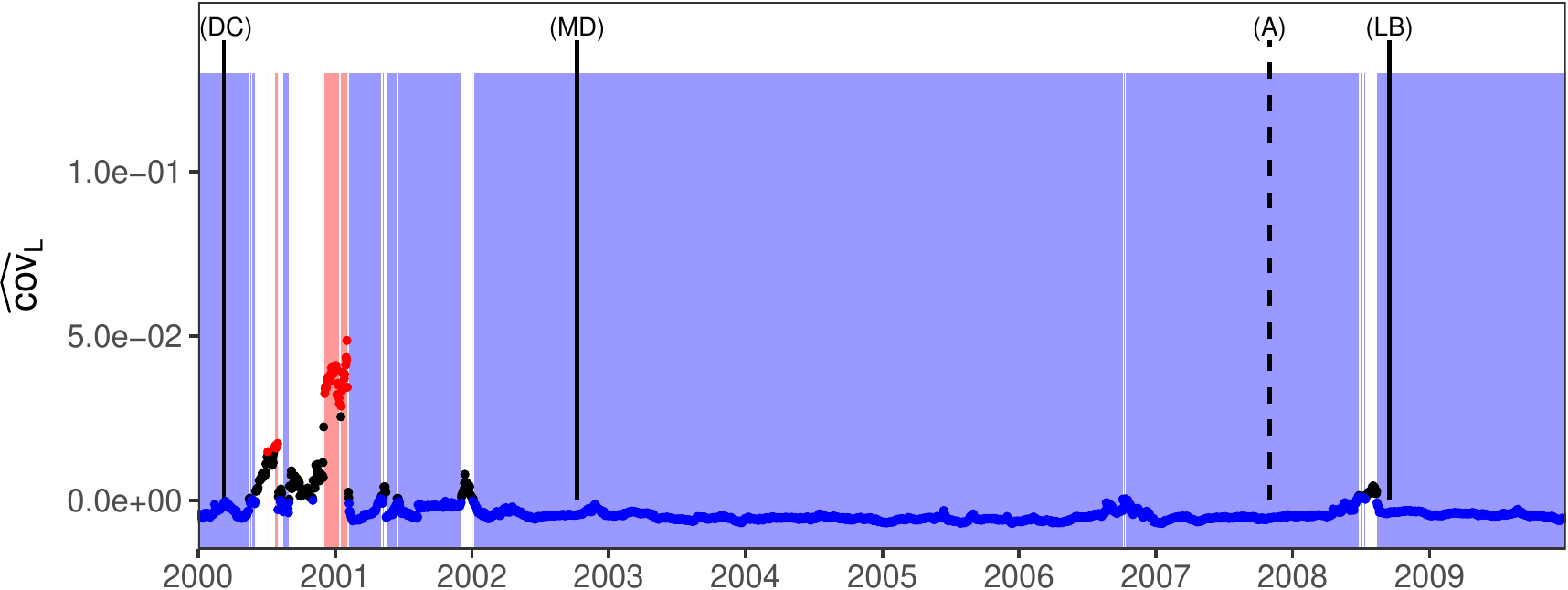}
		}%
	\end{minipage}%
	\caption{\label{subfig:Main:MeanValCovP2_Sub100}For $100$ randomly selected stocks from all 213 stocks: Time evolutions (2000-2009) for the mean reduced-rank covariances $\meanNONDiag{\text{cov}}_{{B}}$ (top) and $\meanNONDiag{\text{cov}}_{{L}}$ (bottom).
		Three criteria (red/blue/green) for absolute and relative collectivity measures are described in Sec.~\ref{sec:AverageSectorCollectivity}. Historical events are listed in Tab.~\ref{tab:FinancialCrises}. Black dots belong to covariance matrices whose collectivities do not fulfill the three criteria.}
\end{figure*}
\begin{figure*}[!htb]
	\centering
	\begin{minipage}{1.0\textwidth}
		\subfloat%
		{\includegraphics[width=1.0\linewidth]{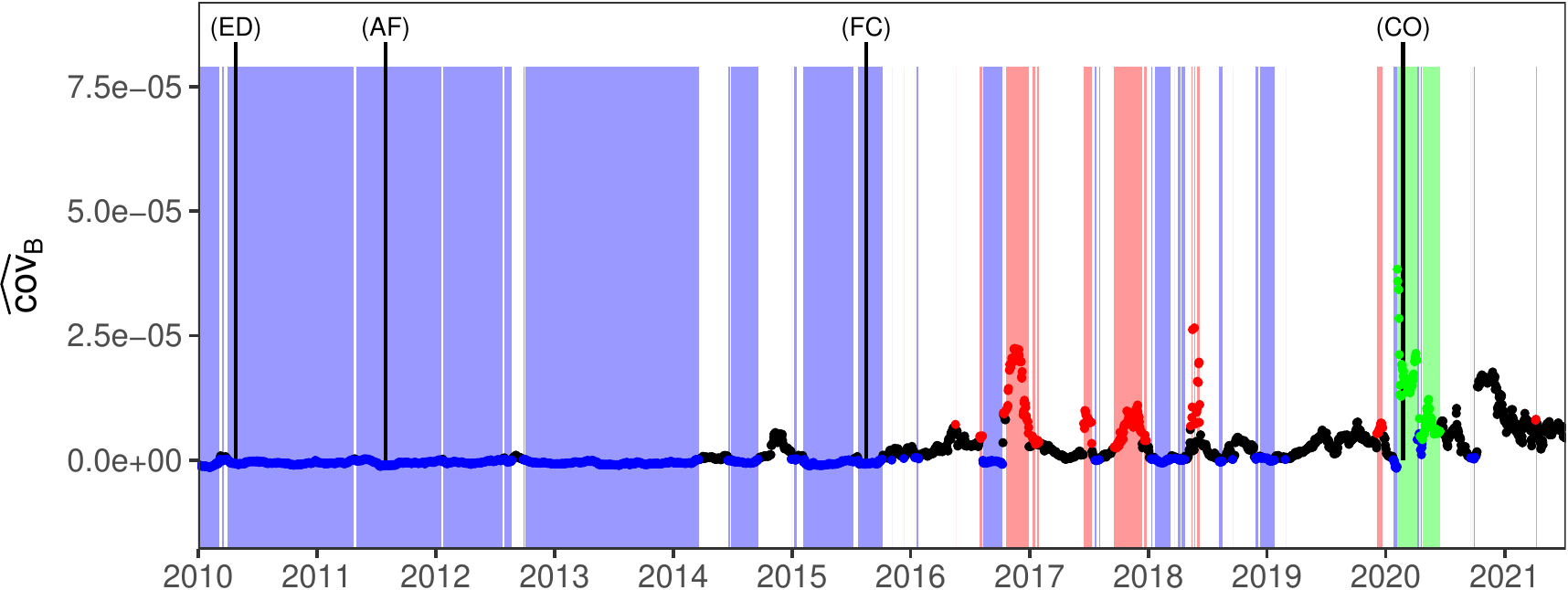}
		}\par
		\subfloat%
		{\includegraphics[width=1.0\textwidth]{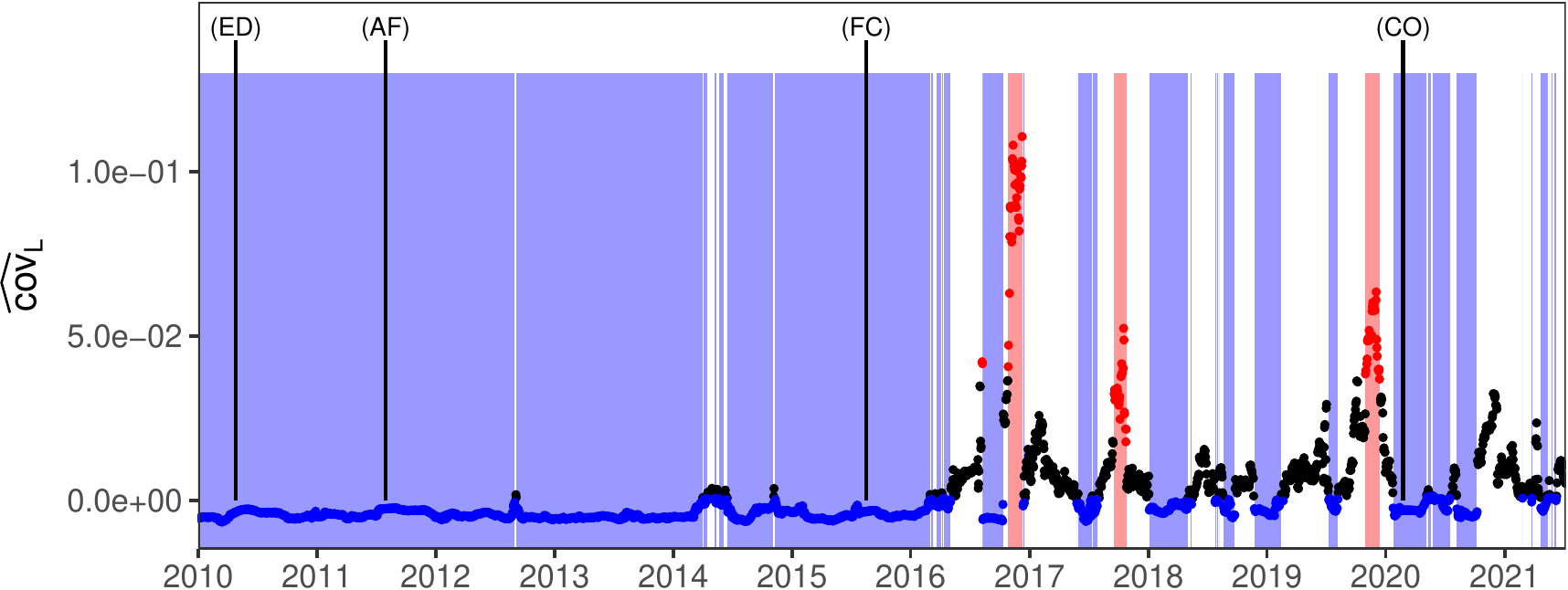}
		}%
	\end{minipage}%
	\caption{\label{subfig:Main:MeanValCovP3_Sub100}For $100$ randomly selected stocks from all 213 stocks: Time evolutions (2010-2021) for the mean reduced-rank covariances $\meanNONDiag{\text{cov}}_{{B}}$ (top) and $\meanNONDiag{{cov}}_{\text{L}}$ (bottom).
		Three criteria (red/blue/green) for absolute and relative collectivity measures are described in Sec.~\ref{sec:AverageSectorCollectivity}. Historical events are listed in Tab.~\ref{tab:FinancialCrises}. Black dots belong to covariance matrices whose collectivities do not fulfill the three criteria.}
\end{figure*}
\begin{figure*}[!htb]
	\centering
	\begin{minipage}{0.5\textwidth}
		\subfloat[\label{subfig:RiskPhaseCovMean_Sub100}]
		{\includegraphics[width=1.0\textwidth]{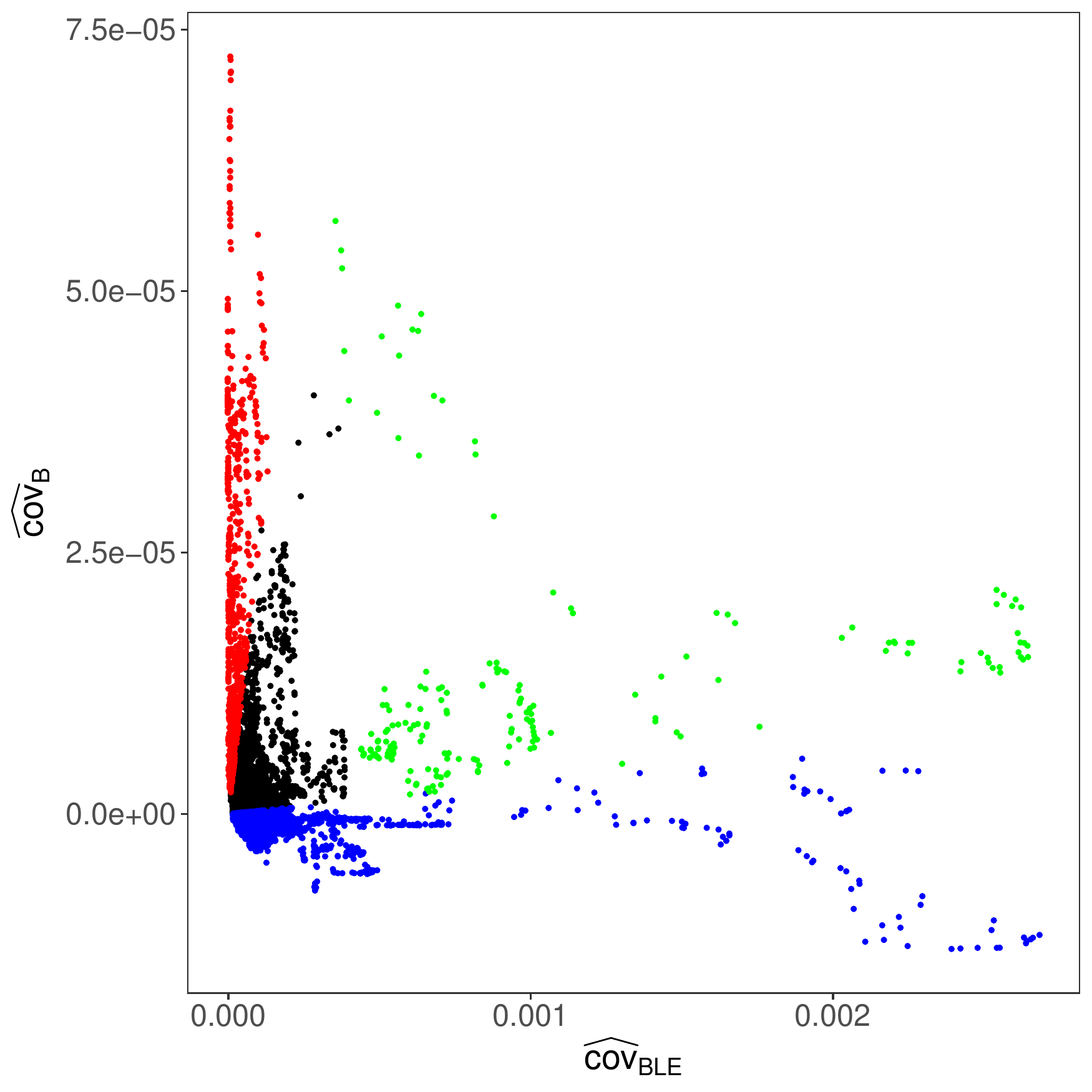}
		}
	\end{minipage}%
	\begin{minipage}{0.5\textwidth}
		\subfloat[\label{subfig:RiskPhaseCorrMean_Sub100}]
		{\includegraphics[width=1.0\textwidth]{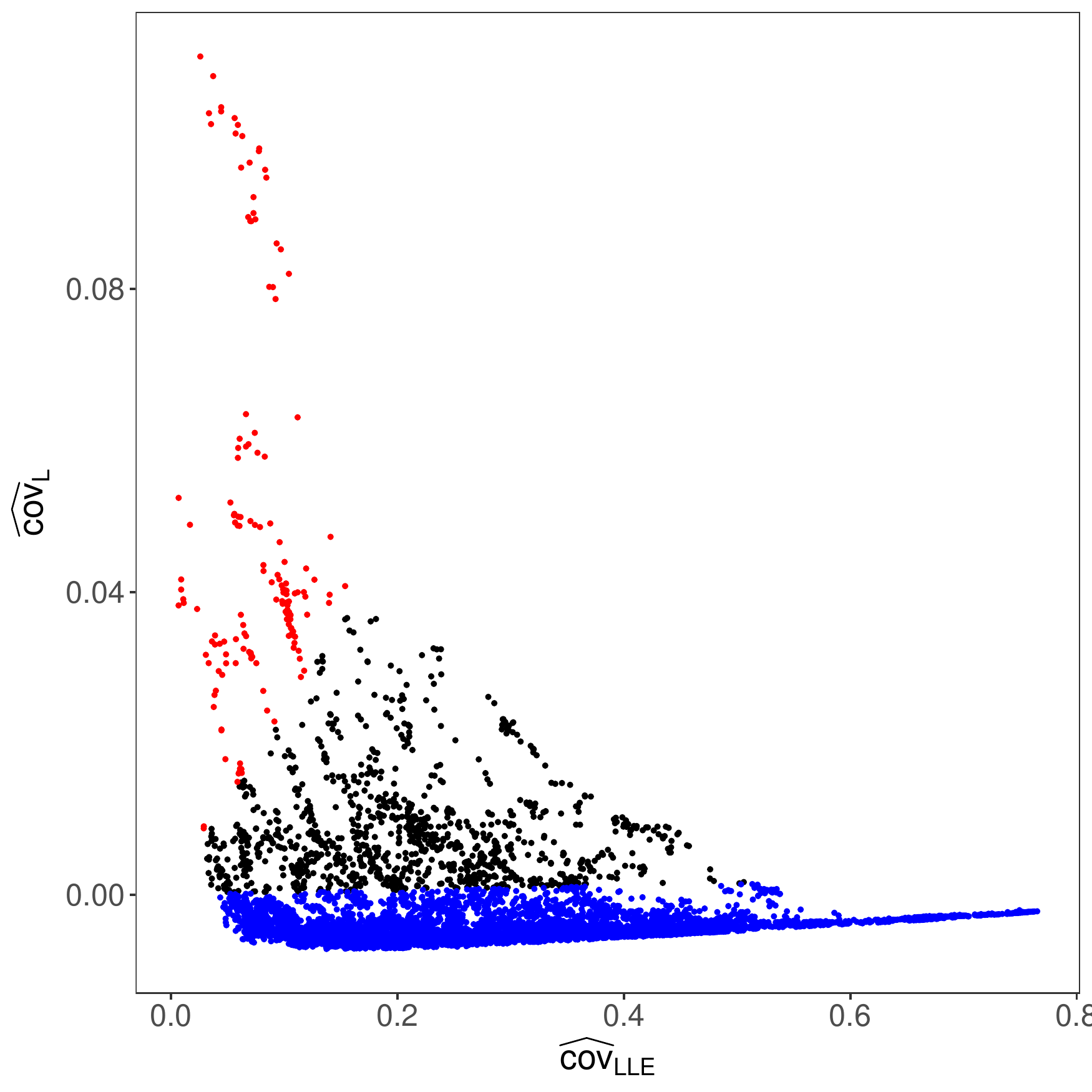}
		}
	\end{minipage}
	\caption{\label{subfig:Main:RiskPhaseMean_Sub100}For $100$ randomly selected stocks from all 213 stocks: Risk-phase diagrams derived from \protect\subref{subfig:RiskPhaseCovMean} covariance matrix $\Sigma$ with $\meanNONDiag{\text{cov}}_{{B}}$ plotted versus $\meanNONDiag{\text{cov}}_{\text{BLE}}$ and \protect\subref{subfig:RiskPhaseCorrMean} correlation matrix $C$ with $\meanNONDiag{\text{cov}}_{{L}}$ plotted versus $\meanNONDiag{\text{cov}}_{\text{LLE}}$.  Each dot corresponds to one of the 7920 intervals of 42 trading days (see~Sec.~\ref{sec:DataSet}). 
		The colored dots (red/blue/green) correspond to the three criteria for absolute and relative collectivity measures described in Sec.~\ref{sec:AverageSectorCollectivity}.
		The green ones are only displayed in \protect\subref{subfig:RiskPhaseCovMean}.
		Black dots belong to covariance matrices whose collectivities do not fulfill the three criteria.}
\end{figure*}

\clearpage

\section{\label{sec:TimeEvoLinRegress}Time evolutions for average sector collectivities obtained by linear regression methods}

\begin{figure*}[!htb]
	\centering
	\begin{minipage}{1.0\textwidth}
		\subfloat%
		{\includegraphics[width=1.0\linewidth]{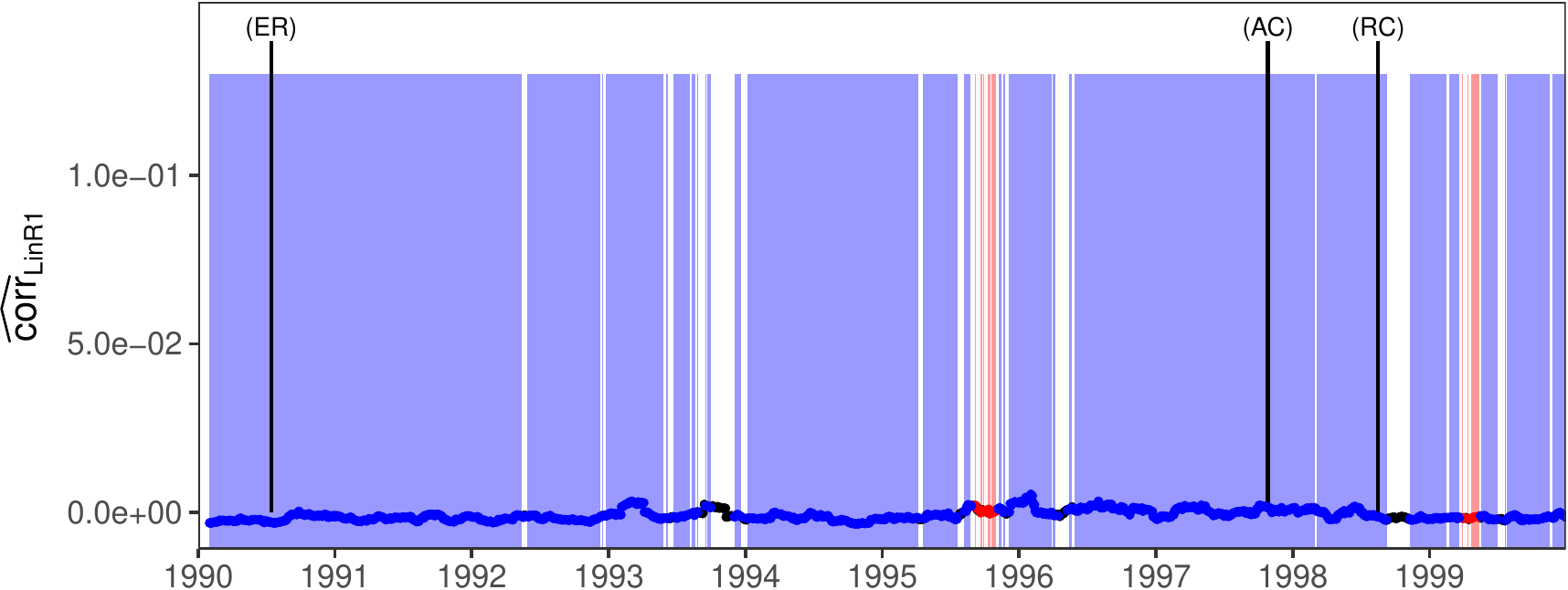}
		}\par
		\subfloat%
		{\includegraphics[width=1.0\textwidth]{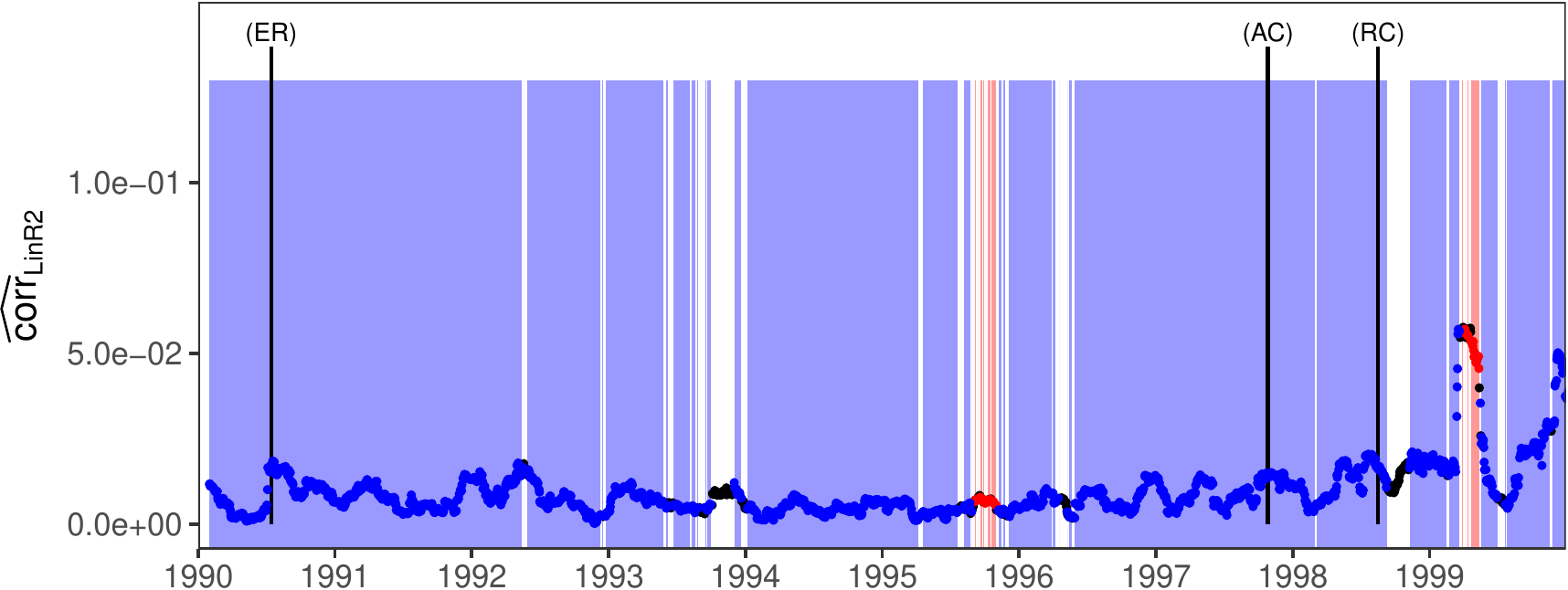}
		}%
	\end{minipage}%
	\caption{\label{subfig:Main:MeanValCovP1_LinR}For the two linear regression methods defined in Sec.~\ref{sec:RiskPhaseLinRegress}: Time evolutions (1990-1999) for $\meanNONDiag{\text{corr}}_{\text{LinR1}}$ (top) and $\meanNONDiag{\text{corr}}_{\text{LinR2}}$ (bottom).
		Three criteria (red/blue) for relative collectivity measures are described in Sec.~\ref{sec:AverageSectorCollectivity}. Historical events are listed in Tab.~\ref{tab:FinancialCrises}. Black dots belong to covariance matrices whose collectivities do not fulfill the three criteria.}
\end{figure*}
\begin{figure*}[!htb]
	\centering
	\begin{minipage}{1.0\textwidth}
		\subfloat%
		{\includegraphics[width=1.0\linewidth]{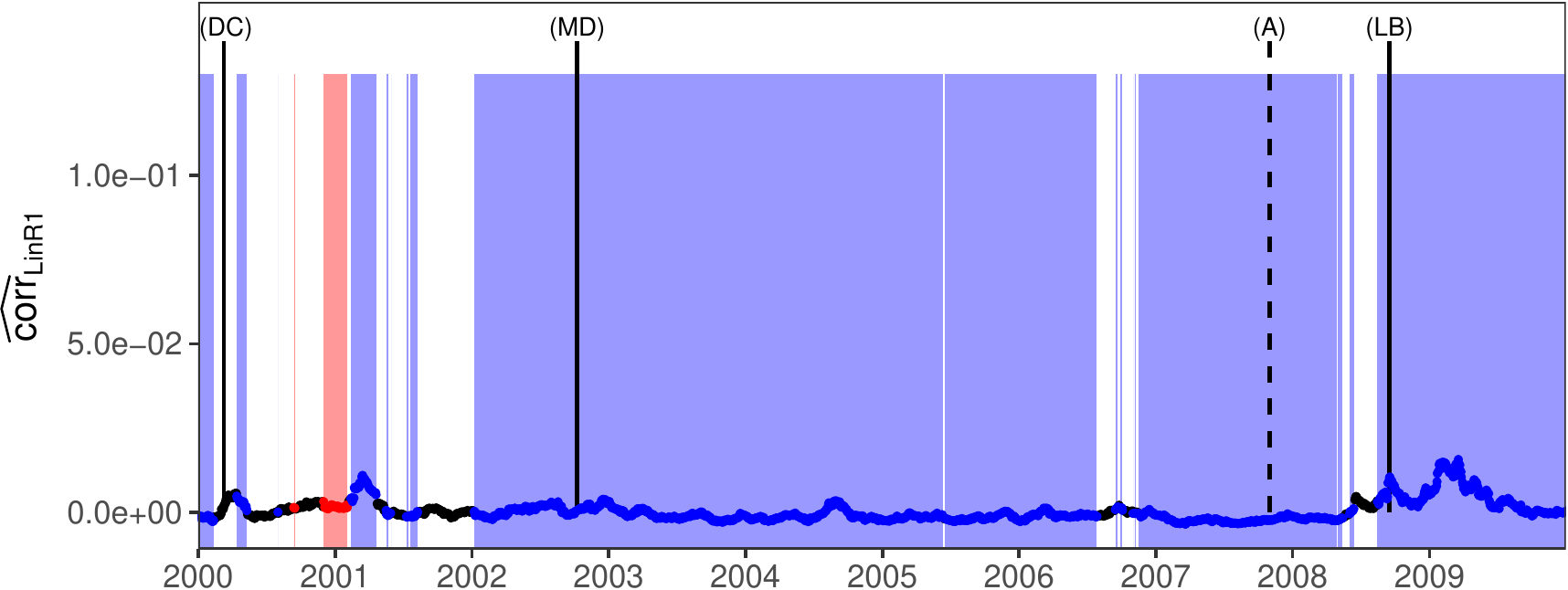}
		}\par
		\subfloat%
		{\includegraphics[width=1.0\textwidth]{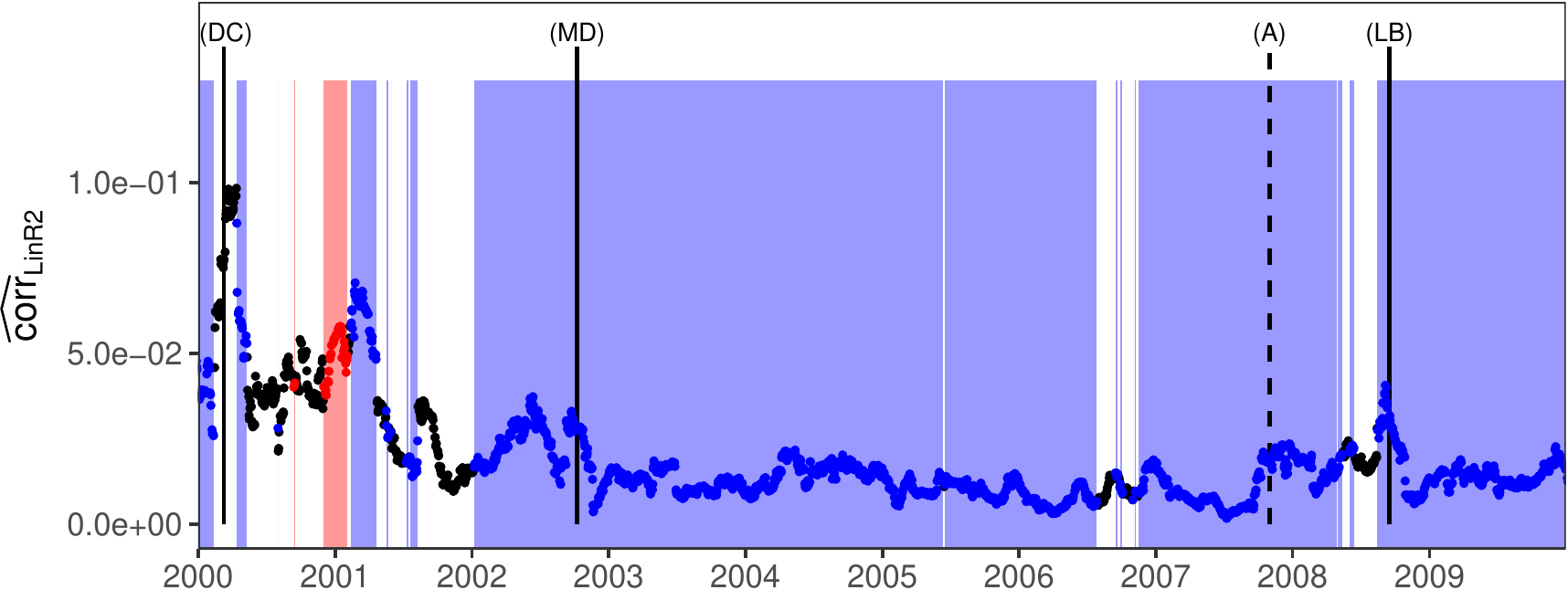}
		}%
	\end{minipage}%
	\caption{\label{subfig:Main:MeanValCovP2_LinR}For the two linear regression methods defined in Sec.~\ref{sec:RiskPhaseLinRegress}: Time evolutions (1990-1999) for $\meanNONDiag{\text{corr}}_{\text{LinR1}}$ (top) and $\meanNONDiag{\text{corr}}_{\text{LinR2}}$ (bottom).
		Three criteria (red/blue/green) for absolute and relative collectivity measures are described in Sec.~\ref{sec:AverageSectorCollectivity}. Historical events are listed in Tab.~\ref{tab:FinancialCrises}. Black dots belong to covariance matrices whose collectivities do not fulfill the three criteria.}
\end{figure*}
\begin{figure*}[!htb]
	\centering
	\begin{minipage}{1.0\textwidth}
		\subfloat%
		{\includegraphics[width=1.0\linewidth]{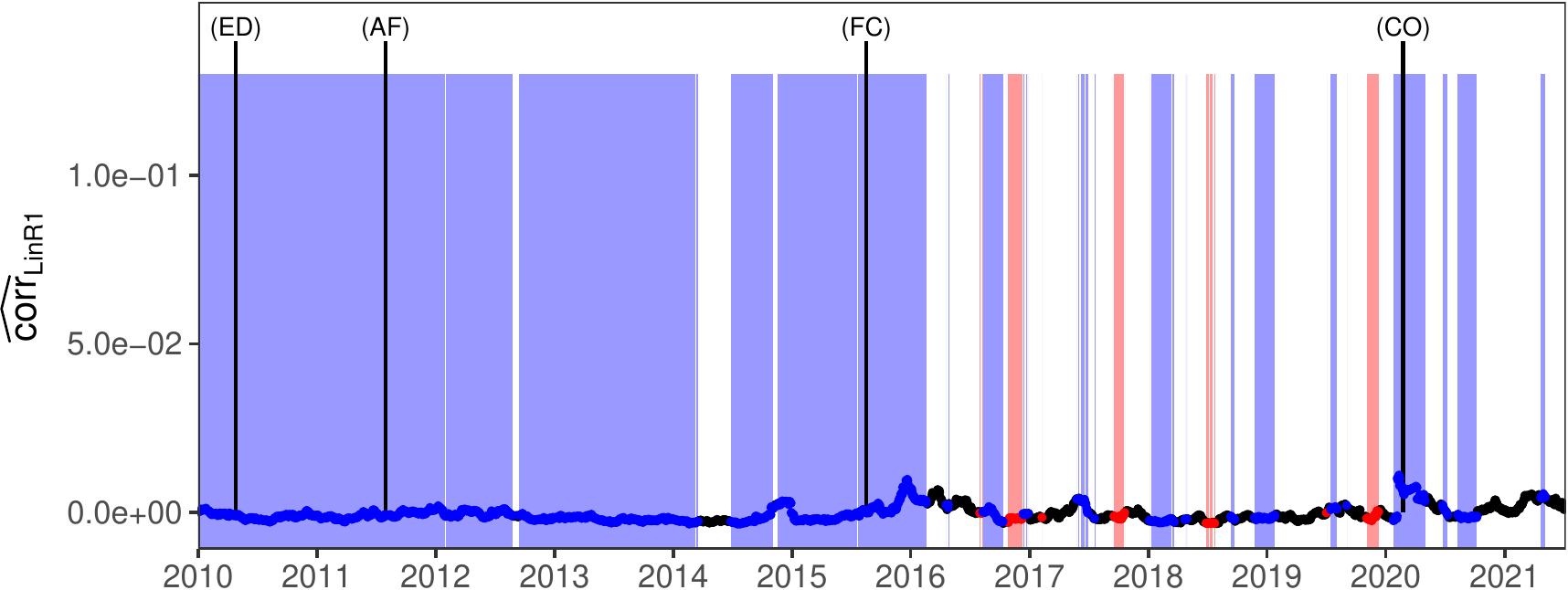}
		}\par
		\subfloat%
		{\includegraphics[width=1.0\textwidth]{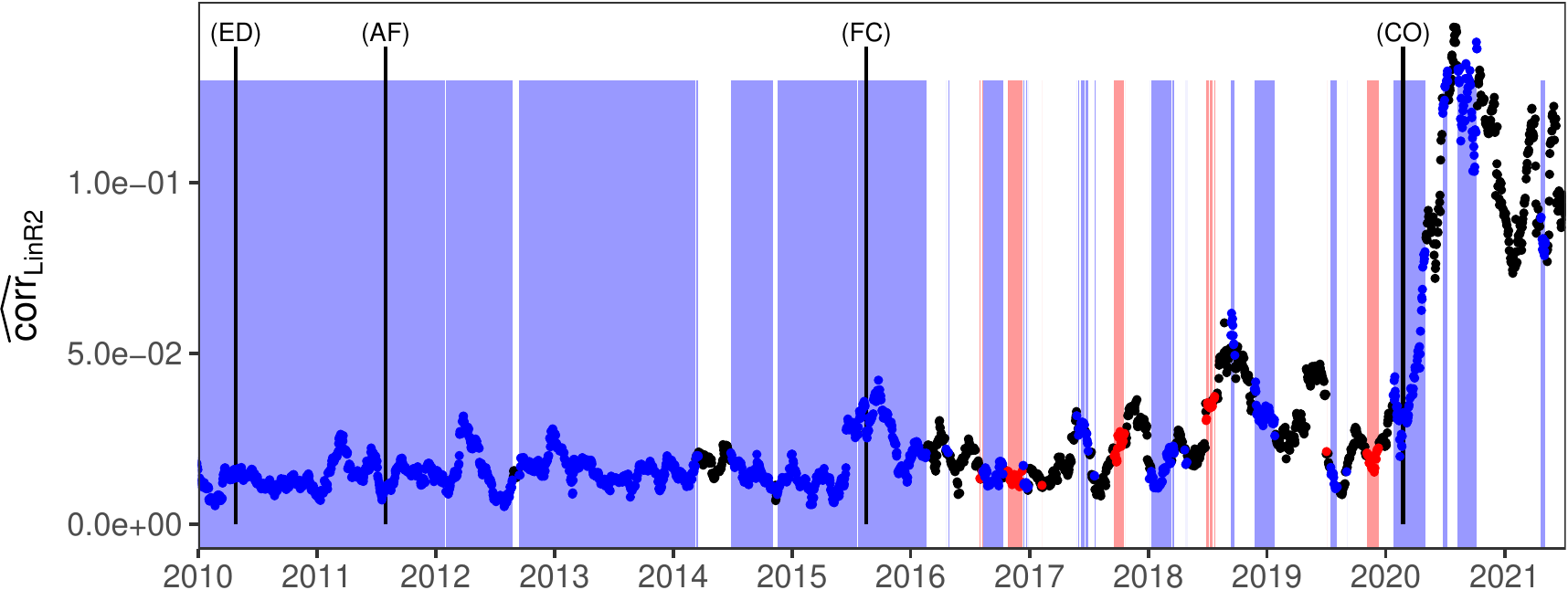}
		}%
	\end{minipage}%
	\caption{\label{subfig:Main:MeanValCovP3_LinR}For the two linear regression methods defined in Sec.~\ref{sec:RiskPhaseLinRegress}: Time evolutions (1990-1999) for $\meanNONDiag{\text{corr}}_{\text{LinR1}}$ (top) and $\meanNONDiag{\text{corr}}_{\text{LinR2}}$ (bottom).
		Three criteria (red/blue/green) for absolute and relative collectivity measures are described in Sec.~\ref{sec:AverageSectorCollectivity}. Historical events are listed in Tab.~\ref{tab:FinancialCrises}. Black dots belong to covariance matrices whose collectivities do not fulfill the three criteria.}
\end{figure*}

\clearpage

\section{\label{sec:TimeEvoHighCollec}Time evolutions of average sector collectivities for higher collective orders}

\begin{figure*}[!htb]
	\centering
	\begin{minipage}{1.0\textwidth}
		\subfloat%
		{\includegraphics[width=1.0\linewidth]{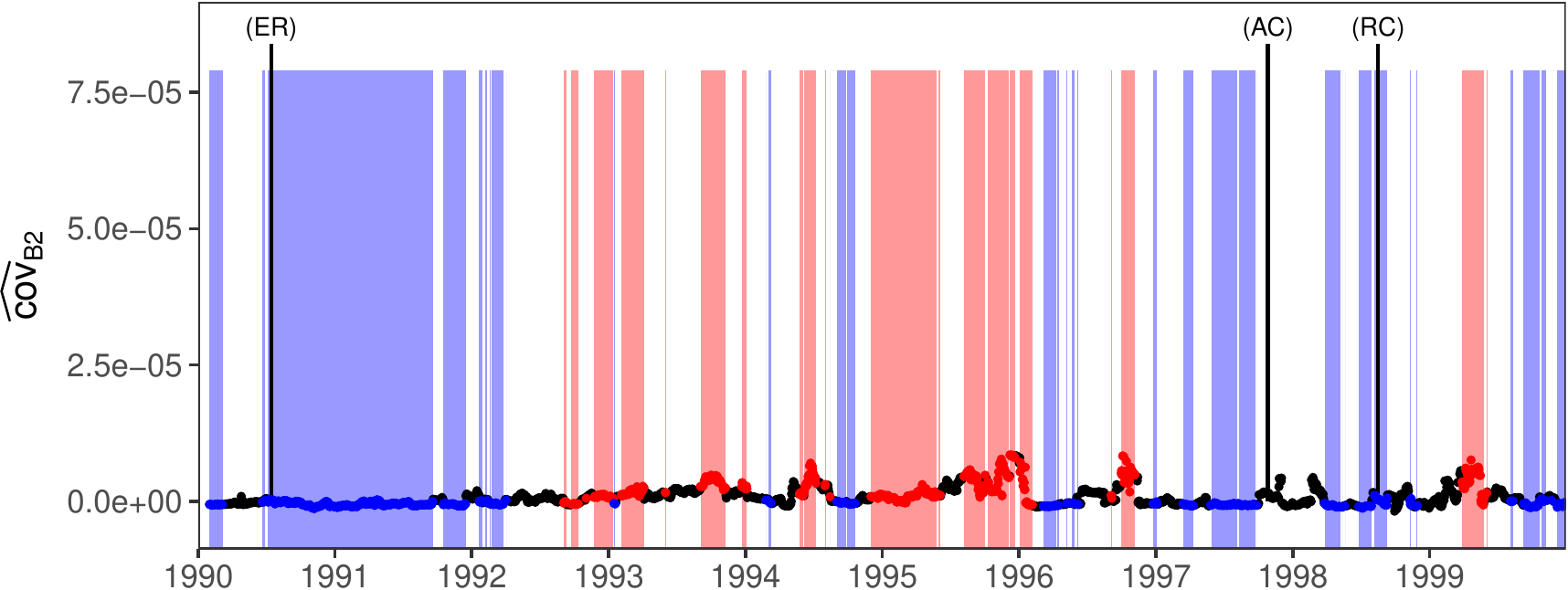}
		}\par
		\subfloat%
		{\includegraphics[width=1.0\textwidth]{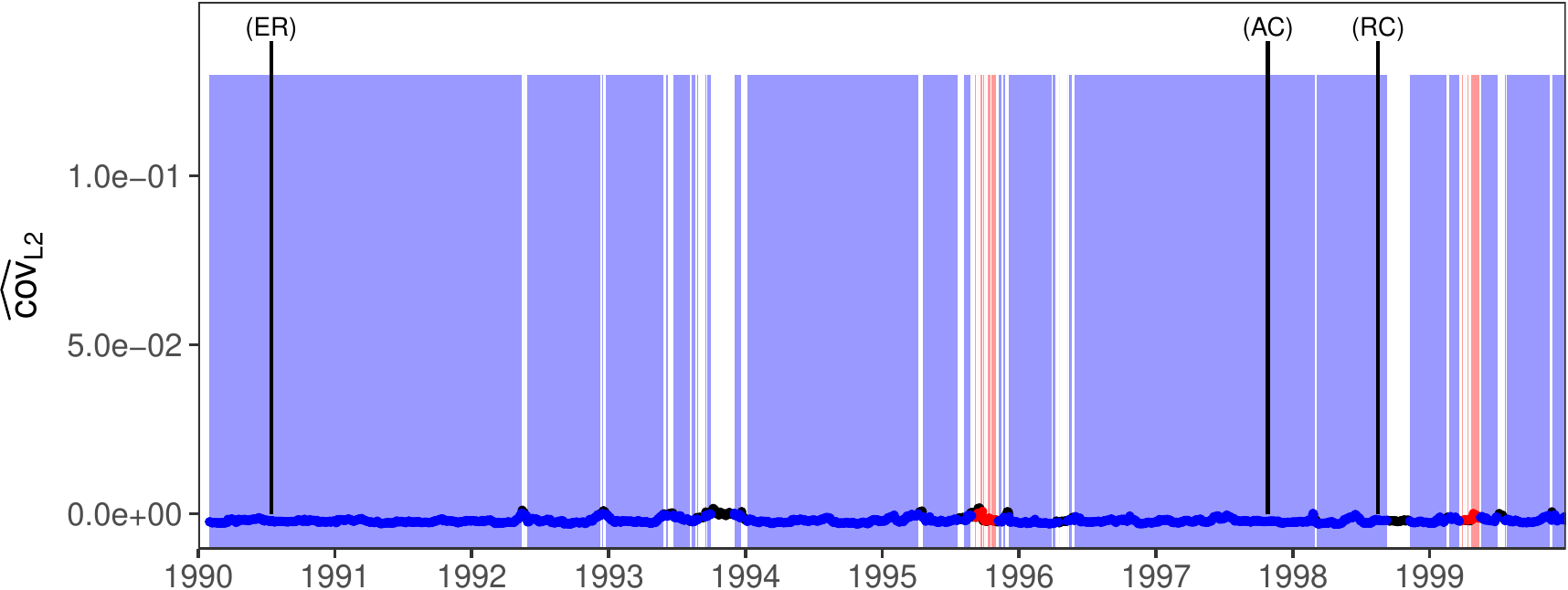}
		}%
	\end{minipage}%
	\caption{\label{subfig:Main:MeanValCovP1_BL2}Higher collective orders defined in Sec.~\ref{sec:HigherOrderEffects}: Time evolutions (1990-1999) for $\meanNONDiag{\text{corr}}_{B2}$ (top) and $\meanNONDiag{\text{corr}}_{L2}$ (bottom).
		Three criteria (red/blue) for relative collectivity measures are described in Sec.~\ref{sec:AverageSectorCollectivity}. Historical events are listed in Tab.~\ref{tab:FinancialCrises}. Black dots belong to covariance matrices whose collectivities do not fulfill the three criteria.}
\end{figure*}
\begin{figure*}[!htb]
	\centering
	\begin{minipage}{1.0\textwidth}
		\subfloat%
		{\includegraphics[width=1.0\linewidth]{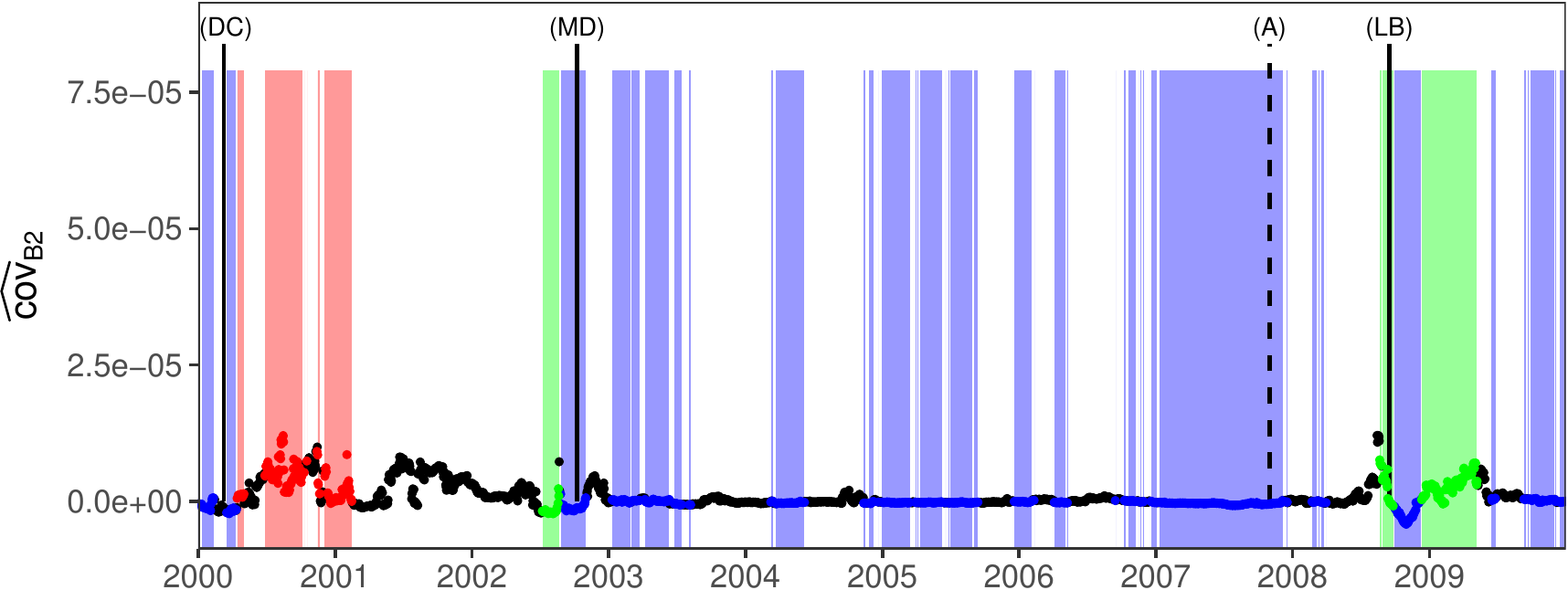}
		}\par
		\subfloat%
		{\includegraphics[width=1.0\textwidth]{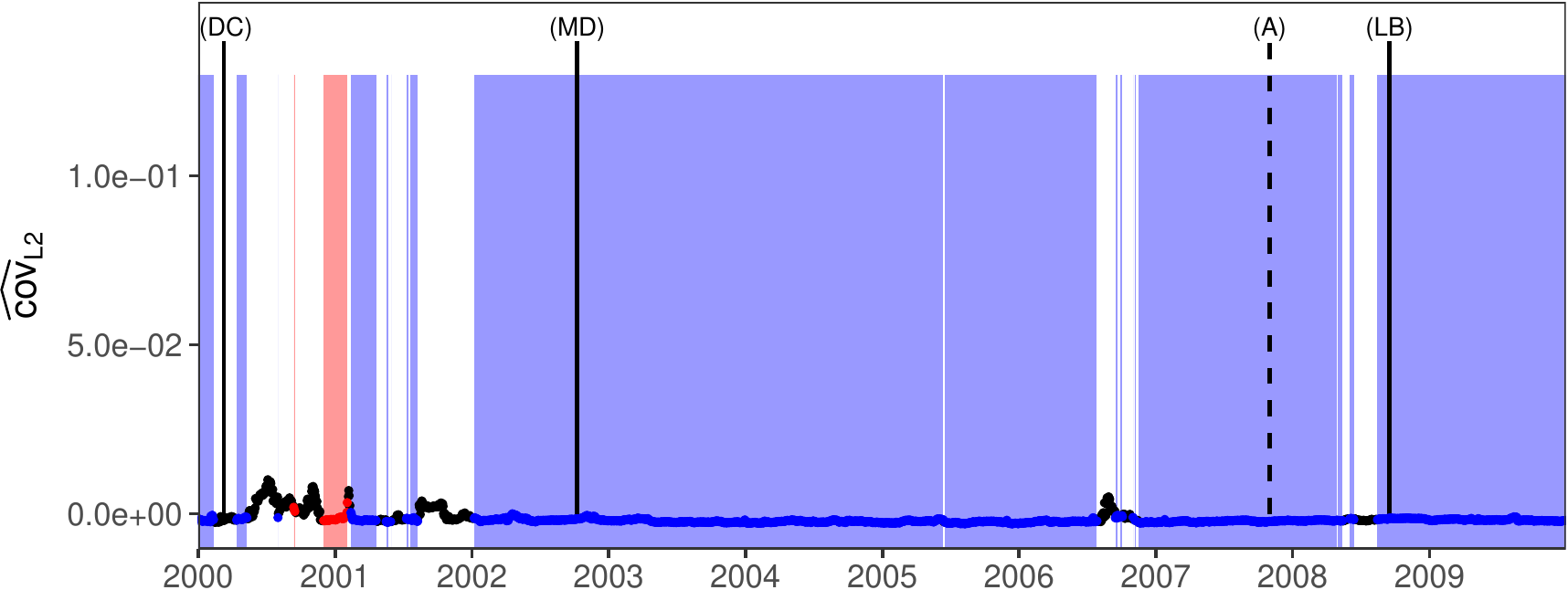}
		}%
	\end{minipage}%
	\caption{\label{subfig:Main:MeanValCovP2_BL2}Higher collective orders defined in Sec.~\ref{sec:HigherOrderEffects}: Time evolutions (1990-1999) for $\meanNONDiag{\text{corr}}_{B2}$ (top) and $\meanNONDiag{\text{corr}}_{L2}$ (bottom).
		Three criteria (red/blue/green) for absolute and relative collectivity measures are described in Sec.~\ref{sec:AverageSectorCollectivity}. Historical events are listed in Tab.~\ref{tab:FinancialCrises}. Black dots belong to covariance matrices whose collectivities do not fulfill the three criteria.}
\end{figure*}
\begin{figure*}[!htb]
	\centering
	\begin{minipage}{1.0\textwidth}
		\subfloat%
		{\includegraphics[width=1.0\linewidth]{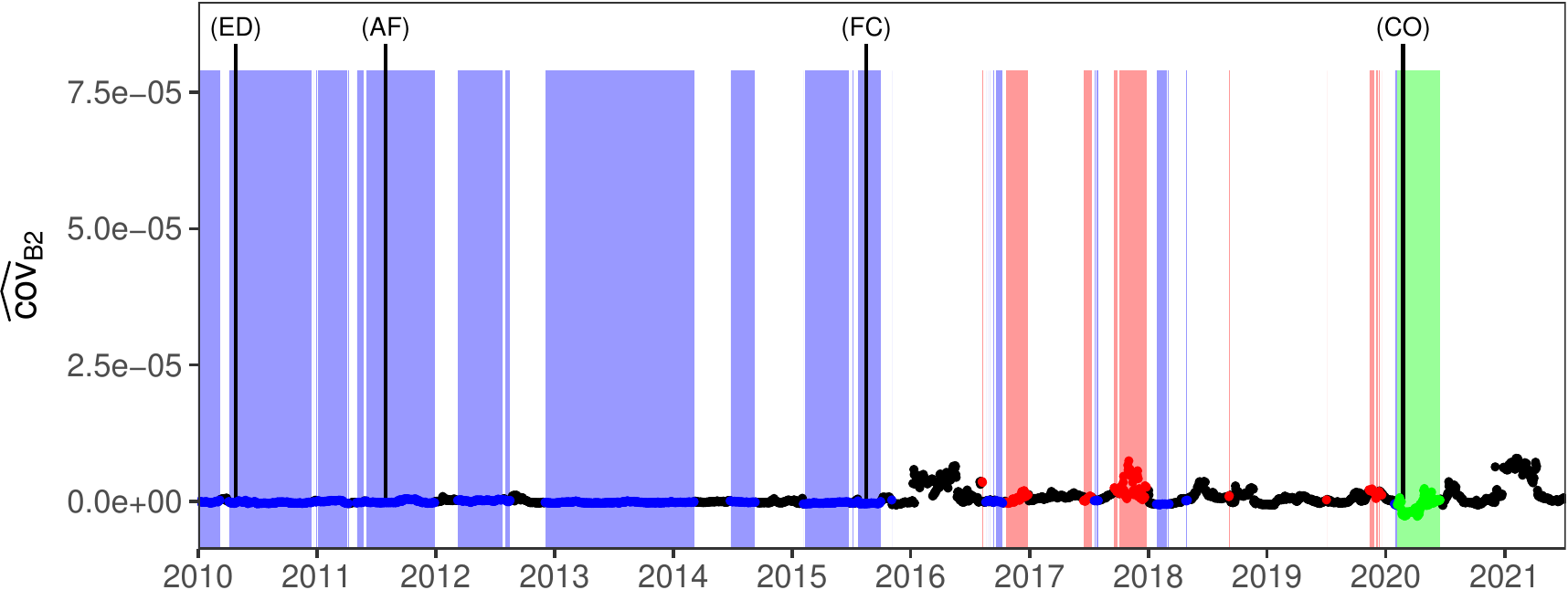}
		}\par
		\subfloat%
		{\includegraphics[width=1.0\textwidth]{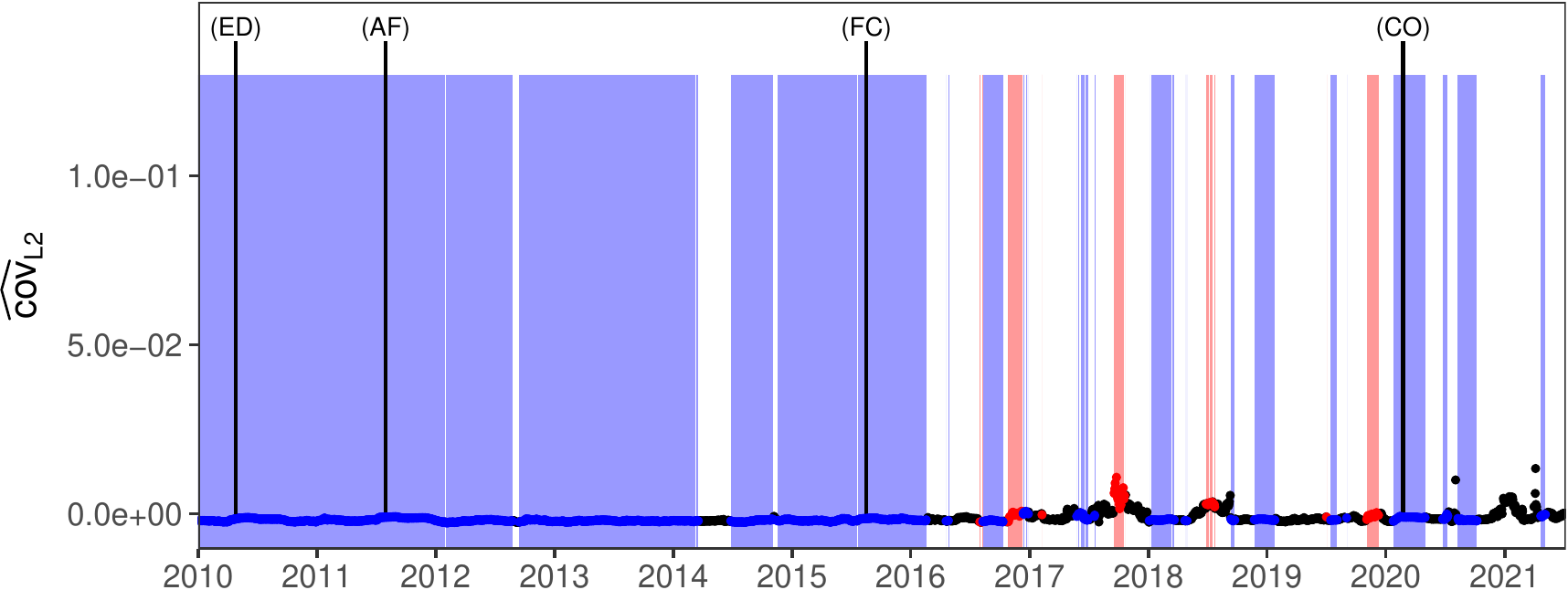}
		}%
	\end{minipage}%
	\caption{\label{subfig:Main:MeanValCovP3_BL2}Higher collective orders defined in Sec.~\ref{sec:HigherOrderEffects}: Time evolutions (1990-1999) for $\meanNONDiag{\text{corr}}_{B2}$ (top) and $\meanNONDiag{\text{corr}}_{L2}$ (bottom).
		Three criteria (red/blue/green) for absolute and relative collectivity measures are described in Sec.~\ref{sec:AverageSectorCollectivity}. Historical events are listed in Tab.~\ref{tab:FinancialCrises}. Black dots belong to covariance matrices whose collectivities do not fulfill the three criteria.}
\end{figure*}

\end{document}